\documentclass[apj]{emulateapj} 
\usepackage{lmodern,blindtext}
\usepackage{mathtools}
\usepackage[colorlinks=true,citecolor=blue]{hyperref}
\usepackage{multirow}
\usepackage{rotating}

\def\apjl{{ApJ}}%
\def\apjs{{ApJS}}%
\def\aap{{A\&A}}%
\shorttitle{X-ray photon index comparison of NLSy1 and BLSy1}
\shortauthors{Ojha et al.}

\begin{document}

\title{A comparison of X-ray photon indices among the narrow  and broad-line Seyfert 1 galaxies}

\author{Vineet Ojha$^{1}$, Hum Chand$^{1, 2}$, Gulab Chand Dewangan$^{3}$, Suvendu Rakshit$^{4}$}

\affil{$^1$Aryabhatta Research Institute of Observational Sciences (ARIES), Manora Peak, Nainital $-$ 263001, India \\
$^{2}$Department of Physics and Astronomical Sciences, Central University of Himachal Pradesh (CUHP), Dharamshala-176215, India \\
$^{3}$Inter-University Centre for Astronomy and Astrophysics (IUCAA), Pune $-$ 411007, India\\ $^{4}$Finnish Centre for Astronomy  with ESO (FINCA), University of Turku, Quantum, Vesilinnantie 5, 20014 University of Turku, Finland\\
   \url{vineetojhabhu@gmail.com}\\ }

\begin{abstract}
We present a detailed comparative systematic study using a sample of
221 Narrow-line Seyfert 1 (NLSy1) galaxies in comparison to a redshift matched sample of  154 Broad-line Seyfert 1 (BLSy1)
galaxies based on their observations using {\it ROSAT} and/or {\it XMM-Newton}
telescopes in soft X-ray band (0.1-2.0 keV). A homogeneous analysis is
carried out to estimate their soft X-ray photon indices
($\Gamma^{s}_{X}$) and its correlations with other parameters of
nuclear activities such as Eddington ratios (R$_\mathrm{Edd}$),
bolometric luminosities (L$_\mathrm{bol}$), black hole masses
(M$_\mathrm{BH}$) and the widths of the broad component of H$\beta$
lines (FWHM(H$\beta$)). In our analysis, we found clear evidence of
the difference in the $\Gamma^{s}_{X}$ and R$_\mathrm{Edd}$
distributions among NLSy1 and BLSy1 galaxies, with steeper
$\Gamma^{s}_{X}$ and higher R$_\mathrm{Edd}$ for the former. Such a
difference also exists in the spectral indices distribution in hard
X-ray ($\Gamma^{h}_{X}$), based on the analysis of  53 NLSy1 and  46
BLSy1 galaxies in the 2-10 keV energy band.  The difference in
R$_\mathrm{Edd}$ distributions does exist even after applying the
average correction for the difference in the inclination angle of
NLSy1 and BLSy1 galaxies. We also estimated R$_\mathrm{Edd}$, based on
SED fitting of  34 NLSy1 and  30 BLSy1 galaxies over the 0.3-10 keV
energy band and found that results are still consistent with
R$_\mathrm{Edd}$ estimates based on the optical bolometric luminosity.
Our analysis suggests that the higher R$_\mathrm{Edd}$ in NLSy1 is
responsible for its steeper X-ray spectral slope compared to the
BLSy1, consistent with the disc-corona model as proposed for the
luminous AGNs.

\end{abstract}


\section{Introduction}
\label{section 1.0}
Narrow-line Seyfert 1 galaxies (NLSy1s) are a peculiar class of
lower-luminosity Active Galactic Nuclei (AGNs), as defined by the
width of the broad component of H$\beta$ (FWHM(H$\beta$)) $\lesssim$ 2000 km s$^{-1}$, flux ratio of [O
  $_{III}]_{\lambda5007}/H\beta\lesssim$ 3 and strong permitted optical/UV Fe~{\sc ii} emission lines~\citep{Shuder-Osterbrock1981ApJ...250...55S,Osterbrock1985ApJ...297..166O, Boroson1992ApJS...80..109B, Grupe1999A&A...350..805G}.~They show steep soft X-ray
spectra and rapid X-ray flux
variability~\citep{Boller1996A&A...305...53B, Wang1996A&A...309...81W, Grupe1998A&A...330...25G, Leighly1999ApJS..125..297L, Komossa-Meerschweinchen2000A&A...354..411K,Miller2000NewAR..44..539M, Gaskell-Hedrick2004ApJ...609...69K}.~Observations suggest that NLSy1s 
tend to have smaller black hole masses (M$_{\mathrm BH}$) and higher Eddington ratios
(defined as the ratio of bolometric-to-Eddington luminosity as
$R_{\mathrm Edd}\equiv L_{\mathrm bol}/L_{\mathrm Edd}$) compared to the broad line AGNs
~\citep{Boroson1992ApJS...80..109B, Pounds1995MNRAS.277L...5P,
  Sulentic2000ApJ...536L...5S, Boroson2002ApJ...565...78B,
  Collin2004A&A...426..797C}.~On the other hand,~\citet{Gayathri2019arXiv190710851G} reported a similarity of R$_{\mathrm Edd}$ and M$_{\mathrm BH}$ among NLSy1s and Broad-line
Seyfert 1 galaxies (BLSy1s)  based on the accretion disc (AD) modeling of their optical spectra.
Comparatively little is known about the
intrinsic emission mechanisms of NLSy1s which are responsible for
their aforementioned properties.  However, since the launch of the
many space telescopes such as ROentgen SATellite ({\it ROSAT}), {\it Chandra},
X-ray Multi-Mirror Mission-Newton ({\it XMM-Newton}), and {\it Fermi} Large Area
Telescope (LAT), many NLSy1s have been detected in the high energy bands
such as X-rays and $\gamma$-rays. These high energy emissions are
thought to be one of the most direct forms of nuclear activity which do
play a crucial role in understanding the accretion process in 
the different types of AGNs.\par For instance, a remarkable correlation has been
found by~\citet{Boller1996A&A...305...53B}
\&~\citet{Wang1996A&A...309...81W} between the soft X-ray photon
indices and the widths of the broad component of H$\beta$ lines (FWHM(H$\beta$)) in the NLSy1s. This is interpreted with 
the variation of accretion rate in different objects
~\citep{Wandel1985ApJ...292..206W, Pounds1994MNRAS.267..193P}. To test
this hypothesis,~\citet{Lu1999ApJ...526L...5L} have compiled a
sample of Seyfert 1 galaxies, QSOs, and found that the soft X-ray
photon indices strongly correlate with the accretion rates. 
Additionally, ~\citet{Laor1997ApJ...477...93L} have found a
correlation between the soft X-ray (0.2$-$2.0 keV) slope and the FWHM
of the H$\beta$ emission line in a sample of 23 low-redshift quasars
suggesting that the physical parameter driving the correlation is the
Eddington ratio.
Many past X-ray studies of the Low luminosity AGNs (LLAGNs, comprising low-ionization nuclear emission-line regions and local Seyfert galaxies), have been carried out to explore any correlation among X-ray photon indices with other parameters of nuclear activities~\citep[e.g., see][]{Gonzalez2006A&A...460...45G,Panessa2006A&A...455..173P,Gu2009MNRAS.399..349G}. For instance,~\citet{Gu2009MNRAS.399..349G} find a significant anticorrelation among the hard X-ray photon indices and the Eddington ratios using a sample of 55 LLAGNs, whose X-ray photon indices are collected from the literature having {\it Chandra} or {\it XMM-Newton} observations.
This anticorrelation resembles the spectra produced from advection dominated accretion flows (ADAFs) model for the X-ray binaries (XRBs) in the low
state~\citep[e.g., see][]{Esin1997ApJ...489..865E}. However it is found in
contrast with the positive correlation reported
by~\citet{Risaliti2009ApJ...700L...6R} for the luminous AGNs. Their
analysis led to an important suggestion that the spectra of LLAGNs
might be produced by Comptonization process in ADAFs, which is similar
to that of XRBs but is different from that in luminous AGNs. As a
result, such analysis has important implications for the physical link
between the accretion efficiency in the (cold) accretion disk of AGNs
and the physical status of the (hot) corona.\par
For the X-ray detected NLSy1 galaxies, their X-ray/$\gamma-$ray emissions can
be either from the jets whose
existence is inferred based on their high variability in short
time scales~\citep{Paliya2014ApJ...789..143P,
  Kshama2017MNRAS.466.2679K, Ojha2019MNRAS.483.3036O}, or it could be
based on the ADAFs mechanism as suggested by
~\citet{Gu2009MNRAS.399..349G} for LLAGNs. Another possibility could
be the accretion-flow/hot-corona system of radiatively efficient
accretion, as suggested by~\citet{Maoz2007MNRAS.377.1696M} where thin
accretion disc may persist at lower accretion rates.
Additionally, it could also be from the widely accepted disk-corona
model. In  this model, UV soft photons from the accretion disc are comptonized
and up-scattered (inverse Comptonization) into the X-ray bands by
a hot corona, existing above the accretion
disc~\citep{Haardt1991ApJ...380L..51H,
  Haardt1993ApJ...413..507H}. 
  To get an insight into the emissions from the central engine of NLSy1s,
  one possibility is to compare the distribution of its key parameters
  such as R$_{\mathrm Edd}$, M$_{BH}$, and X-ray spectral slopes, etc,
  with the control sample of BLSy1s matching in luminosity-redshift
  (L-z) plane. Any observational constraints based on such comparisons,
  can be very useful to probe the above possible mechanisms.  For
  instance, if the R$_{\mathrm Edd}$ of NLSy1s in comparison to BLSy1s   are statistically higher then one would expect an increase
  in the disk temperature, hence the production of more X-ray radiations,
  and at the same time, it can also increase the Compton cooling of the
  corona~\citep{Haardt1991ApJ...380L..51H, Haardt1993ApJ...413..507H,
    Zdziarski2000ApJ...542..703Z, Kawaguchi2001ApJ...546..966K}. This
  can further lead to observable steepening of the X-ray power-law
  more in NLSy1s than BLSy1s.  Therefore, for such an insight
  especially about the X-ray emissions mechanism, X-ray spectral slope
  (both in the soft and hard X-ray energy bands) of a statistical
  large sample of NLSy1s along with its control sample of BLSy1s
  (matching in their L-z plane) is very useful. This can also
  help to parametrize the cooling and heating mechanism of the X-ray
  corona, along with the underlying electrons' energy distribution.\par
However, the main hindrance till now in the aforementioned investigations was the
lack of reasonable statistical homogeneous sample~\citep[see][]{Brandt1997MNRAS.285L..25B} added by a
homogeneous analysis in the soft (0.1-2.0 keV) and the hard (2-10 keV) X-ray bands for the
NLSy1 and BLSy1 galaxies, preferably matching in L-z plane.
This was due to a relatively small available sample size
of a total of 2000 optically detected NLSy1s given
by~\citet{Zhou2006ApJS..166..128Z} based on Sloan Digital Sky Survey
(SDSS) Data Release 3~\citep[SDSS,
  DR-3][]{Schneider2005AJ....130..367S}.  In contrast, based on a 10
fold increase in the number of AGNs in SDSS spectroscopic data release
12~\citep[SDSS DR-12,][]{Alam2015ApJS..219...12A} than the DR-3.
~\citet{Rakshit2017ApJS..229...39R} have recently enlarged the sample
of NLSy1s to a total of 11,101 objects which is about 5 times larger
than the number of previously known NLSy1 galaxies based on~\citet{Zhou2006ApJS..166..128Z} catalog.
This enlarged sample can be used to carry out a systematic and homogeneous analysis of a statistical sample of  NLSy1s (both in optical and  X-ray). This can further also be used to compare its key parameters of nuclear activities, such as R$_{\mathrm Edd}$, M$_{BH}$ and X-ray spectral slopes with a control sample of BLSy1s (preferably match in their L-z plane).
This analysis is also favorable to investigate
whether the steepening reported in the spectral slopes of
NLSy1s~\citep[albeit deduced with small sample size,
  e.g., see][]{Brandt1997MNRAS.285L..25B} as compared to BLSy1s exist
only in soft X-ray band or also extend to the hard X-ray band which is  
 less prone to the soft X-ray excess~\citep{Boller1996A&A...305...53B, Brandt1997MNRAS.285L..25B,
  Vaughan1999MNRAS.309..113V,Boller2002MNRAS.329L...1B,
  Czerny2003A&A...412..317C, Vignali2004MNRAS.347..854V}.\par
Here, we have worked towards the aforementioned goals. For this, we have selected a sample of 
 221 NLSy1s by cross-correlating 11,101 NLSy1s with that of the second {\it ROSAT}
all-sky survey (2RXS) source catalog of~\citet{Boller2016A&A...588A.103B} and
based on any source observation in {\it XMM-Newton}, available on {\scriptsize HEASARC} public data archive\footnote{\scriptsize https://heasarc.gsfc.nasa.gov/db-perl/W3Browse/w3browse.pl} (e.g., Sect.~\ref{section 2.0}).
The corresponding control sample of  154 BLSy1s in the X-ray band, moderately matching in the redshift with that of our NLSy1s (e.g. Sect.~\ref{section 2.0}) sample is used to carry
out the comparative study of these two subclasses.\par
For the homogeneous X-ray analysis of the NLSy1 and BLSy1 galaxies, similar models and
homogeneous methods are adopted for estimating their X-ray spectral slopes both
in the soft and hard X-ray bands. In the same way, a  homogeneous method is also applied
to estimate the black-hole masses for all the members of our samples by careful modeling of the H$\beta$ lines
using their SDSS optical spectra.
This is used to investigate any statistical relationships
among the X-ray photon indices
of NLSy1s and BLSy1s
with their other key parameters of nuclear activities such as
FWHM(H$\beta$), $M_{\mathrm BH}$, bolometric luminosities (L$_{\mathrm bol}$) and R$_{\mathrm Edd}$. This allows us to understand the X-ray emission
mechanisms of NLSy1s as compared to the BLSy1s, along with the
comparison of their properties with other luminous AGNs.\par
The paper is structured as
follows. Sect.~\ref{section 2.0} describes the data sample and selection
criteria. Sect.~\ref{section 3.0} describes observations
and data reduction along with our analysis for X-ray data. Sect.~\ref{section 4.01} gives details of our spectral
analysis. In Sect.~\ref{section 5.0}, we focus on our results while discussion and
conclusion are given in Sect.~\ref{section 6.0}. Finally, we summarize our work in
Sect.~\ref{section 7.0}.  Throughout, we have used a cosmology with
$\Omega_m=0.286$, $\Omega_\lambda=0.714$, and $H_o=69.6
kms^{-1}Mpc^{-1}$ \citep[][]{Bennett2014ApJ...794..135B}.

\begin{figure*}
  \begin{minipage}[]{1.0\textwidth}
  \includegraphics[width=0.5\textwidth,height=0.27\textheight,angle=00]{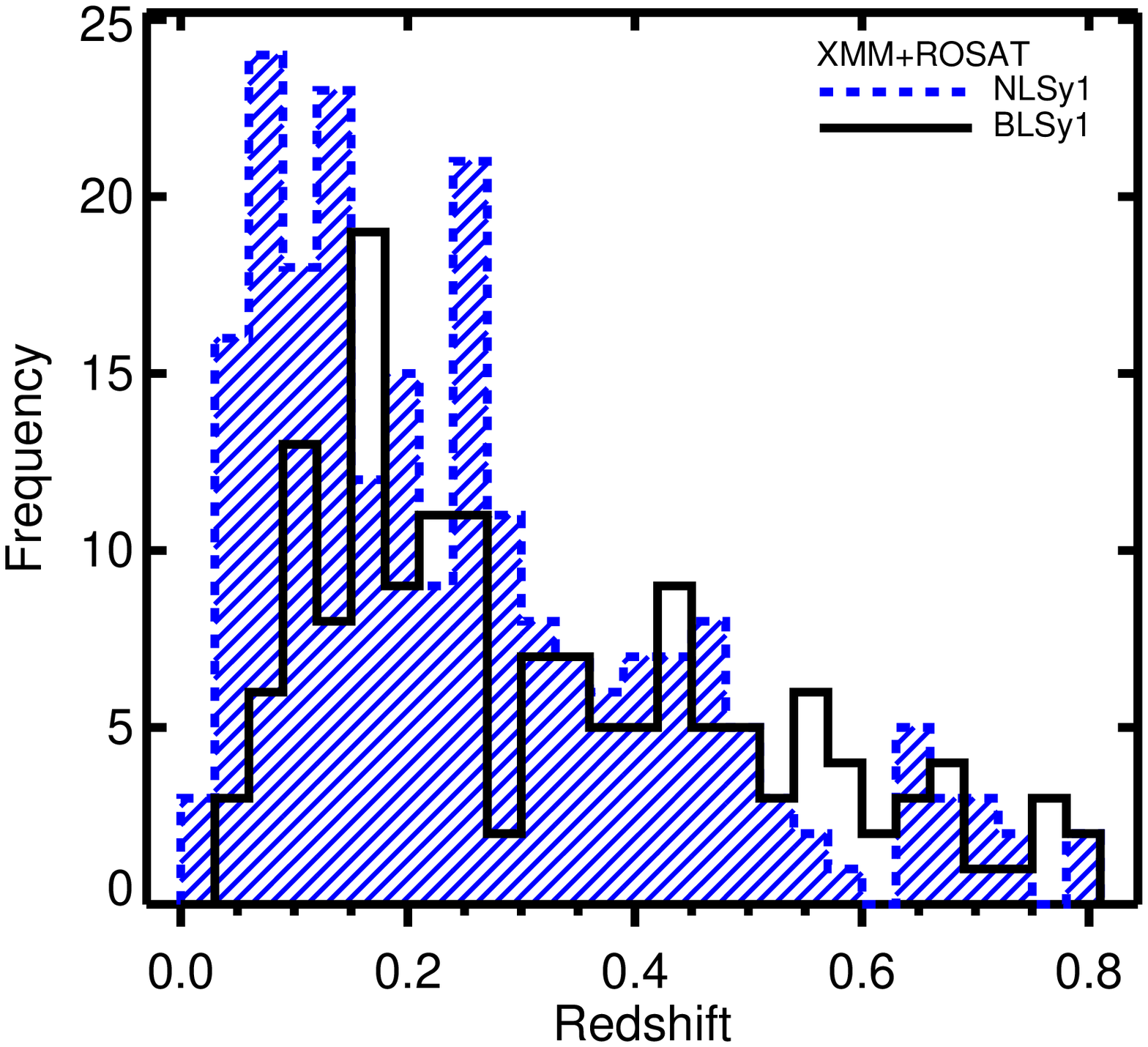}    
  \includegraphics[width=0.5\textwidth,height=0.27\textheight,angle=00]{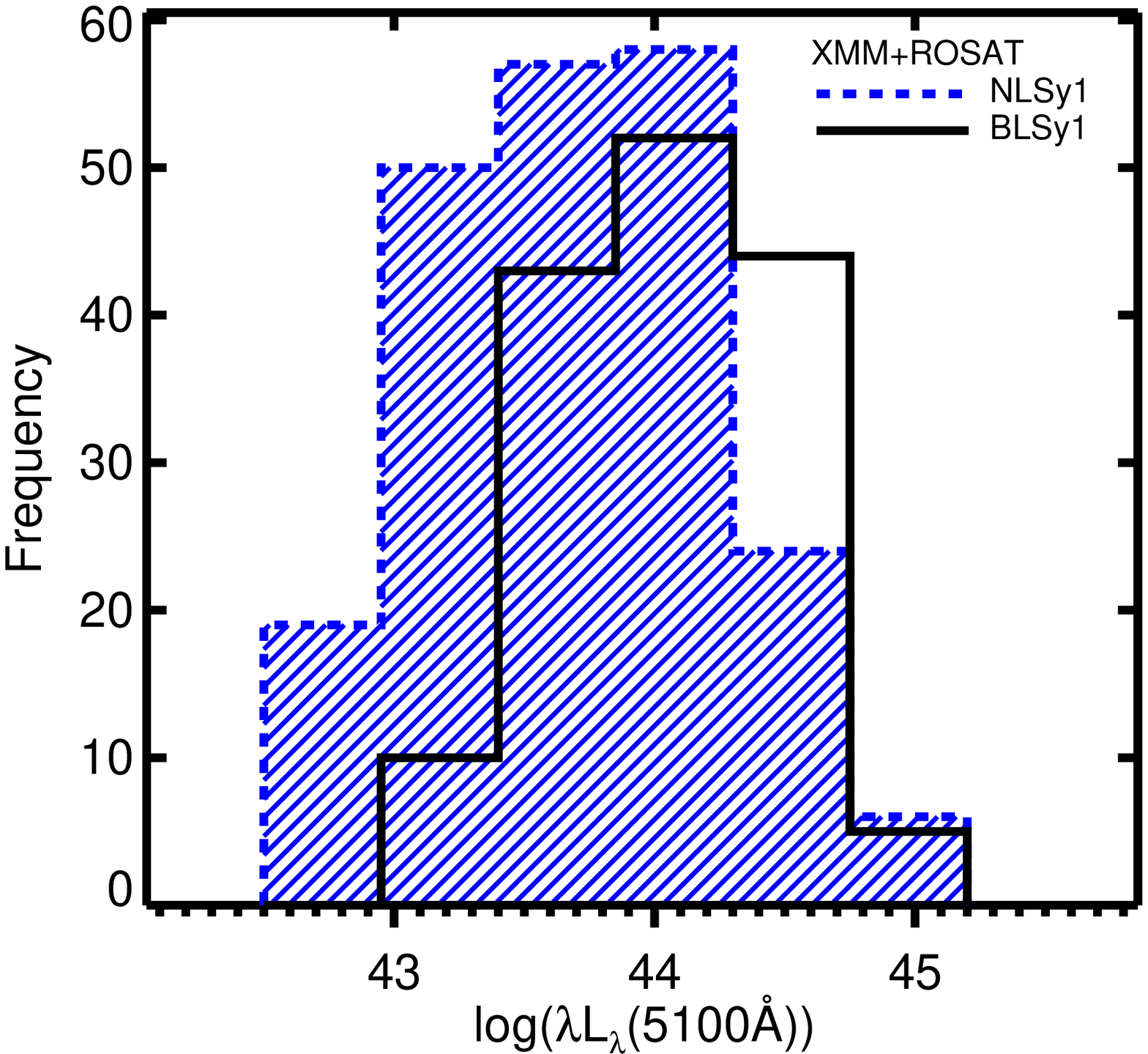}  
  \end{minipage}
  \caption{\scriptsize Distribution of emission redshifts (left) and $\lambda L_{\lambda}$(5100\AA) (right) for our {\it XMM-Newton} and {\it ROSAT} detected combined samples of 221 NLSy1s (blue, filled) and 154 BLSy1s (black, solid line).}
  \label{fig:z_lum_match_xmm_rosat}
\end{figure*}

\begin{figure*}
    \begin{minipage}[]{1.0\textwidth}
    \includegraphics[width=0.51\textwidth,height=0.32\textheight,angle=00]{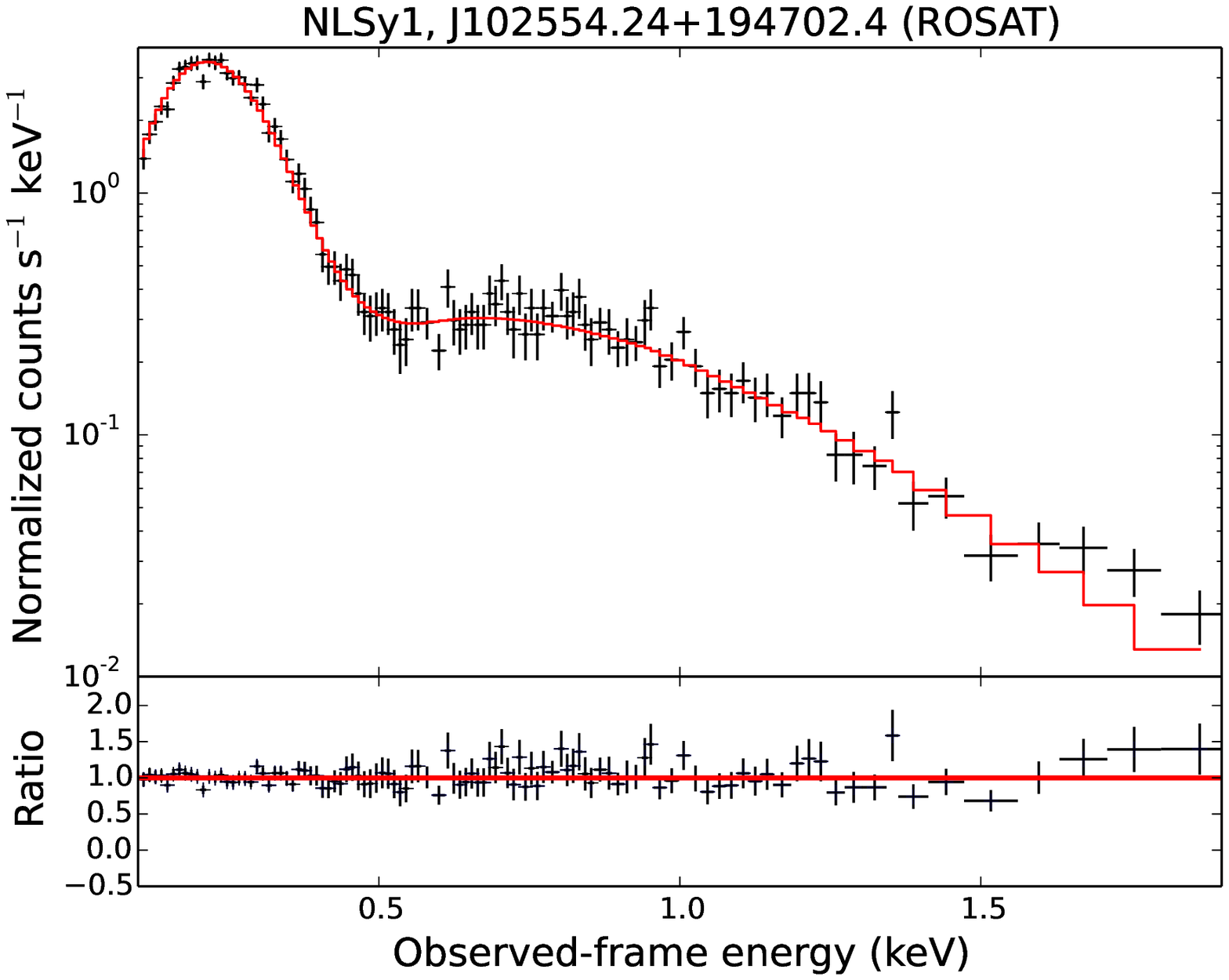}  
    \includegraphics[width=0.51\textwidth,height=0.32\textheight,angle=00]{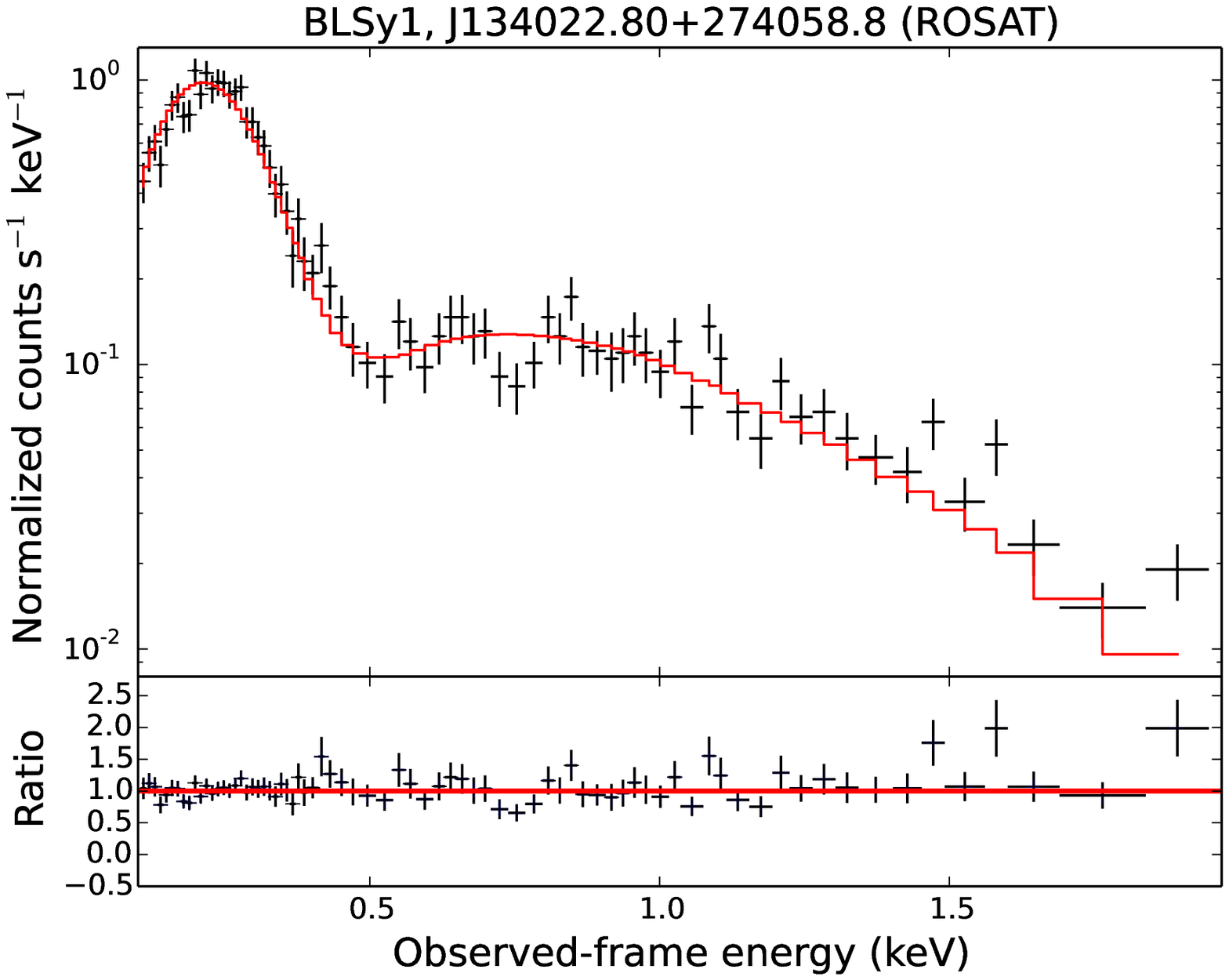}\\\\\\  
     \includegraphics[width=0.51\textwidth,height=0.32\textheight,angle=00]{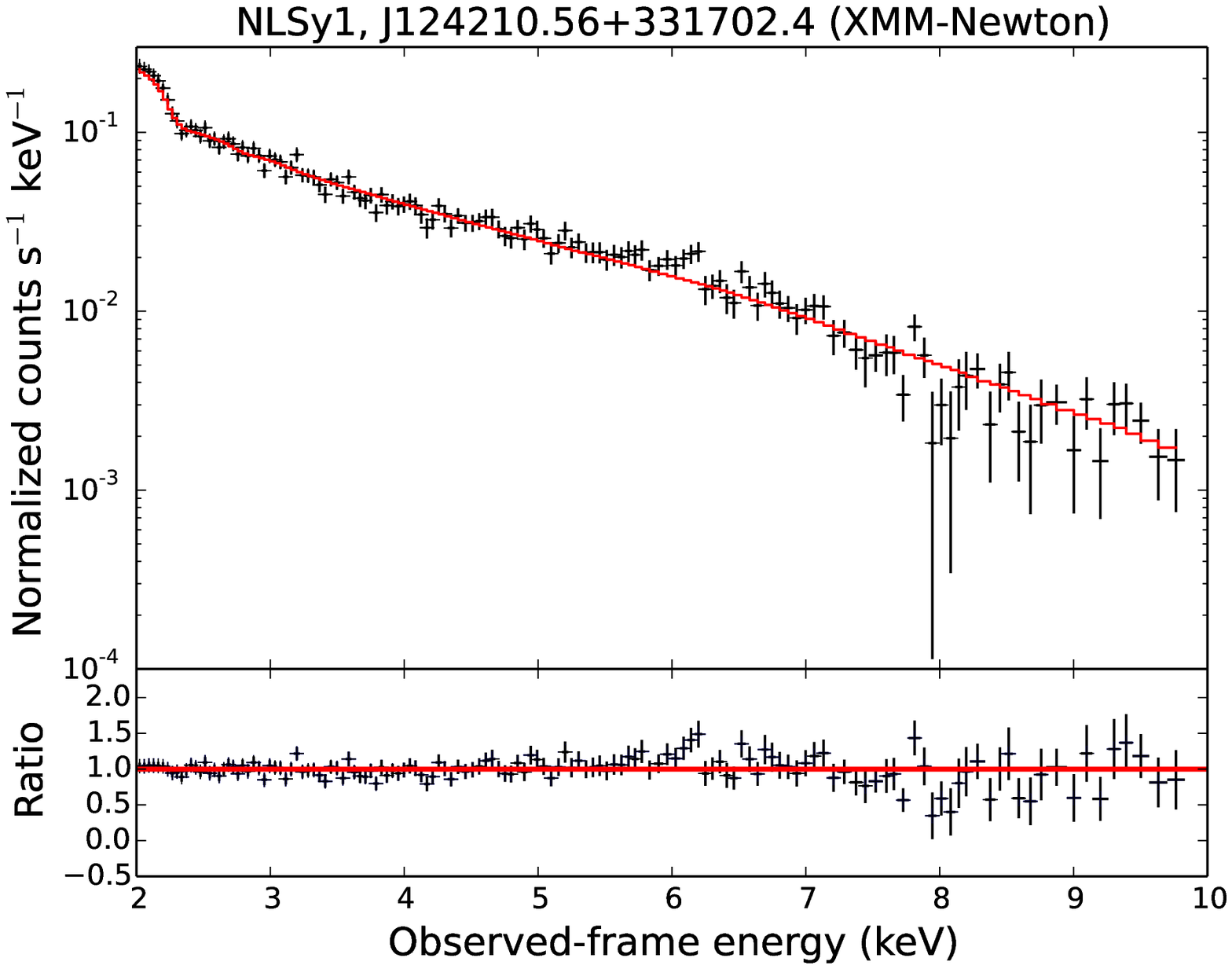}
    \includegraphics[width=0.51\textwidth,height=0.32\textheight,angle=00]{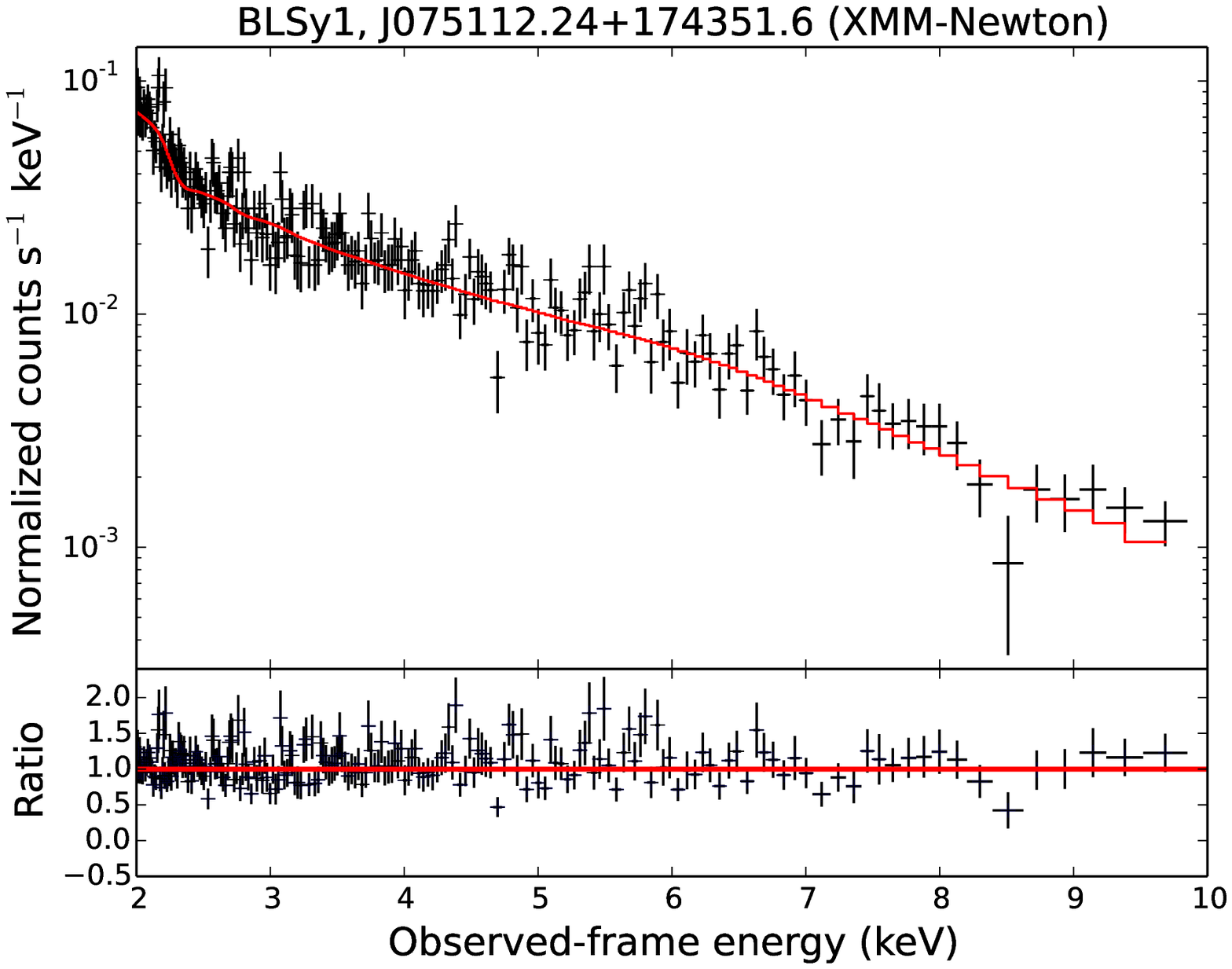}  
    \end{minipage}
    \caption{{\scriptsize \emph{Top}: Representative {\it ROSAT} soft (0.1-2.0 keV)/PSPC data and best fit folded models for the NLSy1 galaxy, viz., J102554.24$+$194702.4 (top left panel) and BLSy1 galaxy, viz., J134022.80$+$274058.8 (top right panel). In each case, our fit was carried out using power-law and double neutral absorption models. \emph{Bottom}: The same as the top panels but for the {\it XMM-Newton} hard (2-10 keV)/PN data of the NLSy1 galaxy, viz., J124210.56$+$331702.4 (bottom left panel) and BLSy1 galaxy, viz., J075112.24$+$174351.6 (bottom right panel).}}
    \label{fig:folded_spec_nlsy1_blsy1}
\end{figure*}

\section{Sample Selection}

\label{section 2.0}

For constructing our sample of NLSy1 galaxies, we used a recent
catalog of NLSy1s given by~\citet{Rakshit2017ApJS..229...39R} in which
they have compiled 11,101 NLSy1s using the SDSS DR-12 database. To
make a sample of X-ray detected NLSy1s, we have cross-correlated these
11,101 NLSy1s with the 2RXS source catalog.  This cross-correlation resulted
in the 1873 matches in the 2RXS catalog within a position offset (in sources RA
and DEC) of 30 arcseconds.  Similarly, we also searched for any {\it XMM-Newton}
telescope based observations for the above sample of 11,101
NLSy1s by using the {\scriptsize HEASARC} public data archive.  This resulted in a
sample of 697 {\it XMM-Newton} observed NLSy1s such that each of the NLSy1
falls within the 27.5 arcmin$^{2}$ offset from the
pointing center of the parent {\it XMM-Newton} observation. Here, for any source
with multiple observation IDs, the repetition is avoided by retaining
only the observation with the largest observing time.
We also noticed that like {\it XMM-Newton} the above energy range is also covered by
{\it Chandra} telescope but due to its much smaller effective area (e.g., $\sim$600 cm$^{2}$)
as compared to {\it XMM-Newton} ($\sim$ 1227
cm$^{2}$), a typical increase in sample size due to the
 observed sources by the {\it Chandra} telescope is found to be nominal (around $\sim$10\%). Therefore, we
have limited our analysis only to {\it XMM-Newton's} covered sources and the {\it ROSAT} 2RXS
catalog's matched sources only. \par
Observations in the {\it ROSAT} were carried out using two detectors, viz., position-sensitive proportional counter (PSPC) and
High-Resolution Imager (HRI). Furthermore, we noted that the HRI
is essentially an imager with very little spectral response.
Therefore, we limited our {\it ROSAT} sample, only to those sources which
were observed with the PSPC instrument. This filter reduces  our sample of {\it ROSAT} detected sources (hereafter {\it ROSAT}) from 1873 to 530 sources. \par
The 0.1-2.0 keV {\it ROSAT} spectrum of each source was extracted using standard
{\scriptsize XSELECT} tasks of the {\scriptsize HEASOFT} software
(version 6.25) with the appropriate circular region around the
source to enhance source signal and reduce the background
noise. This is found to differ for different sources depending on
the number of pixels containing the maximum flux of the source
(e.g. Sect.~\ref{section 3.0}).  The impact of this choice of
the aperture by the eye on the signal-to-noise (S/N), as well as on our analysis (since
for both same aperture is used) is found to be negligible in our
sample. However, a very nominal enhancement in S/N is found due to relatively less
background noise in comparison to a fixed (50 arcseconds) aperture
encircling about 90\% energy fraction.\par
To exclude sources without high quality data, we have put a minimum S/N criterian of 10 on our sample. For computing the S/N, we have estimated $[N_{src}-N_{bkg}]/ \sqrt {\left (N_{src}+N_{bkg} \right)}$, where $N_{src}$ and $N_{bkg}$ refer to total count contributed by aperture around the source and background region. Here the background region is chosen in close proximity to the source with aperture size fixed to its value as used for extracting the source count. This aperture size could either be fixed so that it encircled about 90\% energy fraction or can also be optimized to enhance the S/N, as with an increase in aperture size background noise also increases. We have opted to use the latter, though the increase in S/N using it is found to be nominal in comparison to the former method (e.g. with 50 arcsec fixed aperture to encircle 90\% energy). We also note that the S/N computed by the $[N_{src}-N_{bkg}]/ \sqrt {\left (N_{src}+N_{bkg} \right)}$ method is consistent with that using [count rate]/[error on count rate] as return by XSPEC (version 6.25) for the source `grp' file. For our ROSAT sources, we used 0.1-2.0 keV energy range while computing the S/N using the above [count rate]/[error on count rate] method. The S/N$\ge$10 criterion was satisfied by 83 out of 530 (henceforth also referred to as 83/530) {\it ROSAT/PSPC} (0.1-2.0 keV) detected sources. For the 697 {\it XMM-Newton} detected (hereafter {\it XMM}) NLSy1s,  among its three European Photon Imaging Camera (EPIC) detectors, we have limited our analysis only to PN detector due to its larger effective area (about 1227 cm$^{2}$ at 1 keV). Each source spectrum was extracted over the appropriate circular aperture, selected by the eye around the source, in the same way as we had done for {\it ROSAT} (e.g. Sect.~\ref{section 3.0}). The S/N$\ge$10 criterion was satisfied by 148/697 {\it XMM/PN} sources in (0.3-2.0 keV) energy band.
Further, we noticed that 8 sources are common between the samples of  83 {\it ROSAT} selected NLSy1s and
the 148 {\it XMM} NLSy1s. For these 8 sources, we have used only
{\it XMM-Newton} telescope observations due to its better spectral
resolution and effective area (about 1227 cm$^{2}$ at 1 keV) as compared to the {\it ROSAT} telescope (about 240 cm$^{2}$ at 1 keV).
This led to our final sample of  223 sources with  75 from the {\it ROSAT} and
 148 from the {\it XMM-Newton} telescopes for their further X-ray spectral
fitting.\par

\begin{figure*}
    \begin{minipage}[]{1.0\textwidth}
    \includegraphics[width=0.5\textwidth,height=0.275\textheight,angle=00]{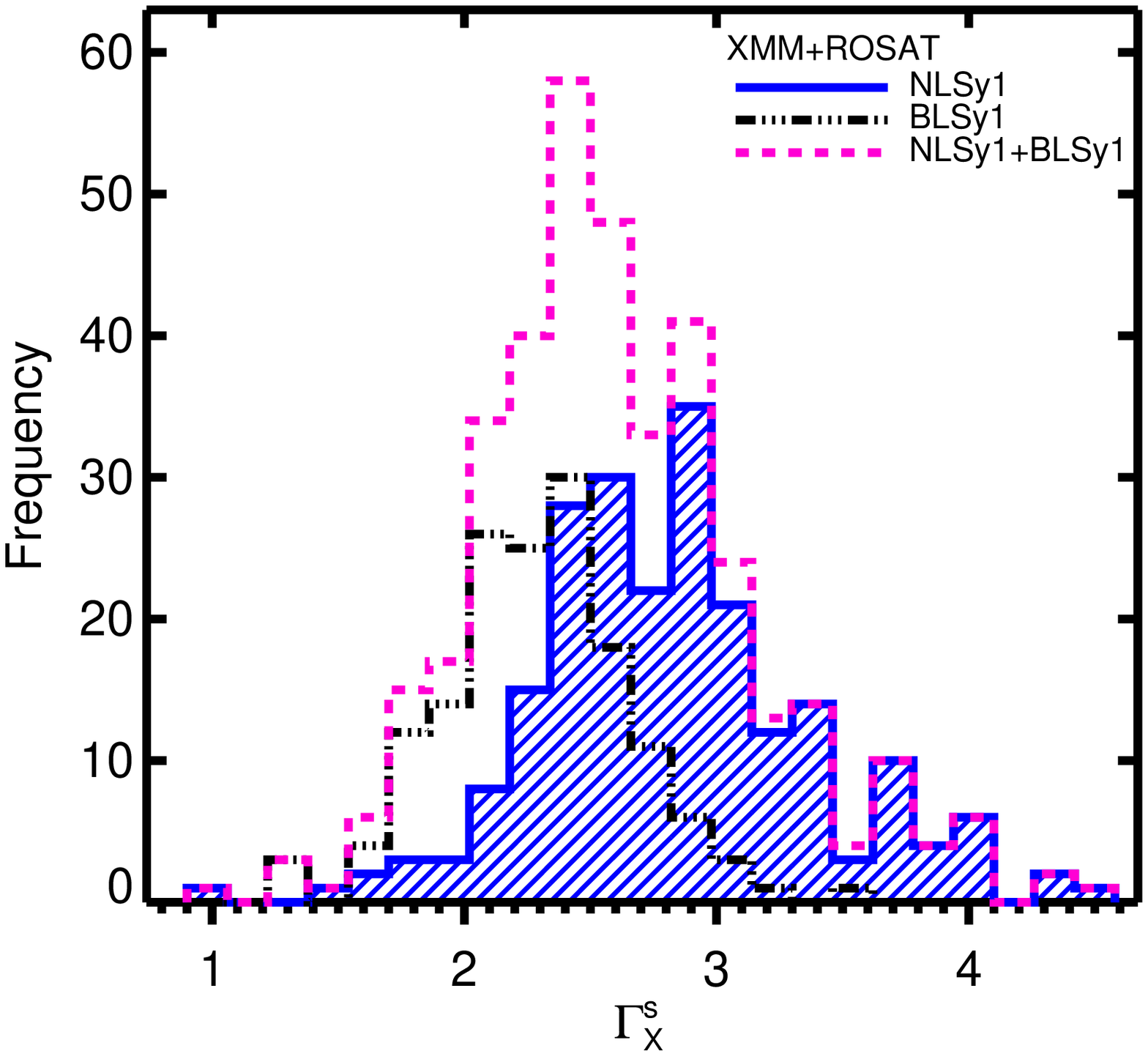}    
  \includegraphics[width=0.5\textwidth,height=0.265\textheight,angle=00]{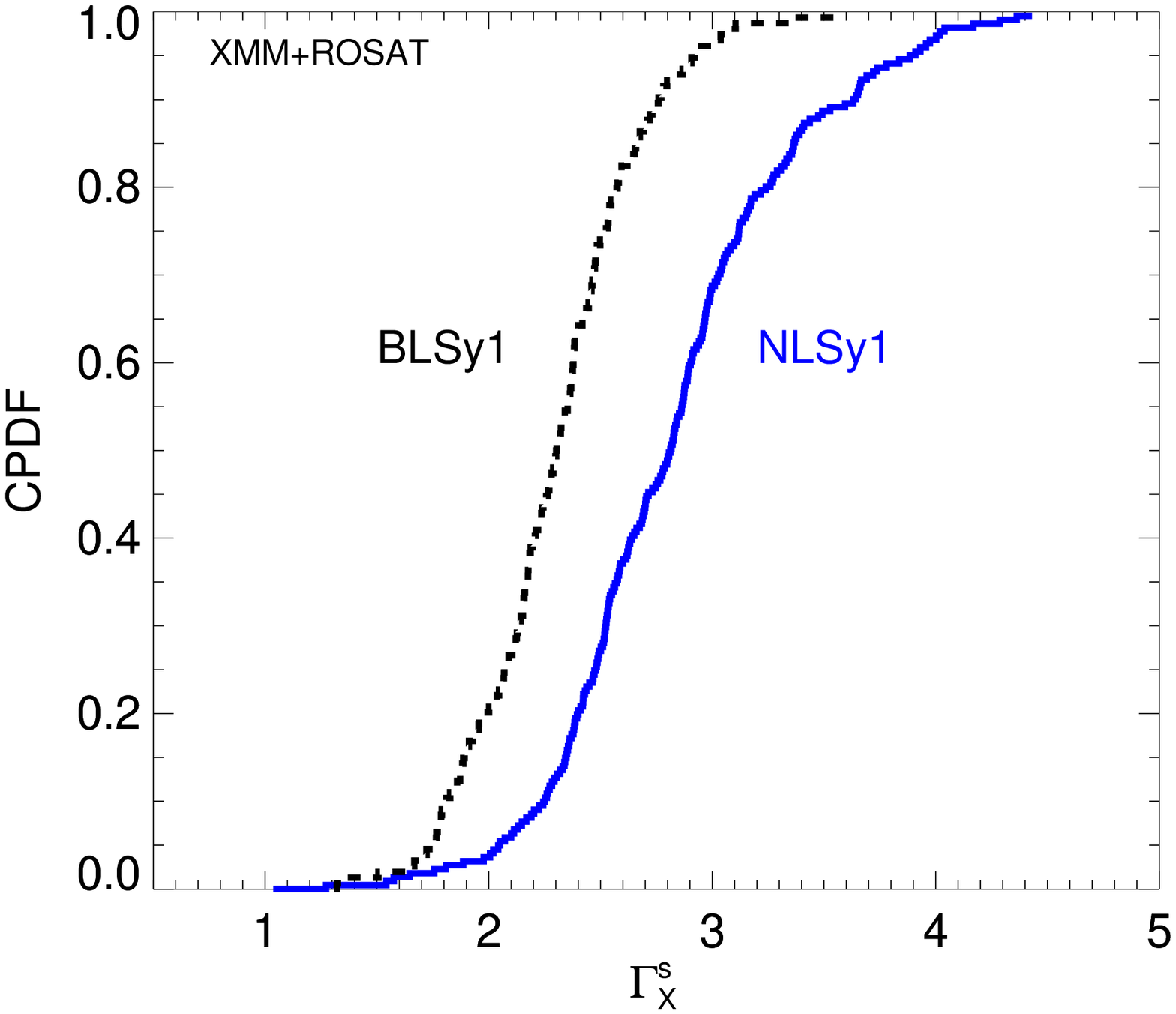}  
    \includegraphics[width=0.5\textwidth,height=0.275\textheight,angle=00]{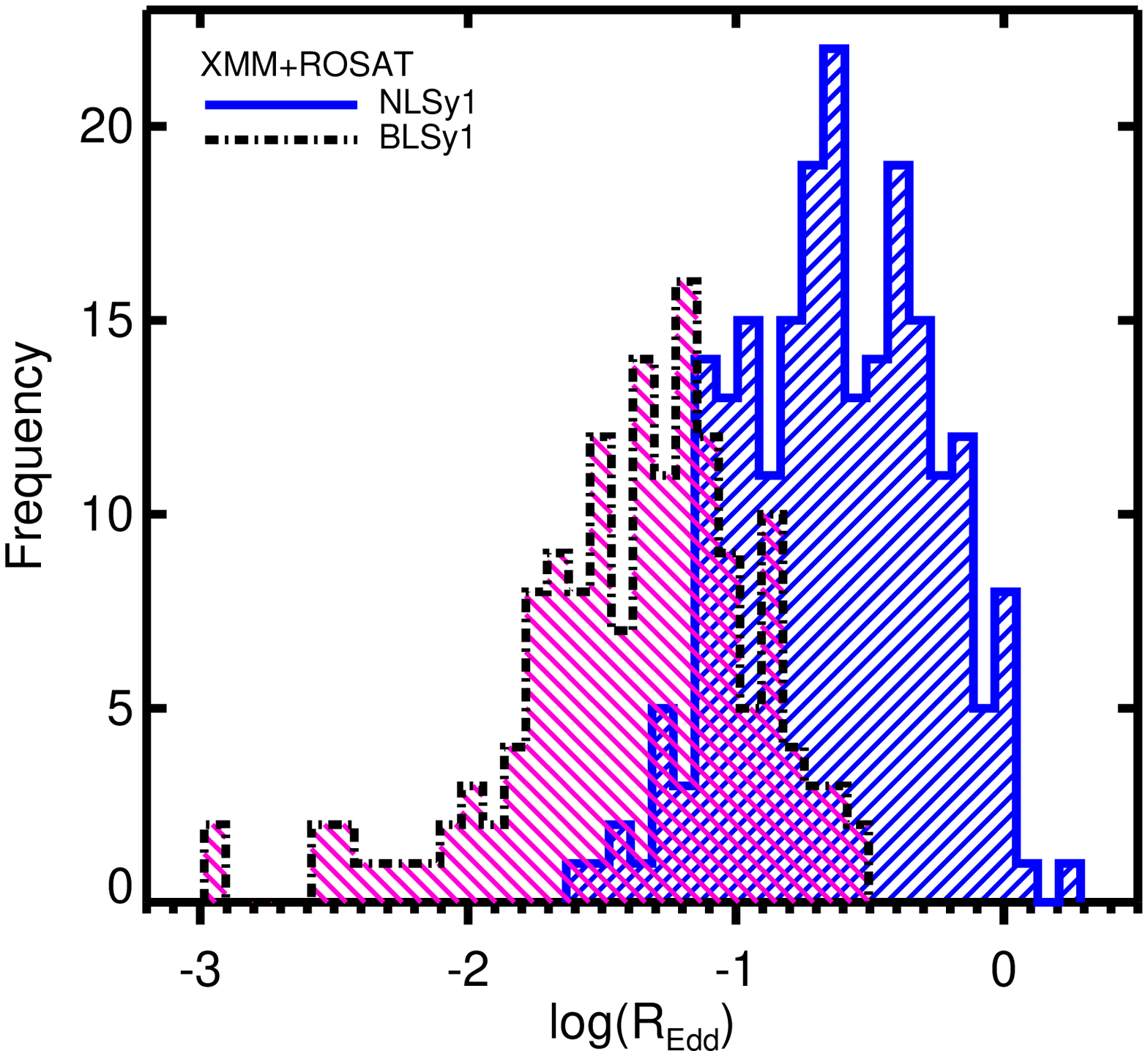}    
    \includegraphics[width=0.5\textwidth,height=0.265\textheight,angle=00]{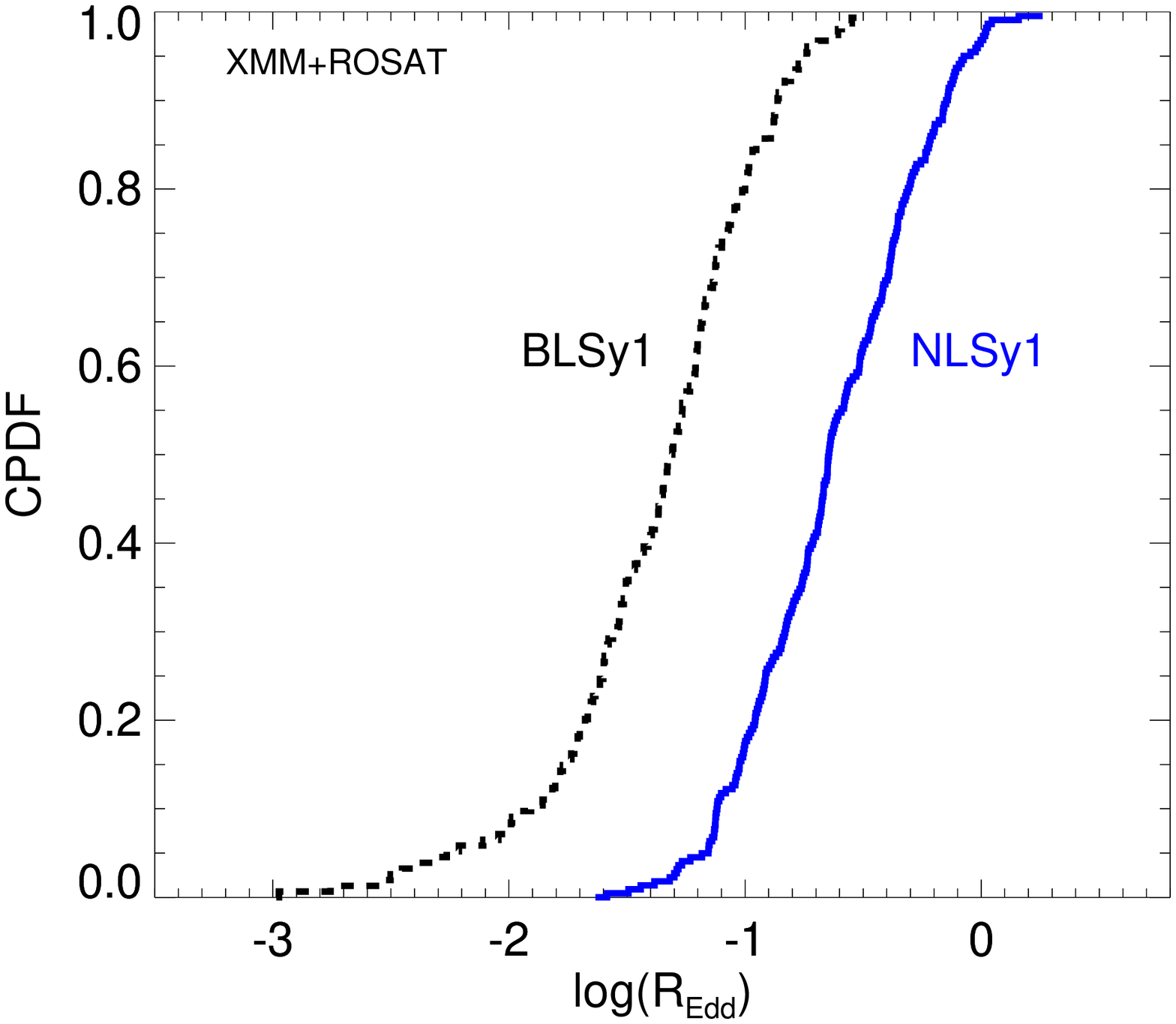}  
      \end{minipage}
  \caption{\scriptsize \emph{Top}: Distribution of 0.1-2.0 keV soft X-ray photon indices (left) and its Cumulative probability distribution function (CPDF, right panel) for our combined {\it ROSAT} and {\it XMM-Newton} soft X-ray (0.1-2.0 keV) detected samples of  221 NLSy1s (blue, filled) and  154 BLSy1s (black, dashed with dotted line) along with joint [NLSy1$+$BLSy1] sample of  375 galaxies (pink, dashed). \emph{Bottom}: The same as the top panels but for the Eddington ratios distribution (left, with blue filled for NLSy1s and pink filled for BLSy1s) and its CPDF (right),  except for the joint [NLSy1$+$BLSy1] sample of 375 galaxies.}
  \label{fig:histo_cpdf_xmm_rosat}
\end{figure*}

\begin{figure*}
    \begin{minipage}[]{1.0\textwidth}
    \includegraphics[width=0.5\textwidth,height=0.275\textheight,angle=00]{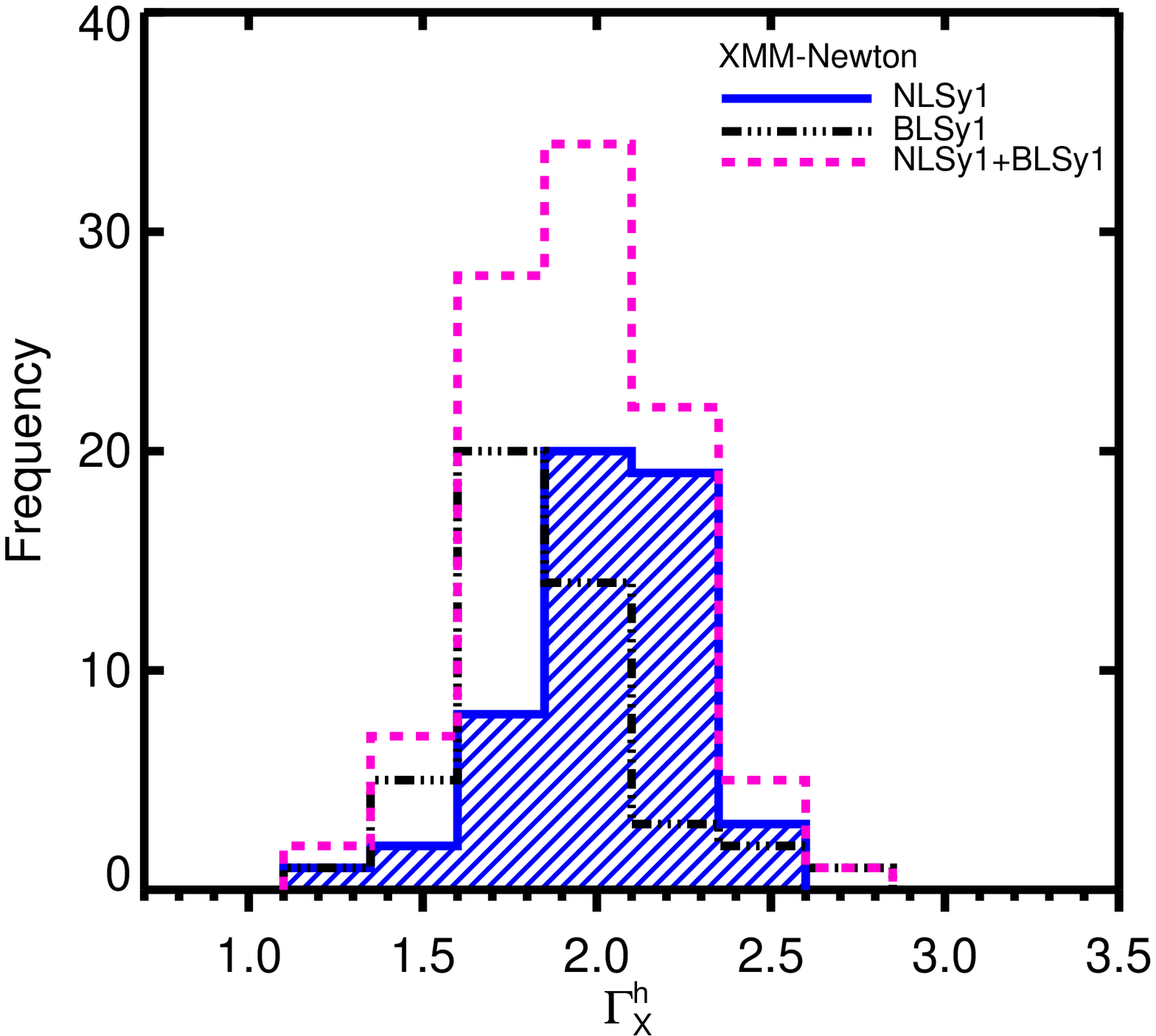}    
  \includegraphics[width=0.5\textwidth,height=0.265\textheight,angle=00]{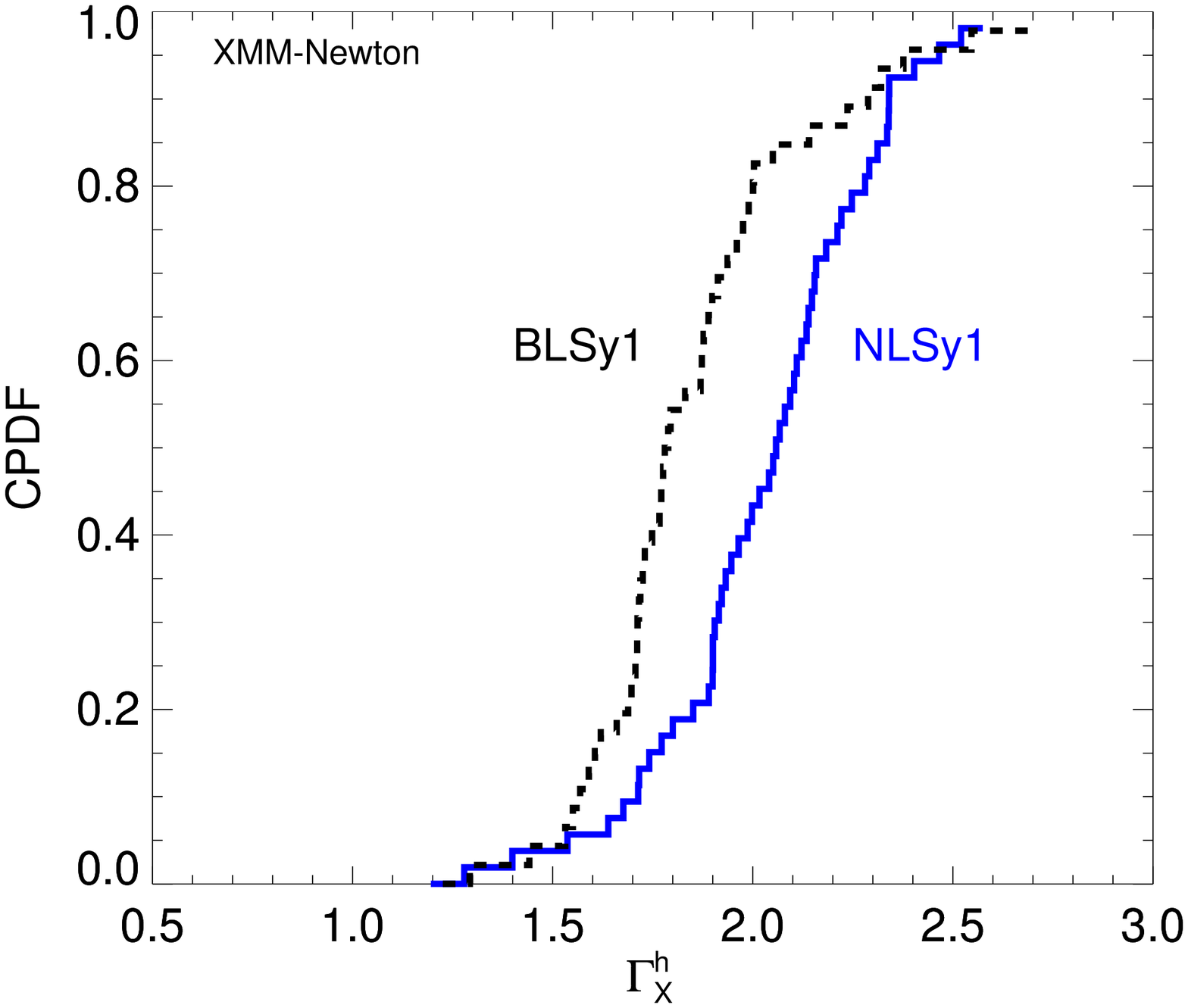}  
    \includegraphics[width=0.5\textwidth,height=0.275\textheight,angle=00]{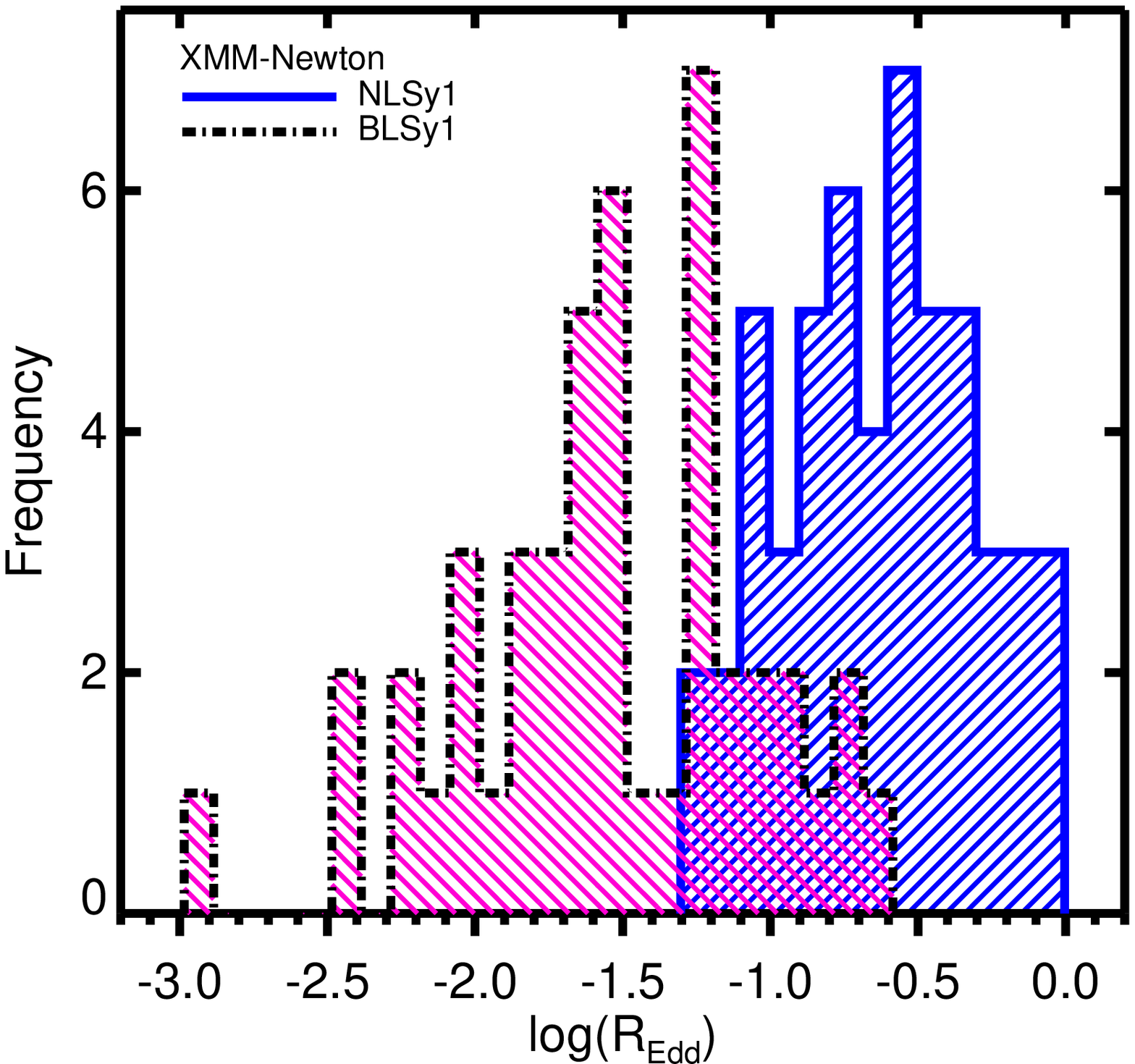}    
    \includegraphics[width=0.5\textwidth,height=0.265\textheight,angle=00]{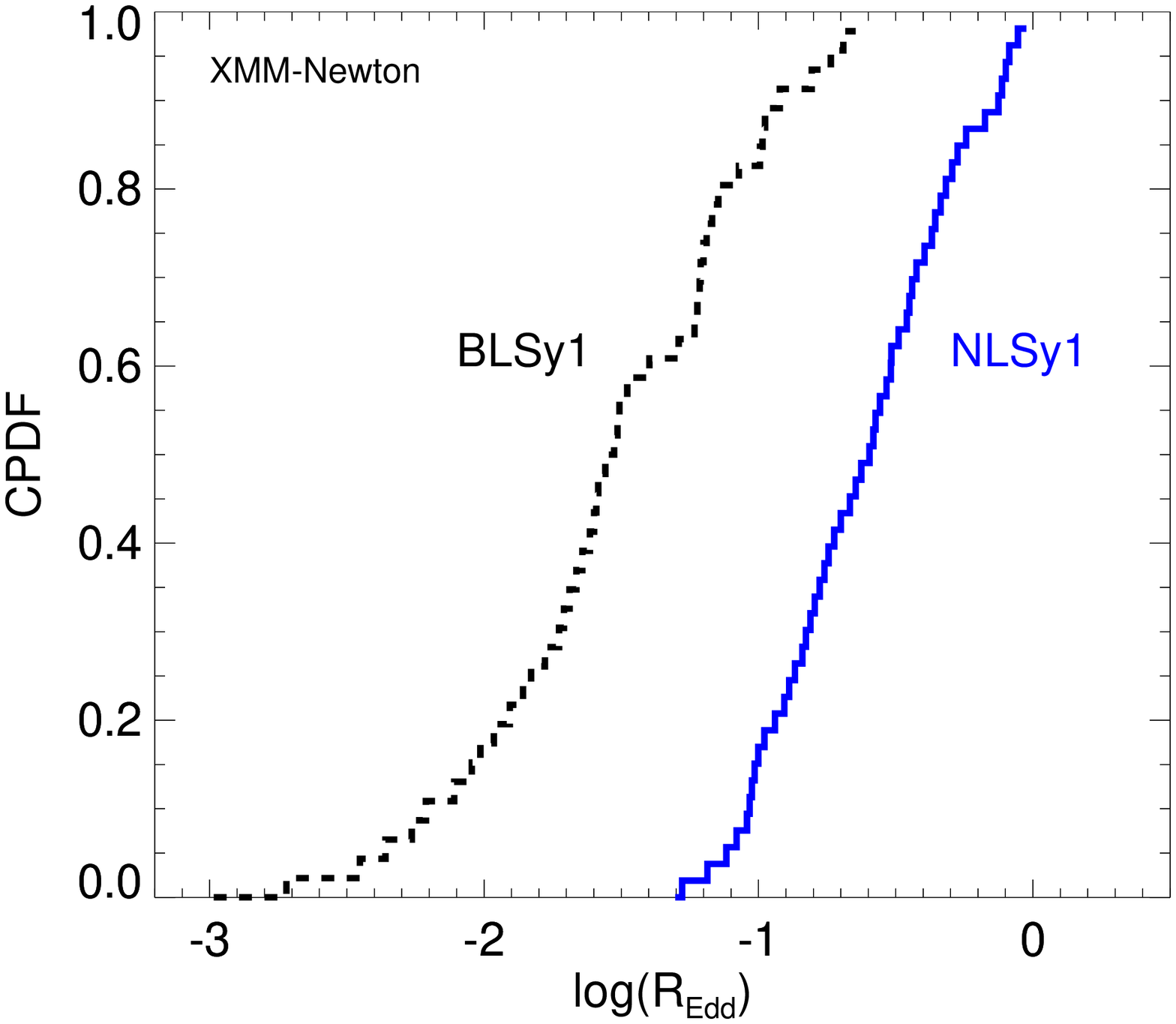}  
    \end{minipage}
    \caption{\scriptsize Same as Fig.~\ref{fig:histo_cpdf_xmm_rosat}, but using the {\it XMM-Newton} hard X-ray (2-10 keV) detected subsamples of  53 NLSy1 and  46 BLSy1 galaxies.}
    \label{fig:histo_cpdf_xmm_hard}
\end{figure*}

\begin{figure*}
    \begin{minipage}[]{1.0\textwidth}
    \includegraphics[width=0.5\textwidth,height=0.275\textheight,angle=00]{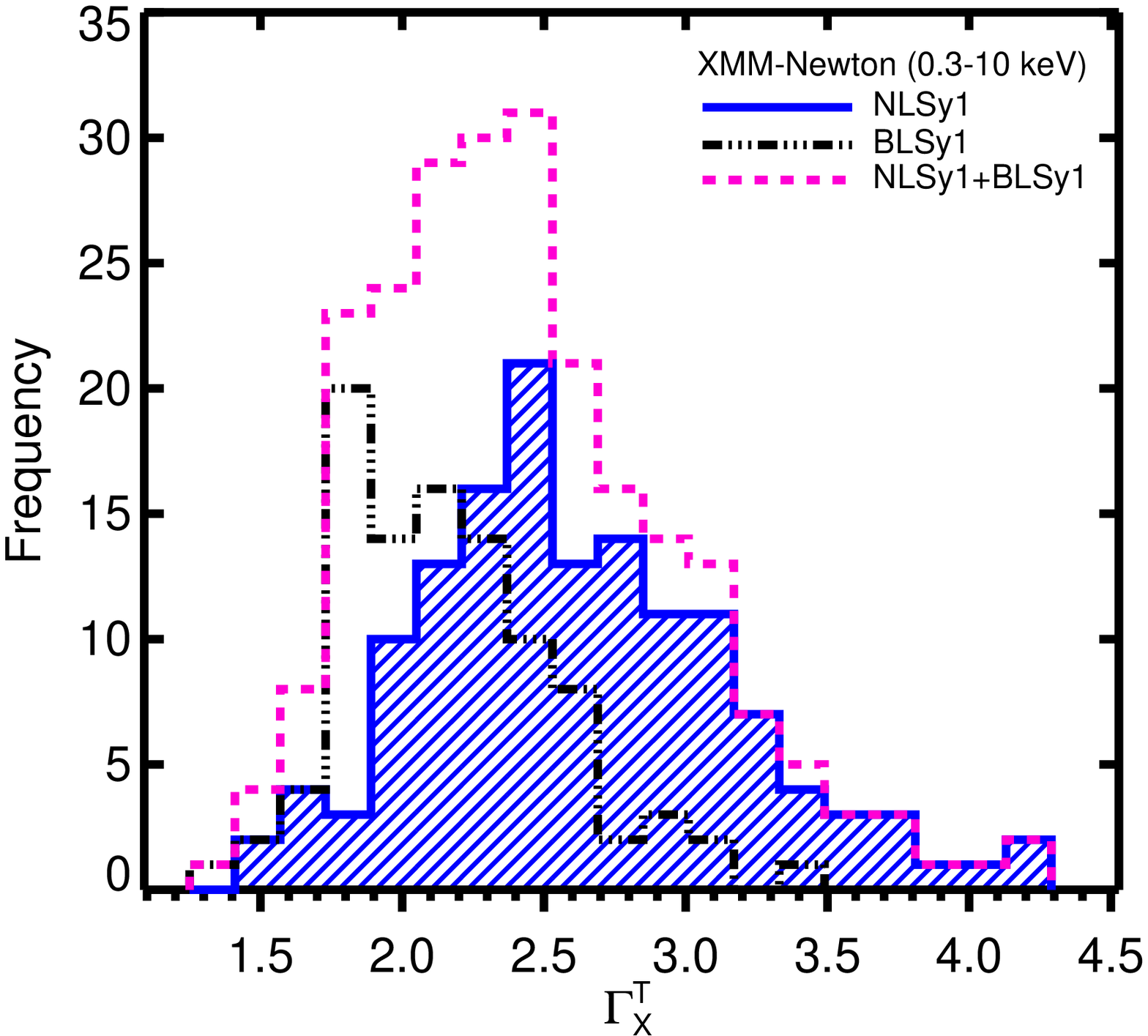}    
    \includegraphics[width=0.5\textwidth,height=0.265\textheight,angle=00]{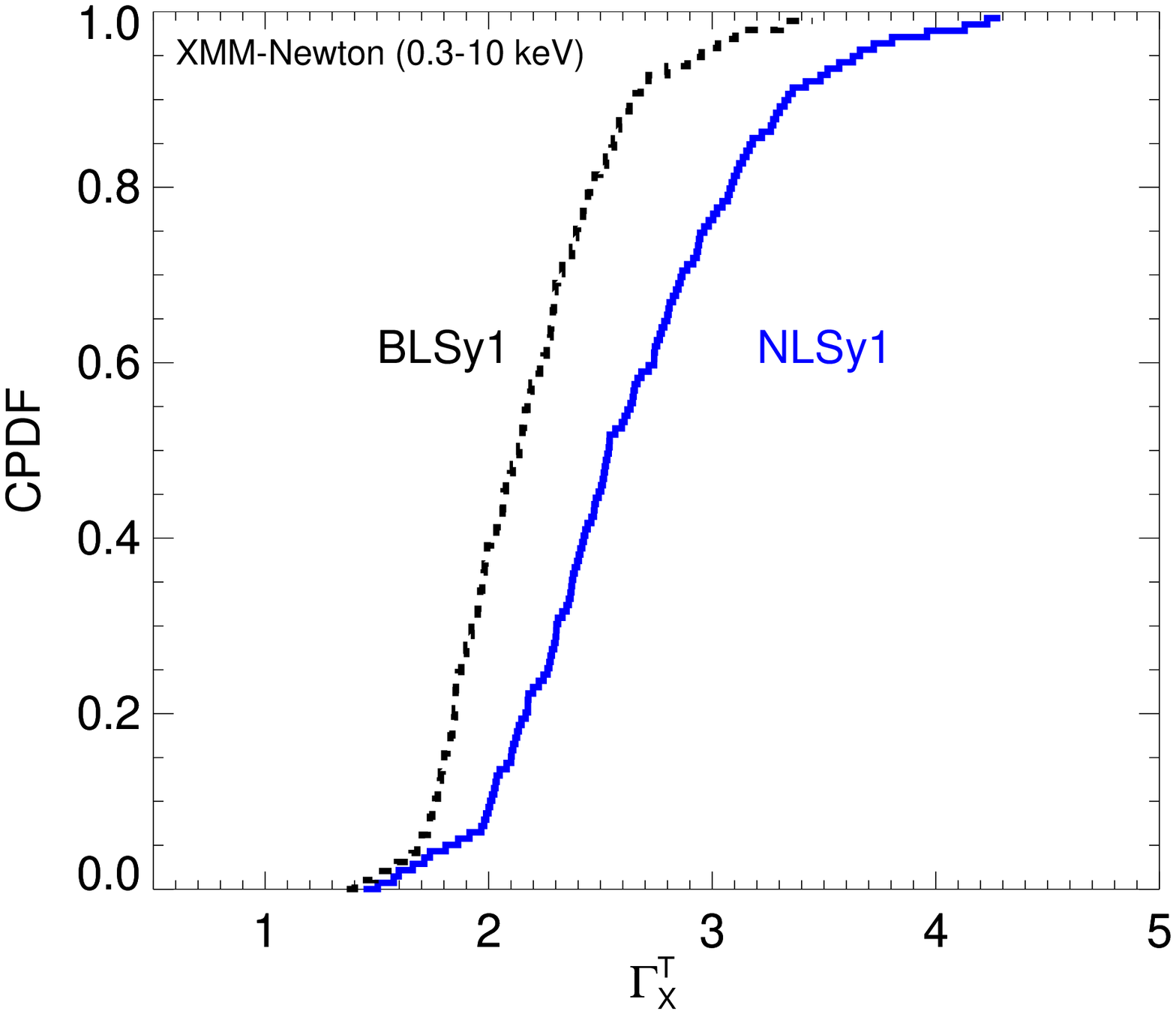}  
    \includegraphics[width=0.5\textwidth,height=0.275\textheight,angle=00]{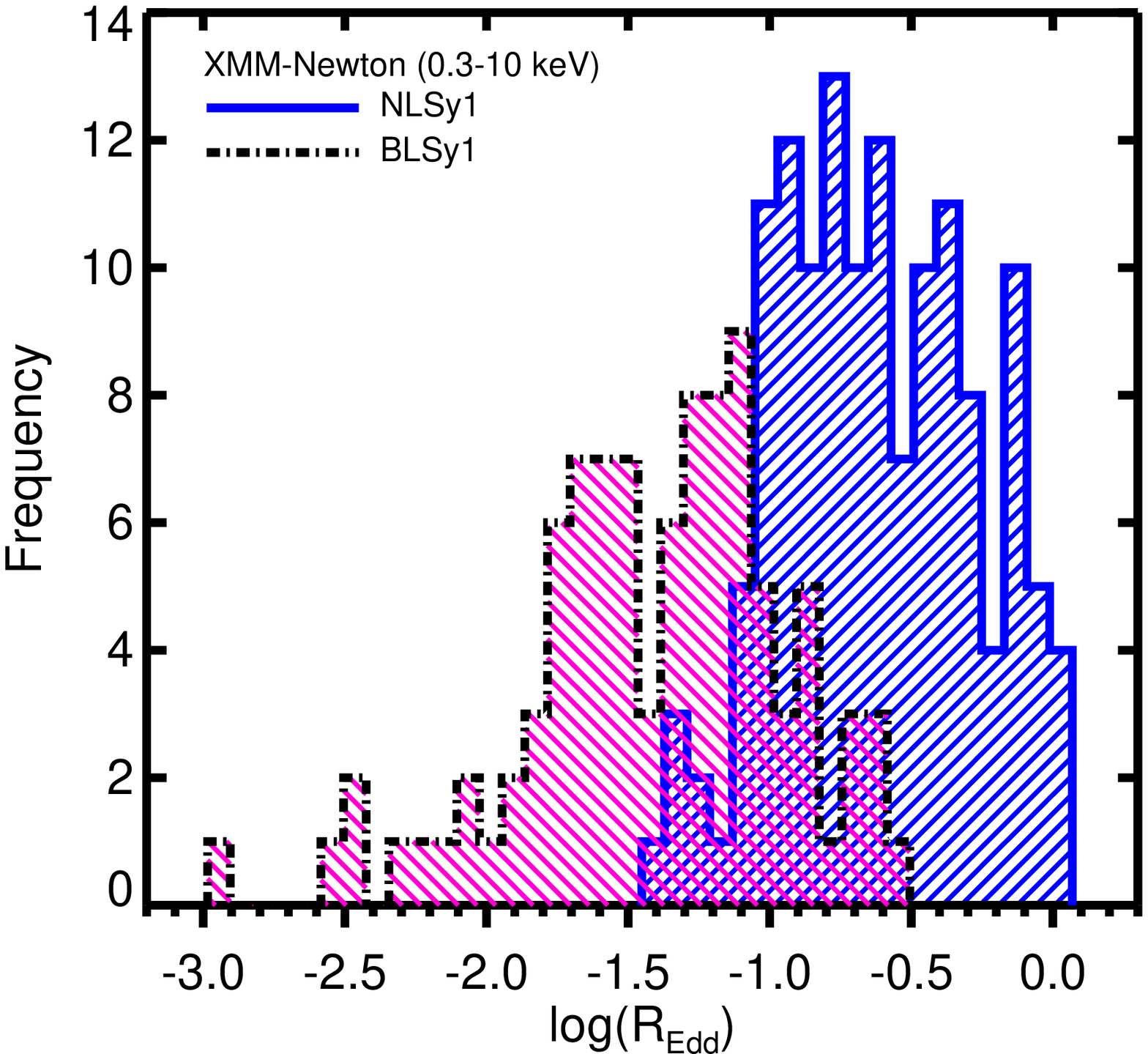}    
    \includegraphics[width=0.5\textwidth,height=0.265\textheight,angle=00]{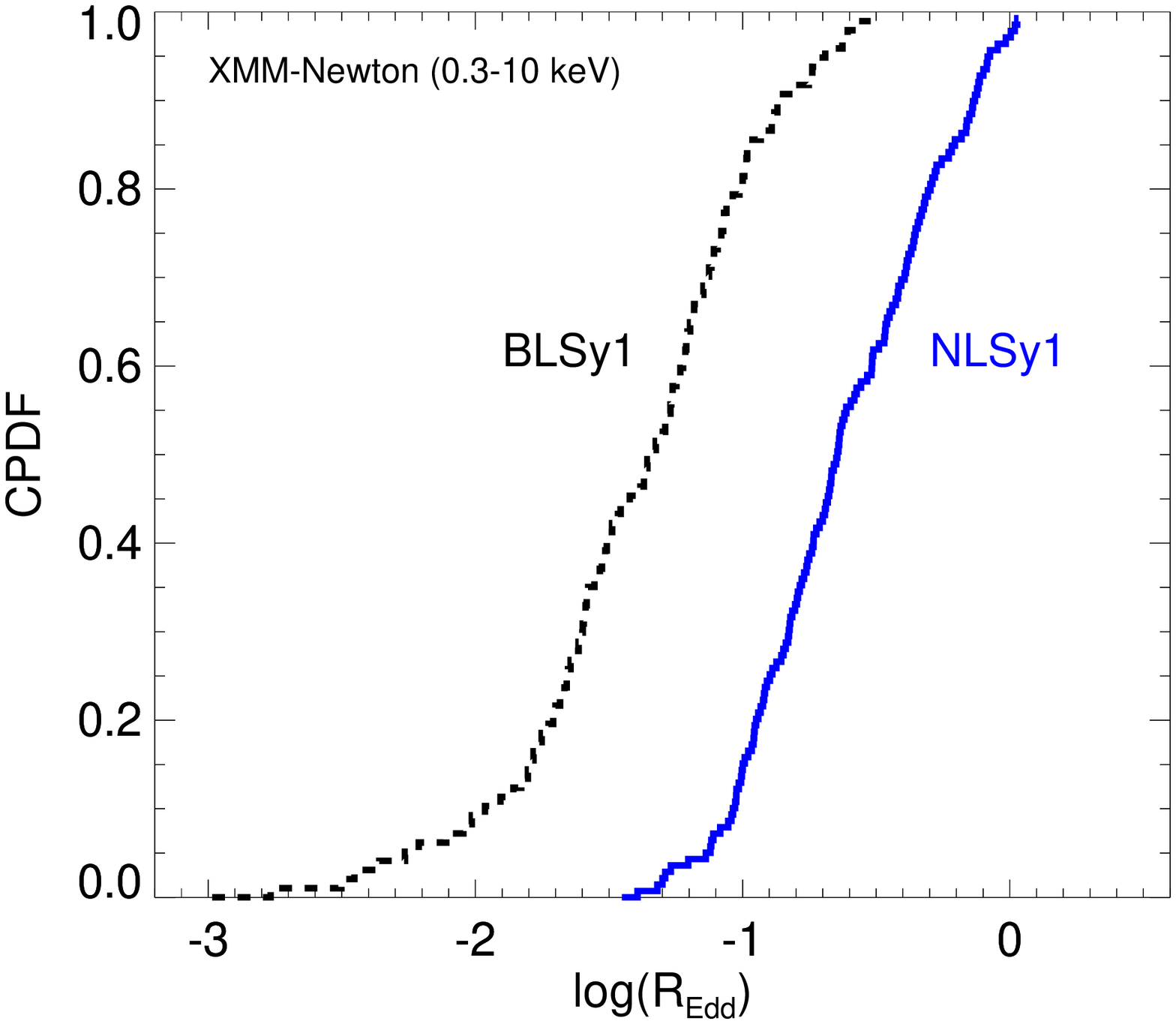}  
    \end{minipage}
    \caption{\scriptsize Same as Fig.~\ref{fig:histo_cpdf_xmm_rosat}, but using the {\it XMM-Newton} X-ray (0.3-10 keV) detected subsamples of 139 NLSy1 and 97 BLSy1 galaxies.}
    \label{fig:histo_cpdf_xmm_total}
\end{figure*}

\begin{figure*}
    \begin{minipage}[]{1.0\textwidth}
    \includegraphics[width=0.5\textwidth,height=0.275\textheight,angle=00]{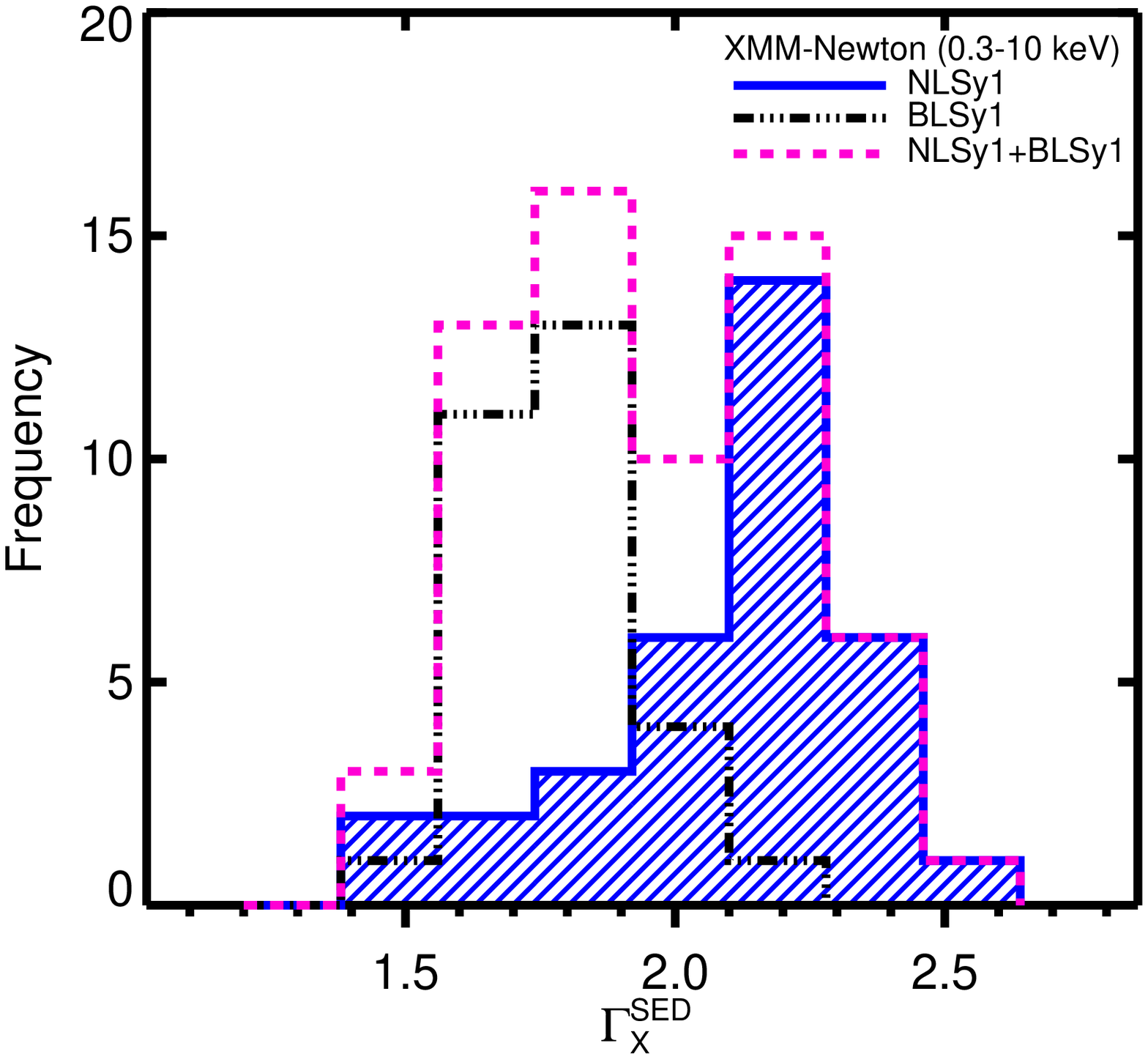}    
  \includegraphics[width=0.5\textwidth,height=0.265\textheight,angle=00]{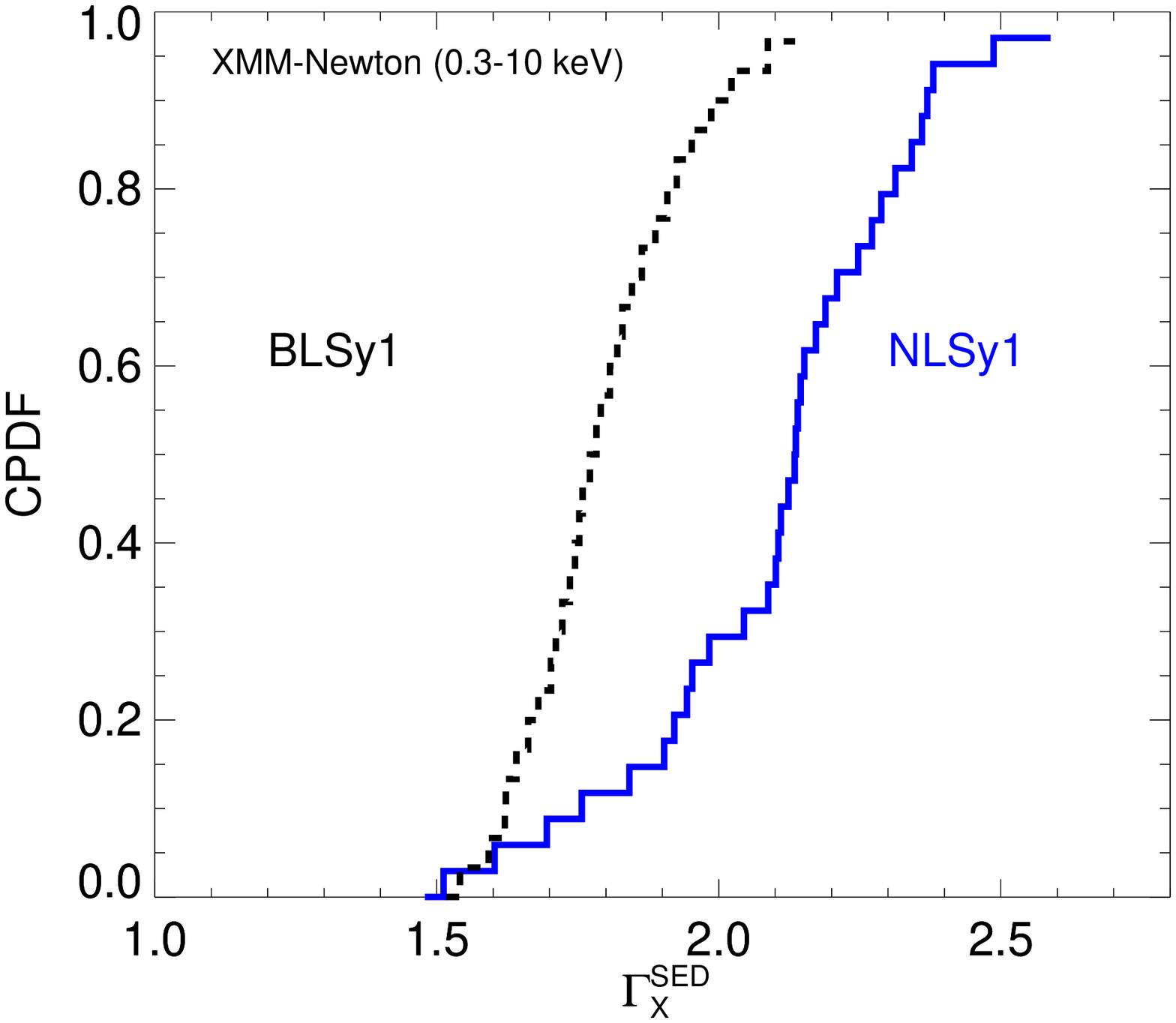}  
    \includegraphics[width=0.5\textwidth,height=0.275\textheight,angle=00]{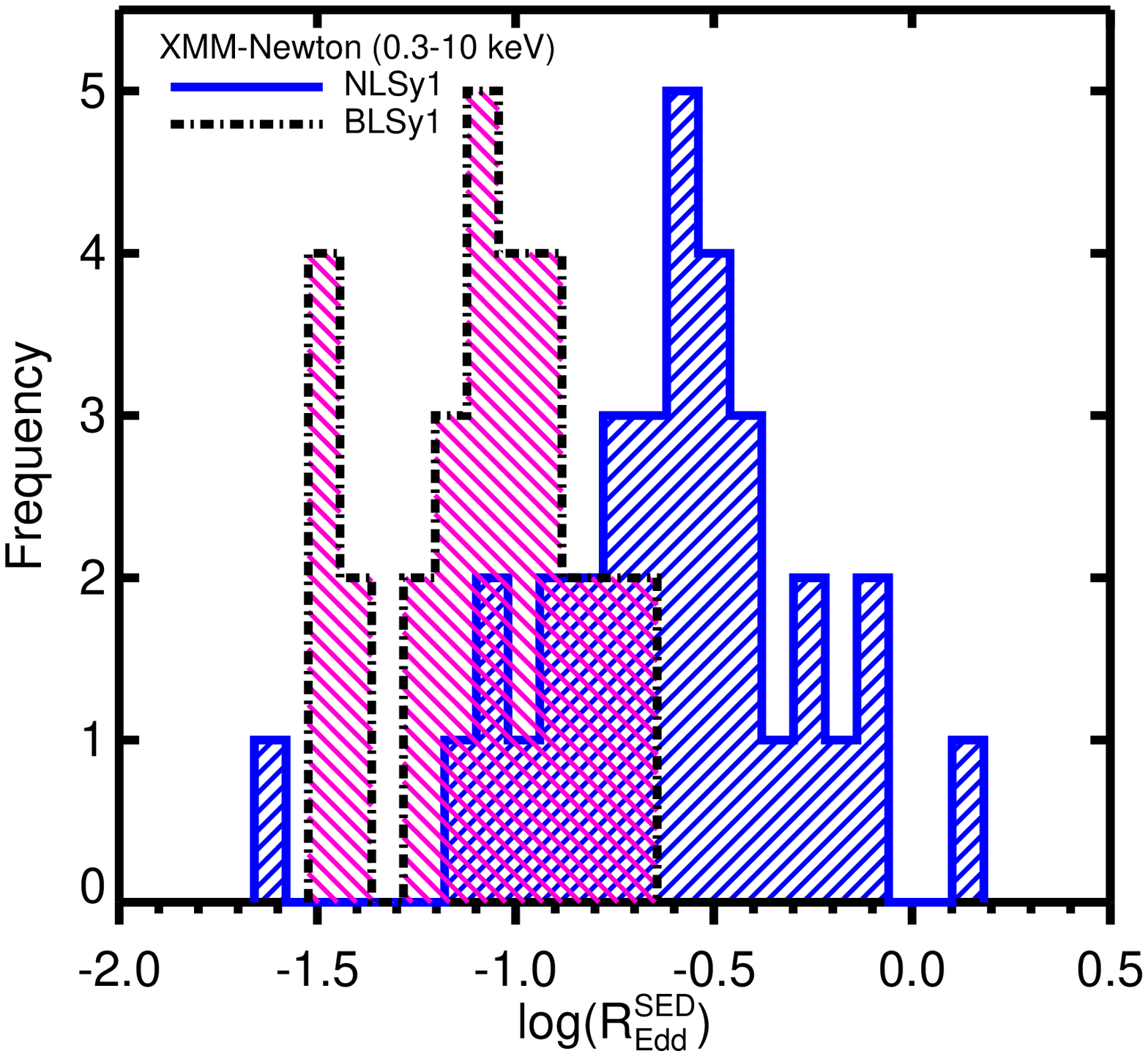}    
    \includegraphics[width=0.5\textwidth,height=0.265\textheight,angle=00]{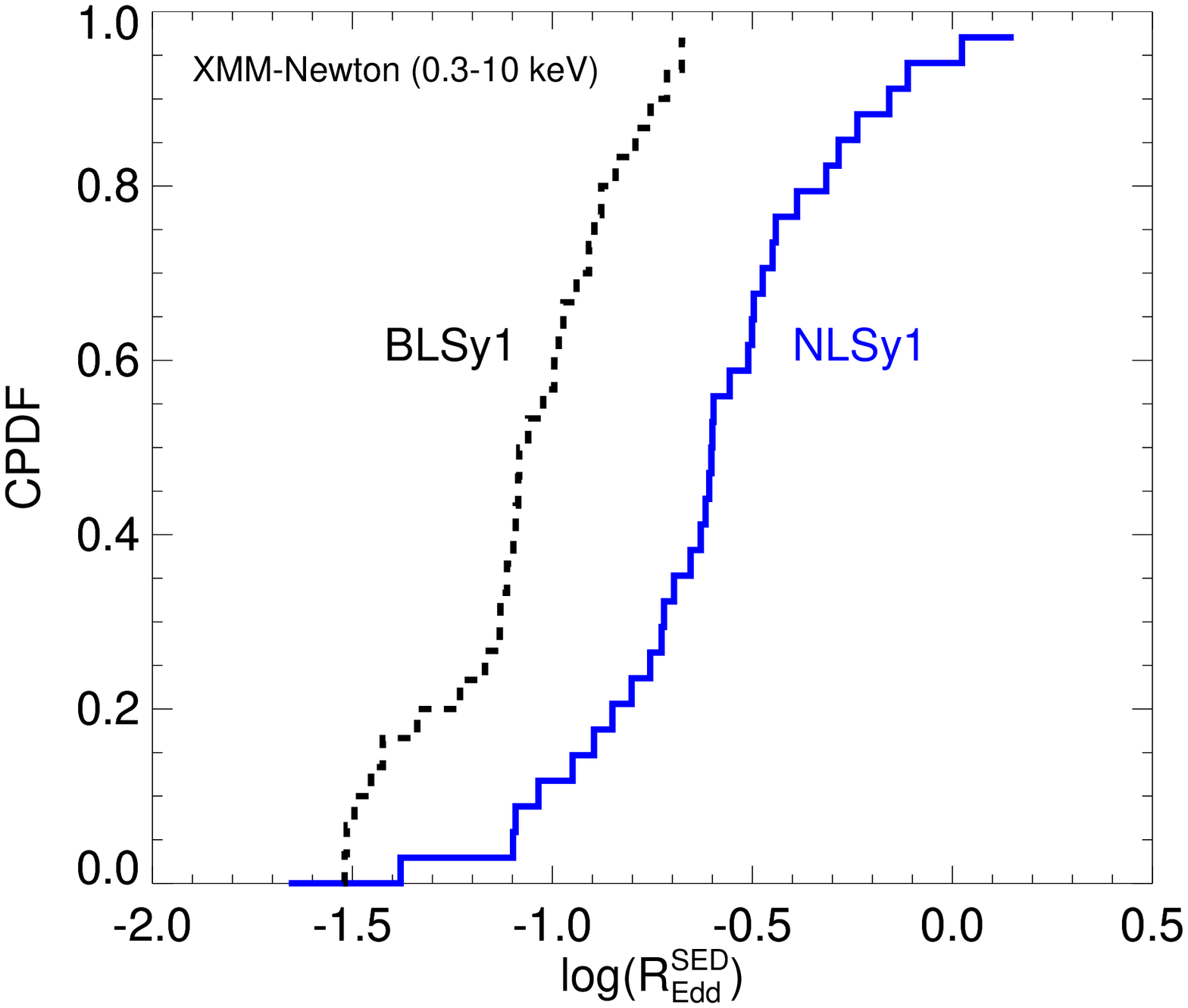}  
      \end{minipage}
    \caption{\scriptsize Same as Fig.~\ref{fig:histo_cpdf_xmm_rosat}, but using the X-ray hot photon indices ($\Gamma_{X}^\mathrm{hot}$) and Eddington ratios (R$_{\mathrm Edd}^{\mathrm SED}$) of  34 NLSy1 and  30 BLSy1 galaxies based on the {\it AGNSED} model~\citep{Kubota2018MNRAS.480.1247K} in the 0.3-10 keV energy band of the {\it XMM-Newton} data.}
      \label{fig:histo_cpdf_xmm_AGNSED}
\end{figure*}

\begin{table*}
   \centering
  \caption{Summary of the sample selection of NLSy1 and BLSy1 galaxies.}
   
  \label{sample_summary}
  
  \begin{tabular}{|c|cc|cc|cc|c|}
    \hline
    
     \multicolumn{1}{|c|}{Telescope used} &\multicolumn{6}{|c|}{Selected (taken$^{\star}$)}\\
    \hline 
  & \multicolumn{2}{|c|}{Soft energy sample} &   \multicolumn{2}{|c|}{Hard energy sample} &   \multicolumn{2}{|l|}{ Total (0.3-10 keV) energy sample} \\ 
    \hline
    
                               & NLSy1  &BLSy1 & NLSy1 & BLSy1 & NLSy1&  BLSy1\\\hline
 {{\it ROSAT}}                 & 530 ({075}) & 289 ({054}) & {---  ---} & {---  ---} & {---  ---}&   {---  ---}  \\\hline
{{\it XMM-Newton}}             & 697 ({146}) & 332 ({100}) & {148 (53)} & {103 (46)} & {148 (139)}&  {103 (97)}  \\\hline
{BOTH ({\it XMM$+$ROSAT})}     & 1227 ({221})& 621 ({154}) & {148 (53)} & {103 (46)} & {148 (139)}&  {103 (97)}  \\\hline
\multicolumn{7}{l}{$^{\star}$After imposing a minimum {S/N$\ge$10} detection criterion (using 0.1-2.0 keV range in {\it ROSAT} and 0.3-10 keV }\\ 
\multicolumn{7}{l}{in {\it XMM}) and  counting the repeated sources only once (retaining only {\it XMM} sources e.g. Sect.~\ref{section 2.0}), along with}\\
\multicolumn{7}{l}{the exclusion of those sources which could not be fitted with the adopted models (e.g. Sect.~\ref{section 4.0}).}\\

  \end{tabular}
  
\end{table*}

To make a sample of BLSy1 galaxies matching in L-z plane with our above sample of NLSy1 galaxies, so as to carry out their
comparative study (e.g. Sect.~\ref{section 1.0}).
We have used a recent compilation of~\citet{Rakshit2017ApJ...842...96R}, where they have
matched the above parent sample of 11,101 NLSy1 galaxies with that of
BLSy1 galaxies catalog both derived using SDSS DR-12. In their
compilation, they found a sample of 5511 NLSy1 and BLSy1 galaxies,
matching in the L-z plane (e.g. their figure 1). We noticed that
out of our 223 NLSy1s, 149 (57 from {\it ROSAT} and  92 from {\it XMM-Newton}) were
indeed the member of this 5511 NLSy1s sample for which L-z matched
sample of 5511 BLSy1s exists.\par
However, due to limited X-ray observations of the aforementioned
samples of NLSy1 and BLSy1 galaxies in {\it ROSAT} and {\it XMM-Newton}, we
found it difficult to construct their exact L-z matched sample for
the X-ray analysis. Nonetheless, by restricting our search for X-ray
observations of BLSy1s, only to the above sample of 5511 BLSy1s, we
can expect to have a closely L-z match in the X-ray detected NLSy1s
and BLSy1s samples. Therefore, we have cross-correlated these 5511
BLSy1s with that of the 2RXS catalog of {\it ROSAT} and also searched their
any {\it XMM-Newton} telescope based observations using a similar procedure
as we had adopted in the case of NLSy1 sample. The cross-correlation
match in {\it ROSAT} resulted in 1156 BLSy1s, among them, 289 were covered
by the PSPC instrument. Similarly, we found observations of 332 BLSy1s
in {\it XMM-Newton} telescope. Further, we also applied the  S/N$\ge$ 10
detection criterion, as had also been used in the sample of NLSy1s
(both for the {\it ROSAT} and {\it XMM} subsamples). This resulted in a sample of
156 BLSy1s consisting of  54 sources from {\it ROSAT} and  103 from
{\it XMM-Newton}. Further, we cross-correlated the  54 {\it ROSAT} BLSy1s with the
 103 {\it XMM} BLSy1s, in order to check for any common sources among them,
but none of the sources were found to be common.\par
 Furthermore, it may be noted that we have used separately
  {\it XMM}-data for the X-ray analysis of total (0.3-10 keV) and hard (2-10
  keV) energy bands as well. Out of 148 NLSy1s, the S/N$\ge$10  criterion is fulfilled by 147/148 in 0.3-10.0 keV and 56/148 in
  2-10 keV. For BLSy1s, all qualify in total 0.3-10 keV energy band,
  but only 51/103 qualify this S/N $\ge$ 10 criterion in hard energy band.\par
  Further, reduction of the samples also occur due to
  non-convergence of the spectral fit (perhaps due to artifact in
  data, see Sect.~\ref{section 4.0}), which in case of NLSy1s allow us
  to use 139/147 in 0.3-10 keV, 146/148 in 0.1-2.0 keV, and 53/56 in 2-10
  keV. Similarly, for BLSy1s we could fit the {\it XMM} sample of 97/103 in
  0.3-10 keV, 100/103 in 0.1-2.0 keV and 46/51 in 2-10 keV, as also
  summarized in Table~\ref{sample_summary}. This led to our final samples of 221 NLSy1s (75 from the {\it ROSAT} and 146 from the {\it XMM-Newton})  and 154 BLSy1s (54 from the {\it ROSAT} and 100 from the {\it XMM-Newton}), for which we have shown the histograms of their redshift and luminosity in Fig.~\ref{fig:z_lum_match_xmm_rosat}. As the figure shows, these two samples of NLSy1 and BLSy1 do moderately match in redshift, having the median redshifts of 0.21 and 0.26, respectively. This gives the Kolmogorov-Smirnov test (K-S test)  based probability of null hypothesis (P$_{null}$) of  $\sim$ 3\%. However, the difference in luminosity is found to be much higher with median values of log$(\lambda L_\lambda$(5100~\AA)) [erg/s/\AA] of 43.67 and 44.08, respectively, and P$_{null} = 2.43\times 10^{-8}$.

\section{OBSERVATIONS AND DATA REDUCTION}

\label{section 3.0}

The X-ray data of our NLSy1 and BLSy1 galaxies were based on
observations taken either with {\it ROSAT/PSPC} or with {\it XMM-Newton/EPIC}
telescopes. The {\it ROSAT} 0.1-2.0 keV spectrum of each NLSy1, BLSy1 galaxy was extracted using
the appropriate circular region around the source (e.g., see 3$^{rd}$ para of Sect.~\ref{section 2.0}).
However, while extracting the corresponding background spectrum for a given source, we had
ensured that its circular aperture is of the same size as taken for
the source and is also in the vicinity of the source, free from any contamination from the other X-ray objects.\par
The standard {\it XMM-Newton} Science Analysis System (SAS) software package (version 16.1.0) was used in data reduction  of PN detector of {\it XMM/EPIC} with updated calibration
files. {\scriptsize EPCHAIN} task was used on EPIC
``Observation Data Files'' for the preliminary processing. 
 Calibrated and concatenated event lists were extracted using the {\scriptsize EVSELECT} task of
SAS. We checked each source's data set for the high
background proton flares by making its light curve in 10 to 12 keV
energy range which is used to make the  good time interval ({\it gti}) file.
Furthermore, pile up was also checked for each source's data set using the {\scriptsize EPATPLOT} task of SAS, with the
appropriate circular region around the source, depending on the number
of pixels containing the maximum flux of the source. If found, then that was 
removed by taking only the annulus region around the source for that
data set. The SAS task {\scriptsize ESPECGET} was used to generate background and background corrected sources spectra.
Further, it may be noted that we have used the $\chi^2$ minimization technique in our analysis for which essential
  criterion is that the data points included in this technique should
be independent. So, while grouping our spectral data of {\it XMM-Newton}, we have taken
care of this point and grouped each spectrum with a minimum of 20 counts
subject to a condition that there should not be more than 4 bins 
per spectral resolution. This was done using the special task, {\scriptsize SPECGROUP} of {\scriptsize SAS} software.

\section{Analysis}
\label {section 4.01}
\subsection{Black hole masses and Eddington ratios measurement}
\label {section 4.1}
To estimate the $M_{\mathrm BH}$ in a homogeneous way as pointed out in Sect.~\ref{section 1.0}, we have opted to use the
single epoch virial method, with improved virial empirical relation
given by~\citet{Vestergaard2006ApJ...641..689V} as;

\begin{equation}
 \begin{multlined}
 \log\, M_{\mathrm BH} = \log\Bigg[\Bigg(\frac{FWHM(H\beta)}{1000\, km\,
     s^{-1}}\Bigg)^{2}\Bigg] + \left(6.91\pm0.02\right ) \\ +
 \log\Bigg(\frac{\lambda L_{\lambda}(5100\AA)}{10^{44}\, erg\,
   s^{-1}}\Bigg)^{0.50\pm0.06}
\end{multlined}
\label{mbh:equation_mbh}
\end{equation} 

where L$_{\lambda}$(5100~\AA) is the monochromatic power-law continuum
luminosity at 5100~\AA \, and FWHM(H$\beta$) is the width of the broad
component of H$\beta$ line. We have taken both these parameters from
the parent catalog of NLSy1s given
by~\citet{Rakshit2017ApJS..229...39R}.  The procedure to obtain these
parameters for BLSy1s was also similar to that used in NLSy1s, as
outlined in ~\citet{Rakshit2017ApJS..229...39R}. In brief, in their
method, they have first carried out a simultaneous fit of an AGN power-law
continuum and host galaxy contribution, by masking the AGN
emission lines. In the second step, a simultaneous fit on the host
galaxy subtracted spectrum is carried out to optimize the best fit
Gaussian profiles for the broad and narrow component of H$\beta$ lines
coming from the AGN broad and narrow line regions, respectively, along with the
underneath local continuum and blends of Fe {\sc ii} emissions~\citep[e.g., see][]{Rakshit2017ApJS..229...39R}.\par
Finally, for the estimations of the Eddington ratio, we have taken L$_{\mathrm bol}$=9.8$\times\lambda
L_{\lambda}$(5100\AA)~\citep{McLure2004MNRAS.352.1390M} and L$_{\mathrm Edd} =
1.45\times10^{38} (M_{\mathrm BH}/M_{\sun})\,erg~s^{-1}$, assuming a mixture
of hydrogen and helium so that the mean molecular weight is $\mu$ =
1.15. The values of log(M$_{\mathrm BH}$) and log(R$_{\mathrm Edd}$) along with
$\Gamma_{X}^{h}$, $\Gamma_{X}^{s}$, and $\Gamma_{X}^{T}$ for each NLSy1 and BLSy1 galaxy are
given in columns 6, 7, 8, 11, and 14 of Table~\ref{nlsy1:table_nlsy1} and
Table~\ref{blsy1:table_blsy1}, respectively.

\begin{table*}
   \centering
  \caption{Summary of the best fit model used for the spectral fitting of NLSy1 and BLSy1 galaxies.}
   
  \label{model_summary}
  
  \begin{tabular}{|c|c|c|c|}
    \hline
    
    \multicolumn{1}{|c|}{Parameter} &   \multicolumn{1}{|c|}{Model} &   \multicolumn{1}{|c|}{NLSy1} &   \multicolumn{1}{|c|}{BLSy1}\\ 
    \hline
   & \multicolumn{3}{|c|}{0.1-2.0 keV}\\
   \hline
\multirow{2}{*}{$\Gamma_{X}^{s}$}   & {\it tbabs$\times ztbabs\times$zpowerlw}                        & 74 (ROSAT) 143 (XMM) & 54 (ROSAT) 98 (XMM)\\
                  &  {\it tbabs$\times ztbabs\times$(zpowerlw$+$zbbody)}            & 01 (ROSAT) 003 (XMM) &  \multicolumn{1}{|r|}{02 (XMM)}\\
   \hline
   & \multicolumn{3}{|c|}{2-10 keV}\\
   \hline
\multirow{2}{*}{$\Gamma_{X}^{h}$}   & {\it tbabs$\times ztbabs\times$zpowerlw}                        & \multicolumn{1}{|r|}{052 (XMM)} & \multicolumn{1}{|r|}{44 (XMM)}\\
                  &  {\it tbabs$\times ztbabs\times$(zpowerlw$+$zbbody+gauss)}      &  \multicolumn{1}{|r|}{001 (XMM)} &  \multicolumn{1}{|r|}{02 (XMM)}\\
   \hline
   & \multicolumn{3}{|c|}{0.3-10 keV}\\
   \hline
\multirow{5}{*}{$\Gamma_{X}^{T}$}   & {\it tbabs$\times ztbabs\times$zpowerlw}                        &  \multicolumn{1}{|r|}{082 (XMM)} &  \multicolumn{1}{|r|}{68 (XMM)}\\
                  &  {\it tbabs$\times ztbabs\times$(zpowerlw$+$zbbody)}            &  \multicolumn{1}{|r|}{045 (XMM)} &  \multicolumn{1}{|r|}{26 (XMM)}\\
                  &  {\it tbabs$\times ztbabs\times$(zpowerlw$+$zbbody+gauss)}      &  \multicolumn{1}{|r|}{009 (XMM)} &  \multicolumn{1}{|r|}{02 (XMM)}\\   
                  &  {\it tbabs$\times ztbabs\times$zbknpower}                      &  \multicolumn{1}{|r|}{002 (XMM)} &  \multicolumn{1}{|r|}{01 (XMM)}\\
                  &  {\it tbabs$\times ztbabs\times$(zbknpower$+$zbbody)}           &  \multicolumn{1}{|r|}{001 (XMM)} &  ---------       --------\\ 
\hline

 \end{tabular}
  
\end{table*}

\begin{sidewaystable}[!htbp]
  \caption{The details of our spectral analysis of 221 NLSy1 galaxies in the soft energy band (0.1-2.0 keV) including the  53 and 139 NLSy1s analyzed also in the hard (2-10 keV) and  total (0.3-10 keV) energy bands,  respectively.}  
\fontsize{3.9}{12.0}\selectfont
\begin{tabular}{ccc ccccccccccccccccccl}
  \hline

{Source Name} & RA  & DEC &z$_{em}$  & S/N &log M$_{\mathrm BH}$ &log R$_{\mathrm Edd}$& $\Gamma_{X}^{h}$  & $\delta\Gamma_{X}^{h-}$ & $\delta\Gamma_{X}^{h+}$ & $\Gamma_{X}^{s}$  & $\delta\Gamma_{X}^{s-}$ & $\delta\Gamma_{X}^{s+}$ &  $\Gamma_{X}^{T}$  & $\delta\Gamma_{X}^{T-}$ & $\delta\Gamma_{X}^{T+}$ & log(${\lambda}L_{\lambda}$) & FWHM(H$\beta$) & $\delta(FWHM(H\beta$)) &  Aperture &\multicolumn{1}{l}{Telescope}\\
       & (deg) &  (deg) &  (redshift)   & (0.3-10 keV)&  & & \multicolumn{3}{c}{(2-10 keV)} & \multicolumn{3}{c}{(0.1-2.0 keV)}   & \multicolumn{3}{c}{(0.3-10 keV)}   & (5100 \AA) & km s$^{-1}$ & km s$^{-1}$ & used&\multicolumn{1}{l}{used}\\
          &           &    &      & & & & &    &     &      &   & &  &   & & erg s$^{-1}$& &&(arsec)&&\\
(1)     &      (2)     &  (3)  &   (4)   &   (5) &   (6)  &   (7)   &   (8) &   (9)  &   (10) &   (11) & (12) &   (13)  &   (14) &   (15) & (16)& (17)& (18) & (19) & (20) &\multicolumn{1}{l}{(21)}&\\

\hline

J010712.00$+$140845.6 & 016.800  & 14.146 & 0.0767 & 054.0 & 6.0221 & $-$0.6422 & 2.0618 & $-$0.2783 & 0.5270 & 2.5856 & $-$0.1670 & 0.2440 & 2.4017 & $-$0.0501 & 0.0797 & 42.55 & 829 & 29 & 12.59&\multicolumn{1}{l}{{\it XMM}}\\

J014644.88$-$004044.4 & 026.687 & $-$00.679 & 0.0824 & 030.2 & 6.7026 & $-$0.6527 & ----- & ----- & ----- & 2.9201 & $-$0.2971 & 0.3178 & ----- & ----- & ----- & 43.22 & 1234 & 20 & 57.25 &\multicolumn{1}{l}{{\it ROSAT}}\\
J081442.00$+$212916.8 & 123.675 & 21.488 & 0.1626 & 175.6 & 7.3140 & $-$0.4641 & 2.0722 & $-$0.0711 & 0.1415 & 2.7822 & $-$0.0454 & 0.0715 & 2.8400 & $-$0.0875 & 0.1012 & 44.02 & 1574 & 22 & 35.87&\multicolumn{1}{l}{{\it XMM}}\\
       .....        &  ..... &.....  & .....  & .....& ..... &.....      &   ..... 	&.....   &.....  &.....  & .....  & .....&   .....  & ..... & ..... &   .....& ..... & .....& .....& \multicolumn{1}{l}{.....}	\\
 
 \hline
 
\multicolumn{22}{l}{Note:$-$ The entire table is available  in the online version. Only a portion of this table is shown here, to display its form and contents.}\\
 
\end{tabular}
\label{nlsy1:table_nlsy1}
\end{sidewaystable}

\begin{sidewaystable}[!htbp]

  \caption{The details of our spectral analysis of  154 BLSy1 galaxies in the soft energy band (0.1-2.0 keV) including the 46 and 97 BLSy1s analyzed also in the hard (2-10 keV) and total (0.3-10 keV) energy bands, respectively.}  
  \label{blsy1:table_blsy1} 
 \fontsize{4.1}{12.0}\selectfont
\begin{tabular}{ccc ccccccccccccccccccl}
\hline

{Source Name} & RA  & DEC &z$_{em}$  & S/N &log M$_{\mathrm BH}$ &log R$_{\mathrm Edd}$& $\Gamma_{X}^{h}$  & $\delta\Gamma_{X}^{h-}$ & $\delta\Gamma_{X}^{h+}$ & $\Gamma_{X}^{s}$  & $\delta\Gamma_{X}^{s-}$ & $\delta\Gamma_{X}^{s+}$ &  $\Gamma_{X}^{T}$  & $\delta\Gamma_{X}^{T-}$ & $\delta\Gamma_{X}^{T+}$& log(${\lambda}L_{\lambda}$) & FWHM(H$\beta$) & $\delta(FWHM(H\beta$)) & Aperture &\multicolumn{1}{l}{Telescope}\\
       & (deg) &  (deg) &  (redshift)   & &  & & \multicolumn{3}{c}{(2-10 keV)}  & \multicolumn{3}{c}{(0.1-2.0 keV)} & \multicolumn{3}{c}{(0.3-10 keV)} & (5100 \AA) & km s$^{-1}$ & km s$^{-1}$ & used &\multicolumn{1}{l}{used}\\
&           &    &      & & & & &    &     &      &   & &  &   & & erg s$^{-1}$& &&(arcsec)& &\\
     (1)     &      (2)     &  (3)  &   (4)   &   (5) &   (6)  &   (7)   &   (8) &   (9)  &   (10) &   (11) & (12) &   (13)  &   (14) &   (15) & (16)& (17)& (18) & (19) & (20) &\multicolumn{1}{l}{(21)}&\\

\hline
J002113.20$-$020115.6 & 005.305 & $-$02.021 & 0.7621 & 024.6 & 8.8620 & $-$1.5821 & ----- & ----- & ----- & 1.8349 & $-$0.2342 & 0.4626 & 1.9144 & $-$0.1726 & 0.2121 & 44.45 & 7303 & 733 &  16.45 &\multicolumn{1}{l}{{\it XMM}}\\
J044759.52$-$043231.2 & 071.998 & $-$04.542 & 0.2569 & 016.2 & 8.5913 & $-$1.6014 & ----- & ----- & ----- & 2.7118 & $-$0.1975 & 0.3120 & ----- & ----- & ----- & 44.16  & 6318 & 131 & 93.79 &\multicolumn{1}{l}{{\it ROSAT}}\\

J075112.24$+$174351.6 & 117.801 & 17.731 & 0.1861 & 169.6 & 7.9220 & $-$1.1421 & 1.7954 & $-$0.0854 & 0.0580 & 2.2096 & $-$0.0355 & 0.0711 & 2.1931 & $-$0.0185 & 0.0595 & 43.95 & 3299 & 54 & 22.77 &\multicolumn{1}{l}{{\it XMM}}\\
 
          .....       &   ..... &   .....  & ..... &  ..... &  &.....      &   ..... 	&.....   &.....  &.....  & .....  & .....&   .....  & ..... & ..... &   .....& ..... & .....&.....&\multicolumn{1}{l}{.....} 	\\
 \hline

 \multicolumn{22}{l}{Note:$-$ The entire table is available in the online version. Only a portion of this table is shown here, to display its form and contents.}\\
\end{tabular}
\end{sidewaystable}

\subsection{X-ray spectral analysis}
\label{section 4.0}

For the spectral analysis of {\it ROSAT} detected  75 NLSy1 and  54 BLSy1 galaxies,
we have used {\scriptsize XSPEC} version 12.10.1~\citep{Arnaud1996ASPC..101...17A, Dorman2001ASPC..238..415D} tasks of {\scriptsize HEASOFT}. The response matrices files (RMFs) required for the spectral
fitting were obtained from the latest calibration database available
publicly on the {\scriptsize HEASARC} calibration database\footnote{https://heasarc.gsfc.nasa.gov/docs/heasarc/caldb/} and the ancillary response files (ARFs) were generated with
the {\scriptsize PCARF} task of {\scriptsize HEASOFT}. The extracted spectrum (e.g. Sect.~\ref{section 3.0}) of each NLSy1, BLSy1 was grouped
with a minimum of 20 counts per bin using the {\scriptsize GRPPHA} routine of the {\scriptsize XSELECT} task, which
permitted us to use $\chi^{2}$ minimization for spectral fitting. To
obtain the soft X-ray (0.1-2.0 keV) photon indices (hereafter $\Gamma_{X}^{s}$),
we have used the physically motivated model consisting of basic power
law and the double neutral absorption (i.e., {\it
  tbabs$\times ztbabs\times$zpowerlw}) in {\scriptsize XSPEC}
software to the spectral data in the observed frame energy range of
0.1-2.0 keV. Special cares were taken to properly fit the absorption of
soft X-rays during the fitting of NLSy1 and BLSy1 galaxies.
During the fitting, the redshift of the source was kept fixed to its precise
value known from its optical spectrum and also Galactic hydrogen column
density in the direction of the source was kept fixed based on the value 
given by~\citet{Dickey1990ARA&A..28..215D}. However, the normalization, host galaxy absorption, intrinsic absorption, and
$\Gamma_{X}^{s}$ of the source were the free parameters of the fit.
In one of our NLSy1 galaxy, viz., J162901.20$+$400758.8, we noted soft X-ray
excess below 0.5 keV. For fitting this source, we added a 
blackbody component to our basic model
(i.e., {\it tbabs$\times ztbabs\times$(zpowerlw+zbbody)}).  
In summary, we could estimate using {\it ROSAT} data in 0.1-2.0 keV range, the $\Gamma_{X}^{s}$ of 75 NLSy1 and
54 BLSy1 galaxies, as listed in Table~\ref{nlsy1:table_nlsy1} and
Table~\ref{blsy1:table_blsy1} (see columns 11-13), respectively.\par
For spectral analysis of the {\it XMM/EPIC-PN} sample, we have used physically motivated models as had been employed above for {\it ROSAT/PSPC} sample.
Our model fittings converged well to estimate soft X-ray (0.3-2.0 keV) photon indices for the 146/148 NLSy1 and 100/103 BLSy1 galaxies, and for the 
 hard X-ray (2-10 keV) photon indices (hereafter
$\Gamma^{h}_{X}$) in 53/56 NLSy1 and 46/51 BLSy1 galaxies. Among them,  51
NLSy1 and  44 BLSy1 galaxies were fitted in both hard and
soft energy bands of {\it XMM-Newton}, while  2 NLSy1 and   2 BLSy1 galaxies were only
fitted in hard X-ray energy band. 
We also noticed the soft X-ray
excesses in 3 NLSy1 and 2 BLSy1 galaxies and fitted them by adding black body component to our basic model. However, one NLSy1 galaxy (viz., J105128.32$+$335851.6) and 2 BLSy1 (viz., J125553.04$+$272403.6 and J161745.6$+$060350.4) galaxies also showed emission line feature.
Therefore, to fit these sources,  we had added the black body and
Gaussian components  to our basic model, viz., {\it tbabs$\times ztbabs\times$(zpowerlw+zbbody+gauss)}. A summary of these best fit models is given in Table~\ref{model_summary}.\par
Representative data and best fit folded models for one member of
NLSy1 and BLSy1 galaxies samples, each in {\it ROSAT} and {\it XMM-Newton}
are shown in the top and bottom panels of Fig.~\ref{fig:folded_spec_nlsy1_blsy1},
respectively. The typical range of the X-ray photon index
varies for our samples of NLSy1 and BLSy1 galaxies in the soft
energy band from 1.1-4.4 and 1.3-3.6, respectively, while in
the hard energy band varies from 1.2-2.6 and 1.2-2.7,
respectively.\par
In summary, based on our combined sample of 223 NLSy1s (75 from the
{\it ROSAT} and 148 from the {\it XMM}), we got $\Gamma^{s}_{X}$
measurements for the 221 NLSy1s (75 from the {\it ROSAT} and  146 from the
{\it XMM}) as listed in Table~\ref{nlsy1:table_nlsy1}. Similarly, out
of the sample of  156 BLSy1s (54 from the {\it ROSAT} and  103 from the
{\it XMM}), we got $\Gamma^{s}_{X}$ for  154 BLSy1s (54 from the {\it ROSAT}
and 100 from the {\it XMM}) as listed in
Table~\ref{blsy1:table_blsy1}. The histograms and cumulative probability distribution functions (CPDFs) of
$\Gamma^{s}_{X}$ for the samples of 221  NLSy1 and 154 BLSy1 galaxies
along with their combined sample (hereafter [NLSy1$+$BLSy1])  are shown in the top panels of
Fig.~\ref{fig:histo_cpdf_xmm_rosat}. However, for the $\Gamma^{h}_{X}$
measurements, we could use only  53 NLSy1 and  46 BLSy1 galaxies
based on their {\it XMM} subsamples of  56 NLSy1 and  51 BLSy1
galaxies, as listed in Table~\ref{nlsy1:table_nlsy1} and Table~\ref{blsy1:table_blsy1}, respectively. The
histograms and CPDFs of $\Gamma^{h}_{X}$ are shown in the top panels of
Fig.~\ref{fig:histo_cpdf_xmm_hard}.

\subsection{X-ray spectral analysis in 0.3-10 keV energy range}
\label{section 4.2}

The spectral coverage of the {\it XMM} data also allows us to estimate the 0.3-10 keV photon indices (hereafter $\Gamma_{X}^{T}$). This is also useful to compare $\Gamma_{X}^{T}$ distribution with the $\Gamma_{X}^{s}$ and $\Gamma_{X}^{h}$ distributions, both for NLSy1 and BLSy1 galaxies. Our fittings in the 0.3-10 keV could converge for 
82/146 NLSy1s and 68/100 BLSy1s with our basic model, i.e., {\it tbabs$\times ztbabs\times$(zpowerlw)}. However, in case of 
45 NLSy1 and 26 BLSy1 galaxies, we had to add black body component to our basic model, i.e., {\it tbabs$\times ztbabs\times$(zpowerlw+zbbody)}. There were requirement of an additional Gaussian emission component for the fittings of 8 NLSy1 and 2 BLSy1 galaxies., i.e.,
{\it tbabs$\times ztbabs\times$(zpowerlw+zbbody+gauss)}.
 For 2 NLSy1s (viz., J150506.48$+$032631.2, J151312.48$+$001937.2) and 1 BLSy1  (viz., J135435.76$+$180516.8), it was found that they can not be fitted with single powerlaw, therefore, they were accounted by broken power law model 
 i.e., {\it tbabs$\times ztbabs\times$(zbknpower)}. For one NLSy1, viz., J12410.56$+$331702.4, similar feature with additional soft X-ray excess was accommodated by the inclusion of black-body emission, e.g.,  as 
 {\it tbabs$\times ztbabs\times$(zbknpower+zbbody)}.
 The above combination of models, as also summarized in Table~\ref{model_summary}, have allowed us to estimate $\Gamma_{X}^\mathrm{T}$ of 139/148 NLSy1  and 97/100 BLSy1 galaxies, whose distributions are shown in Fig.~\ref{fig:histo_cpdf_xmm_total}. From this distributions, it can be seen that the typical range of $\Gamma_{X}^\mathrm{T}$ varies for our subsamples of NLSy1 and BLSy1 galaxies in the total energy band from 1.4-4.3 and 1.4-3.4, respectively which is consistent with the range $\Gamma_{X}^\mathrm{s}$ and $\Gamma_{X}^\mathrm{h}$ (e.g see Sect.~\ref{section 4.0} above).

\subsection{X-ray spectral analysis in 0.3-10 keV energy range using {\it AGNSED} model}
\label{section 4.3}
  
  Another physically motivated model consisting of the spectral energy
  distribution of the AGN~\citep[hereafter, {\it AGNSED}, or `agnsed',][]{Kubota2018MNRAS.480.1247K} and the Galactic absorption
  (i.e., {\it tbabs$\times$agnsed}), can also be employed using {\scriptsize XSPEC} on the {\it XMM} spectra in 0.3-10 keV energy range.
  Here, we have limited ourself only to those 53 NLSy1s and 46 BLSy1s
  {\it XMM} sources, for which their hard energy data has also enabled us
  (due to sufficient S/N) to estimate their $\Gamma^{h}_{X}$ as well
  (e.g. Sect.~\ref{section 4.0}).
  Special cares were also taken to properly fit the absorption of soft
  X-rays, by fixing the Galactic hydrogen column density 
  to its value given by~\citet{Dickey1990ARA&A..28..215D}. During the fitting black hole mass,
  redshift, and comoving (proper) distance of the source were kept
  fixed to their precise value known from the optical spectra. We also
  kept fix, the black hole spin, inclination angle
  i (for the warm comptonising component and the outer disc), electron
  temperature (for the hot comptonization component), reprocessing,
  and normalization parameters to 0.5, 30$^{\circ}$, 100 keV, 0, and
  1, respectively. However, the other parameters of this model, viz.,
  Eddington ratio (R$_{\mathrm Edd}^{\mathrm SED}=\dot{m}$), electron temperature for the warm
  comptonization component (kT$_{e}^\mathrm{warm}$), hot photon index
  ($\Gamma^\mathrm{hot}_{X}$), warm photon index
  ($\Gamma^\mathrm{warm}_{X}$), outer radius of the hot comptonization
  component, and outer radius of the warm comptonization component
  were kept free. \par
  In 3 NLSy1 and 2 BLSy1 galaxies, we noted soft X-ray feature or warm
  absorption below 2 keV. To carry out the fitting of these 5 sources,
  we considered a warm absorber model, namely {\it zxipcf} in addition to the above model (i.e., {\it
    tbabs$\times zxipcf\times$agnsed}). The above combination of
  models allowed us to estimate $\Gamma_{X}^\mathrm{hot}$ and R$_{\mathrm Edd}^{\mathrm SED}$ of 34 NLSy1 and 30
  BLSy1 galaxies, whose distributions are shown in Fig.~\ref{fig:histo_cpdf_xmm_AGNSED}.
  However, the fit of the remaining  19 (out of a total of 53)
  NLSy1 and 16 BLSy1 (out of a total of 46) galaxies did not converge with either of the
  above models due to either their bad data coverage up to 10 keV or
  non-convergence of their fit parameters.\par

\begin{sidewaystable}[!htbp]

  \caption {Results of the correlation analysis for the samples of soft X-ray energy selected  221 NLSy1 and  154 BLSy1 galaxies and the subsamples of 53 NLSy1s,  46 BLSy1s  and 139 NLSy1s, 97 BLSy1s of hard and total X-ray energies selected, respectively. The photon indices (in soft, hard, and total energy bands) are estimated using a simple power-law and absorption model.}
 \label{table_nlsy1_blsy1_corr}
 \fontsize{6.7}{7.0}\selectfont
  \begin{tabular}{l r r r l r r r l r r r l}\\
\hline
   
    \multicolumn{1}{c}{{\it ROSAT/XMM} (0.1-2.0 keV)} &  \multicolumn{4}{c}{NLSy1 (221 sources)} &\multicolumn{4}{c}{BLSy1 (154 sources)}  &\multicolumn{4}{c}{NLSy1$+$BLSy1 (375 sources)}\\
   \hline
  {Correlation} & \multicolumn{1}{c}{m$^{\dagger}$} & \multicolumn{1}{c}{c$^{\ddagger}$}&  \multicolumn{1}{r}{$\rho^{\star}$} &   \multicolumn{1}{c}{P$_{null}^{\ast}$}  & \multicolumn{1}{c}{m$^{\dagger}$} &  \multicolumn{1}{c}{c$^{\ddagger}$}& \multicolumn{1}{r}{$\rho^{\star}$} &   \multicolumn{1}{c}{P$_{null}^{\ast}$} & \multicolumn{1}{c}{m$^{\dagger}$} &  \multicolumn{1}{c}{c$^{\ddagger}$}& \multicolumn{1}{r}{$\rho^{\star}$} &   \multicolumn{1}{c}{P$_{null}^{\ast}$}\\

       \hline

$\Gamma_{X}^{s}-\log(R_{\mathrm Edd})$ & {  0.67$\pm$0.04  }& {  3.27$\pm$0.02  }& {  0.44  }& {  1.28$\times$10$^{-11}$       }& {  0.37$\pm$0.03 }& {  2.79$\pm$0.04 }& {  0.41 }& {  1.71$\times$10$^{-7}$    }& {  0.62$\pm$0.02 }& {  3.19$\pm$0.02 }& {  0.62 }& {  1.16$\times$10$^{-40}$}\\
$\Gamma_{X}^{s}-\log(L_{\mathrm bol}$)& {  0.07$\pm$0.02   }& { $-0.31\pm$0.88 }& {  0.36 }& {  4.10$\times$10$^{-8}$          }& {  0.24$\pm$0.04 }& {  $-$8.36$\pm$1.56 }& {  0.08 }& {  0.30                  }& {  $-$0.09$\pm$0.02 }& {  6.81$\pm$0.71 }& {  0.04 }& {  0.47}\\
$\Gamma_{X}^{s}-\log(M_{\mathrm BH}$) & { $-$0.23$\pm$0.02}& {  4.50$\pm$0.18   }& {  0.15  }& {   0.03                       }& {  $-$0.27$\pm$0.03 }& {  4.51$\pm$0.28 }& { $-$0.25}& {   2.03$\times$10$^{-3}$  }& { $-$0.40$\pm$0.01 }& {  5.68$\pm$0.09 }& {  $-$0.40 }& {  1.73$\times$10$^{-15}$}\\ 
$\Gamma_{X}^{s}-\log(FWHM (H{\beta}$)) & {  $-$1.66$\pm$0.08 }& {  8.14$\pm$0.27 }& {  $-$0.21 }& {  1.36$\times$10$^{-3}$  }& {  $-$0.81$\pm$0.07 }& {  5.27$\pm$0.25 }& {   $-$0.35 }& {  9.91$\times$10$^{-6}$   }& {  $-$1.18$\pm$0.03 }& {  6.61$\pm$0.10 }& {  $-$0.55 }& {  2.87$\times$10$^{-31}$}\\
 \hline

    \multicolumn{1}{c}{{\it XMM (2-10 keV)}} &  \multicolumn{4}{c}{NLSy1 ({ 53} sources)} &\multicolumn{4}{c}{BLSy1 ({ 46} sources)} &\multicolumn{4}{c}{NLSy1$+$BLSy1 ({ 99} sources)}\\     
 \hline

$\Gamma_{X}^{h}-\log(R_{\mathrm Edd})$ & {  0.29$\pm$0.06  }& {  2.33$\pm$0.03  }& {  0.42  }& {  1.64$\times$10$^{-3}$  }& {  0.17$\pm$0.09 }& {  2.03$\pm$0.14 }& {  0.43 }& {  2.65$\times$10$^{-3}$   }& {  0.35$\pm$0.03 }& {  2.34$\pm$0.03 }& {  0.56 }& {  2.48$\times$10$^{-9}$}\\
$\Gamma_{X}^{h}-\log(L_{\mathrm bol}$)& {  0.10$\pm$0.07   }& { $-$2.27$\pm$2.99   }& {  0.23  }& {  0.10               }& {  0.11$\pm$0.09 }& {  $-$3.23$\pm$3.51 }& {  0.34 }& {  0.02                }& {  $-$0.06$\pm$0.05 }& {  4.94$\pm$2.15 }& {  0.09 }& {  0.38}\\
$\Gamma_{X}^{h}-\log(M_{\mathrm BH}$) & { $-$0.27$\pm$0.06}& {  4.03$\pm$0.43   }& {  $-$0.04  }& {   0.79              }& {  $-$0.02$\pm$0.08 }& {  1.98$\pm$0.65 }& { $-$0.20}& {   0.18              }& {  $-$0.28$\pm$0.03 }& {  4.09$\pm$0.19 }& {  $-$0.37 }& {  1.52$\times$10$^{-4}$}\\ 
$\Gamma_{X}^{h}-\log(FWHM (H{\beta}$)) & {  $-$0.70$\pm$0.13 }& {  4.35$\pm$0.41 }& {  $-$0.37 }& {  6.47$\times$10$^{-3}$      }& {  $-$0.26$\pm$0.20 }& {  2.71$\pm$0.73 }& {   $-$0.36 }& {  1.54$\times$10$^{-2}$   }& {  $-$0.68$\pm$0.06 }& {  4.27$\pm$0.19 }& {  $-$0.48 }& {  3.57$\times$10$^{-7}$}\\

  \hline

 \multicolumn{1}{c}{ {\it XMM (0.3-10 keV)}} &  \multicolumn{4}{c}{ NLSy1 (139 sources)} &\multicolumn{4}{c}{ BLSy1 (97 sources)} &\multicolumn{4}{c}{ NLSy1$+$BLSy1 (236 sources)}\\     
 \hline

{ $\Gamma_{X}^{T}-\log(R_{\mathrm Edd})$ }& {  0.51$\pm$0.03  }& {  2.91$\pm$0.03  }& {  0.42  }& {  3.84$\times$10$^{-7}$  }& {  0.34$\pm$0.02 }& {  2.56$\pm$0.04 }& {  0.47 }& {  1.11$\times$10$^{-6}$   }& {  0.50$\pm$0.01 }& {  2.86$\pm$0.02 }& {  0.56 }& {  2.76$\times$10$^{-21}$}\\
{ $\Gamma_{X}^{T}-\log(L_{\mathrm bol}$)}& {  0.06$\pm$0.02   }& { $-$0.18$\pm$0.88 }& {  0.23 }& {  7.32$\times$10$^{-3}$  }& {  0.02$\pm$0.02 }& {  1.14$\pm$1.05 }& {  0.14 }& {  0.17                }& {  $-$0.22$\pm$0.01 }& {  12.23$\pm$0.59 }& {  $-$0.03 }& {  0.68}\\
{ $\Gamma_{X}^{T}-\log(M_{\mathrm BH}$) }& { $-$0.22$\pm$0.02}& {  4.08$\pm$0.18   }& {  $-$0.01  }& {   0.93              }& {  $-$0.28$\pm$0.02 }& {  4.35$\pm$0.17 }& { $-$0.31}& {   2.14$\times$10$^{-3}$        }& {  $-$0.34$\pm$0.01 }& {  4.91$\pm$0.06 }& {  $-$0.41 }& {  9.05$\times$10$^{-11}$}\\ 
{ $\Gamma_{X}^{T}-\log(FWHM (H{\beta}$)) }& {  $-$1.32$\pm$0.08 }& {  6.73$\pm$0.25 }& {  $-$0.30 }& {  2.84$\times$10$^{-4}$      }& {  $-$0.76$\pm$0.05 }& {  4.87$\pm$0.17 }& {   $-$0.41 }& {  2.62$\times$10$^{-5}$   }& {  $-$0.91$\pm$0.02 }& {  5.42$\pm$0.07 }& {  $-$0.52 }& {  4.49$\times$10$^{-18}$}\\

  \hline
    
       \multicolumn{13}{l}{$^{\dagger,\ddagger}$Slope ($m$) and intercept ($c$) of a best-fit linear correlation of the form {\it y=mx$+$c}. In all the correlations y=$\Gamma_{X}$. The independent variables X are: X=log(R$_{\mathrm Edd}$) for the $\Gamma_{X}-\log(R_{\mathrm Edd}$) corelation,}\\
       \multicolumn{13}{l}{X=log(L$_{\mathrm bol}$) for the $\Gamma_{X}-\log(L_{\mathrm bol}$) corelation, X=log(M$_{\mathrm BH}$/M$_{\sun}$) for the $\Gamma_{X}-\log(M_{\mathrm BH}$/M$_{\sun}$) corelation, and X=log(FWHM(H${\beta}$)) for the $\Gamma_{X}-\log(FWHM(H{\beta}$)) corelation.}\\

       \multicolumn{13}{l}{$^{\star}$Spearman's correlation coefficient ($\rho$).}\\
       \multicolumn{13}{l}{$^{\ast}$Probability of a null correlation from Spearman's test (P$_{null}$).}\\

 \end{tabular}
  \end{sidewaystable}

\section{Results}
\label {section 5.0}
\subsection{Comparison of $\Gamma_{X}$ and R$_{\mathrm Edd}$ among the samples of BLSy1 and NLSy1 galaxies}
\label {section 5.1}

We used a sample of 221 NLSy1s to compare its physical parameters with 154 BLSy1s (e.g., see
Table~\ref{sample_summary}, Sect.~\ref{section 2.0}) moderately
matching in the redshift plane (e.g., see Fig.~\ref{fig:z_lum_match_xmm_rosat}).
The histograms of our homogeneous analysis (e.g. Sect.~\ref{section 4.0}) of 0.1-2.0 keV photon indices for both the
samples are shown in the top left panel of
Fig.~\ref{fig:histo_cpdf_xmm_rosat}. As can be seen from these
histogram plots that there is a clear difference in the photon indices distributions
with the median values of  2.81 and  2.30 for the samples of NLSy1 and
BLSy1 galaxies, respectively, with former being systematically
steeper. This is also evident from the CPDF plots of $\Gamma_{X}^{s}$ as shown in the
top right panel of Fig.~\ref{fig:histo_cpdf_xmm_rosat}. To quantify
this difference statistically, we have carried out the
Kolmogorov-Smirnov test (K-S test) giving the probability of the null
hypothesis (i.e., two distribution are similar) as  P$_{null}$ =
4.02$\times$10$^{-19}$, suggesting a clear significant
difference. Similarly, we have also plotted distributions of R$_{\mathrm Edd}$
for both NLSy1 and BLSy1 galaxies in the bottom panels of
Fig.~\ref{fig:histo_cpdf_xmm_rosat}.  As can be seen from its
histograms (bottom left panel) and CPDFs (bottom right panel)
that R$_{\mathrm Edd}$ of the sample of  221 NLSy1s  is
systematically higher as compare to the sample of  154 BLSy1s, with the median values of
0.23 and 0.05, respectively, resulting in K-S
test based  $P_{null}$ of 2.66$\times10^{-35}$.\par
Furthermore, we also did the above comparison in the
hard energy band (2-10 keV) using the $\Gamma_{X}^{h}$ estimated for the
subsamples of  53 NLSy1 and  46 BLSy1 galaxies with their distributions as
shown in the top left panel of Fig.~\ref{fig:histo_cpdf_xmm_hard}. The
$\Gamma_{X}^{h}$ for the subsamples of NLSy1 and BLSy1 galaxies having
the median values of 2.06 and 1.78, respectively, and the K-S test based
$P_{null}$ of  5.13$\times10^{-5}$, suggesting a smaller difference in their
photon indices in comparison to the difference found in the soft energy band.\par
Additionally, we also carried out a similar comparison in the
total energy band (0.3-10 keV) using the $\Gamma_{X}^{T}$ of 139 NLSy1 and 97 BLSy1 galaxies with their distributions as
shown in the top left panel of Fig.~\ref{fig:histo_cpdf_xmm_total}.
The median values of $\Gamma_{X}^{T}$ for the subsamples of NLSy1 and BLSy1 galaxies are
found to be 2.53 and 2.13, respectively, resulting in K-S test based
$P_{null}$ of 4.50$\times10^{-9}$.  
This still suggests a significant difference  in their photon indices 
though it is smaller than the difference found in the soft energy band.\par
Furthermore, quantification of any
such physical differences among NLSy1 and BLSy1 galaxies can also be
obtained based on the correlations of spectral indices  (in the soft 
hard and total 0.3-10 keV X-ray energy bands) with the other parameters of nuclear activities of
AGNs such as R$_{\mathrm Edd}$, L$_{\mathrm bol}$, M$_{\mathrm BH}$, and FWHM(H$\beta$), as we discuss in
next subsection.

\begin{table*}
  
 \centering
 \caption { Results of the correlation analysis in 0.3-10 keV for the samples of 34 NLSy1 and 30 BLSy1 galaxies using the {\it AGNSED} model.}
 
 \label{table_nlsy1_blsy1_corr_Gamma_hot_warm}
 \small

 \begin{tabular}{l r r r l r r r l }\\
   
\hline
   
    \multicolumn{1}{c}{{\it XMM (0.3-10.0 keV)}} &  \multicolumn{3}{c}{NLSy1 (34 sources)} &\multicolumn{5}{c}{BLSy1 (30 sources)}\\
   \hline
  {Correlation$^{\dagger}$} & \multicolumn{1}{c}{m} & \multicolumn{1}{c}{c}&  \multicolumn{1}{c}{$\rho$} &   \multicolumn{1}{c}{P$_{null}$}  & \multicolumn{1}{c}{m} &  \multicolumn{1}{c}{c}& \multicolumn{1}{c}{$\rho$} &   \multicolumn{1}{c}{P$_{null}$}\\
       \hline
$\Gamma_{X}^\mathrm{hot}-\log(R_{\mathrm Edd}^{\mathrm SED})$ & 0.44$\pm$0.09  & 2.42$\pm$0.04  & 0.62  & 9.71$\times$10$^{-5}$  & 0.40$\pm$0.11 & 2.22$\pm$0.11 & 0.54 & 1.87$\times$10$^{-3}$  \\
$\Gamma_{X}^\mathrm{hot}-\log(L_{\mathrm bol}$)& 0.13$\pm$0.04   &$-3.74\pm$1.83  & 0.42  & 1.42$\times$10$^{-2}$  & 0.10$\pm$0.04 & $-$2.92$\pm$1.77 & 0.53 &2.40$\times$10$^{-3}$   \\
$\Gamma_{X}^\mathrm{hot}-\log(M_{\mathrm BH}$) &$-$0.07$\pm$0.03& 2.69$\pm$0.24   & 0.16  & 3.73$\times$10$^{-1}$  & 0.04$\pm$0.03 & $-$2.08$\pm$0.28 & 0.01& 9.47$\times$10$^{-1}$ \\ 
$\Gamma_{X}^\mathrm{hot}-\log(FWHM (H{\beta}$)) & $-$0.48$\pm$0.10 & 3.65$\pm$0.30 & $-$0.18 & 3.03$\times$10$^{-1}$ & $-$0.24$\pm$0.08 & 2.67$\pm$0.32 &  $-$0.22 & 2.49$\times$10$^{-1}$   \\

        \hline
$\log(R_{\mathrm Edd}^{\mathrm SED})-\log(R_{\mathrm Edd})$ & 0.30$\pm$0.10 & $-$0.42$\pm$0.02 & 0.47 & 2.44$\times$10$^{-3}$ & 0.14$\pm$0.12  & $-$0.82$\pm$0.20  & 0.51  & 7.28$\times$10$^{-4}$\\

       \hline
       \multicolumn{9}{l}{$^{\dagger}$The correlation parameters are the same as Table~\ref{table_nlsy1_blsy1_corr}, but in $y=mx+c$ fit, for y=$\Gamma_{X}^\mathrm{hot}$ in all the correlations and X=$\log(R_{\mathrm Edd}^{\mathrm SED})$ }\\
       \multicolumn{9}{l}{ in the $\Gamma_{X}^\mathrm{hot}-\log(R_{\mathrm Edd}^{\mathrm SED}$) correlation. In the last row y and x are $\log(R_{\mathrm Edd}^{\mathrm SED})$ and $\log(R_{\mathrm Edd})$, respectively.}\\
                 \end{tabular}  
 
\end{table*}

\subsection{Comparison of $\Gamma_{X}$ correlations with AGN parameters among the samples of NLSy1 and BLSy1 galaxies}
\label{section 5.2}
\begin{figure*}[!t]
\begin{minipage}[]{1.0\textwidth}
 \includegraphics[width=0.37\textwidth,height=0.215\textheight,angle=00]{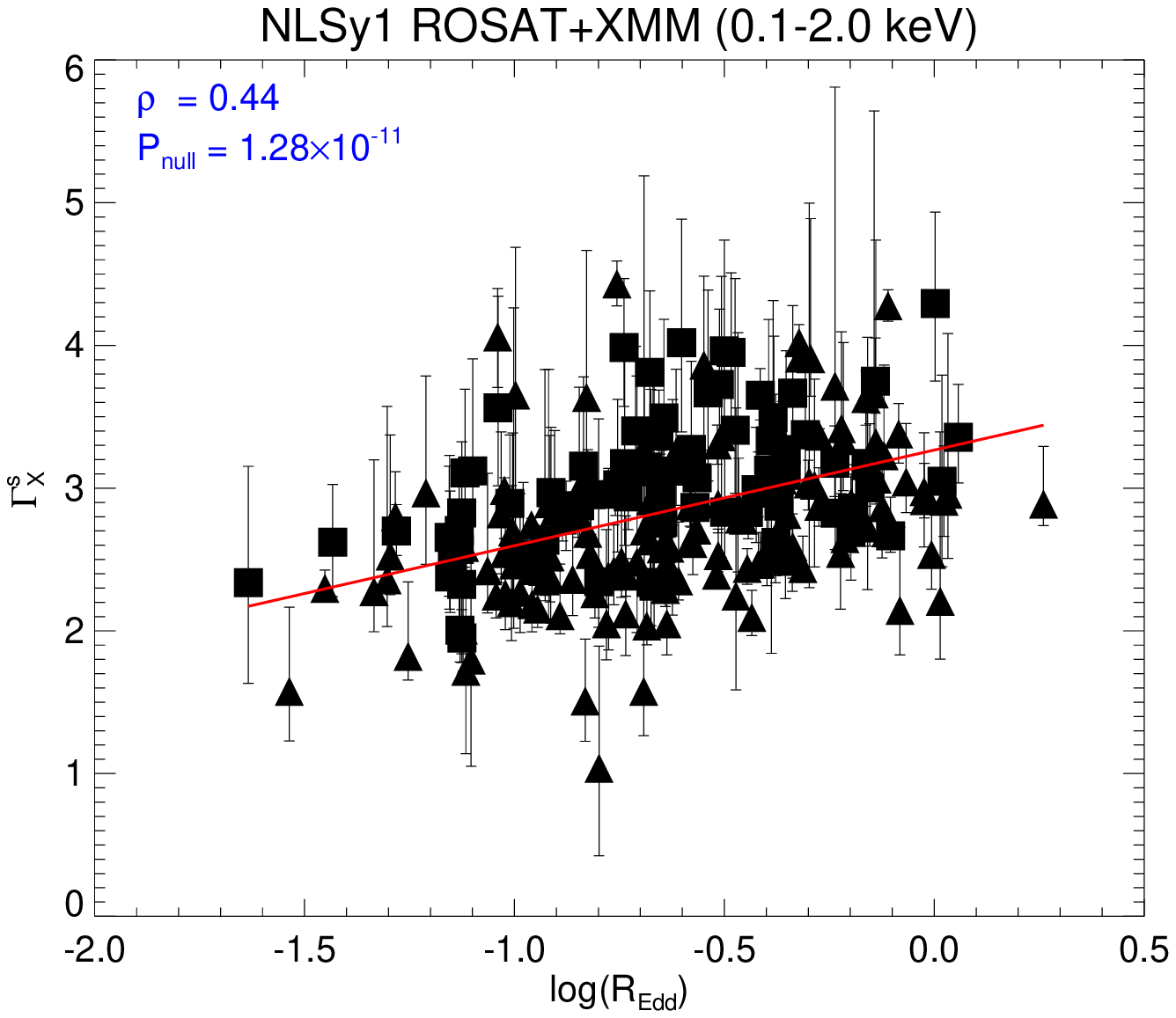} 
      \hspace{-1.10cm}
 \includegraphics[width=0.37\textwidth,height=0.215\textheight,angle=00]{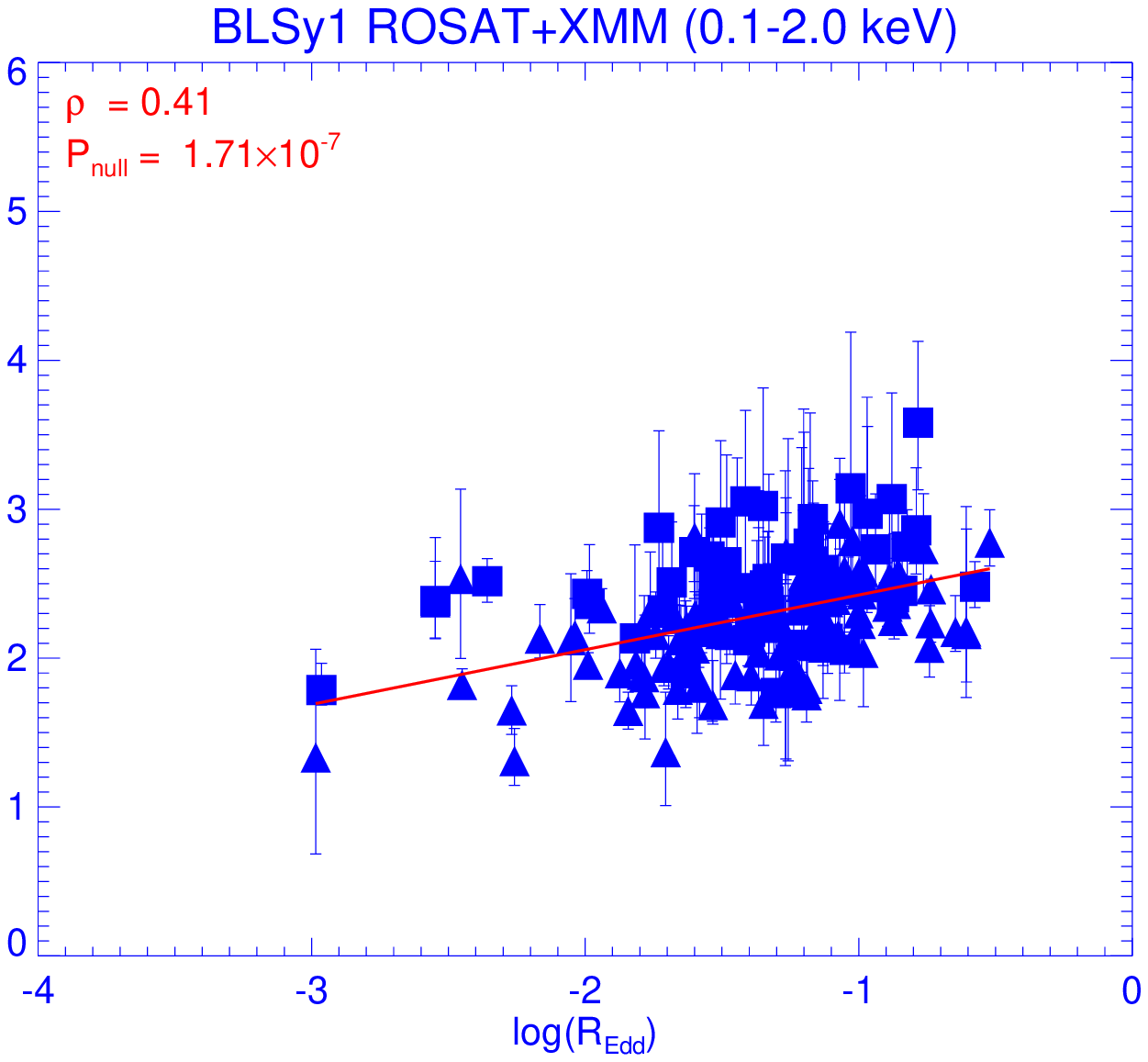} 
      \hspace{-1.10cm}
 \includegraphics[width=0.37\textwidth,height=0.215\textheight,angle=00]{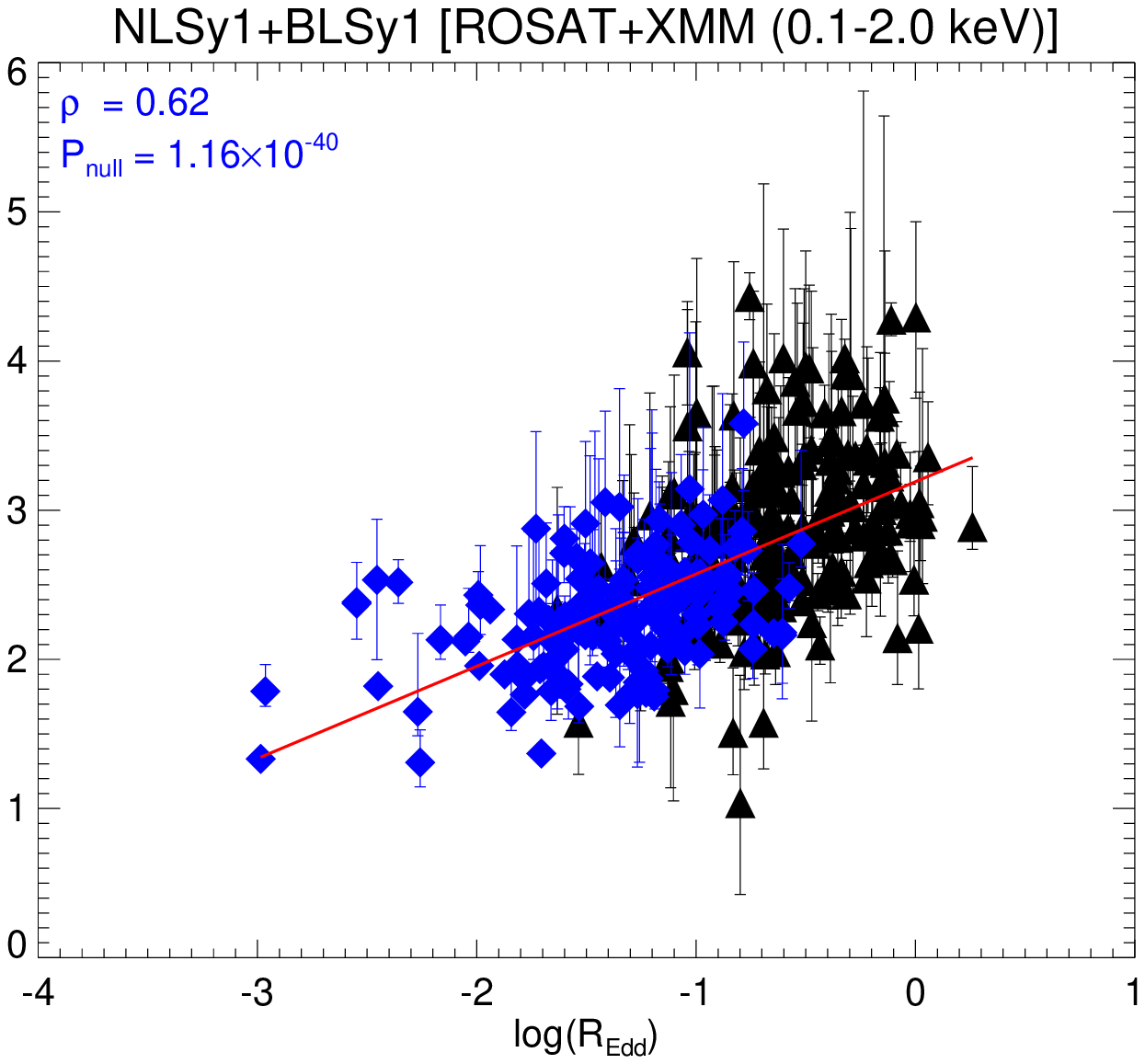}
 \includegraphics[width=0.37\textwidth,height=0.215\textheight,angle=00]{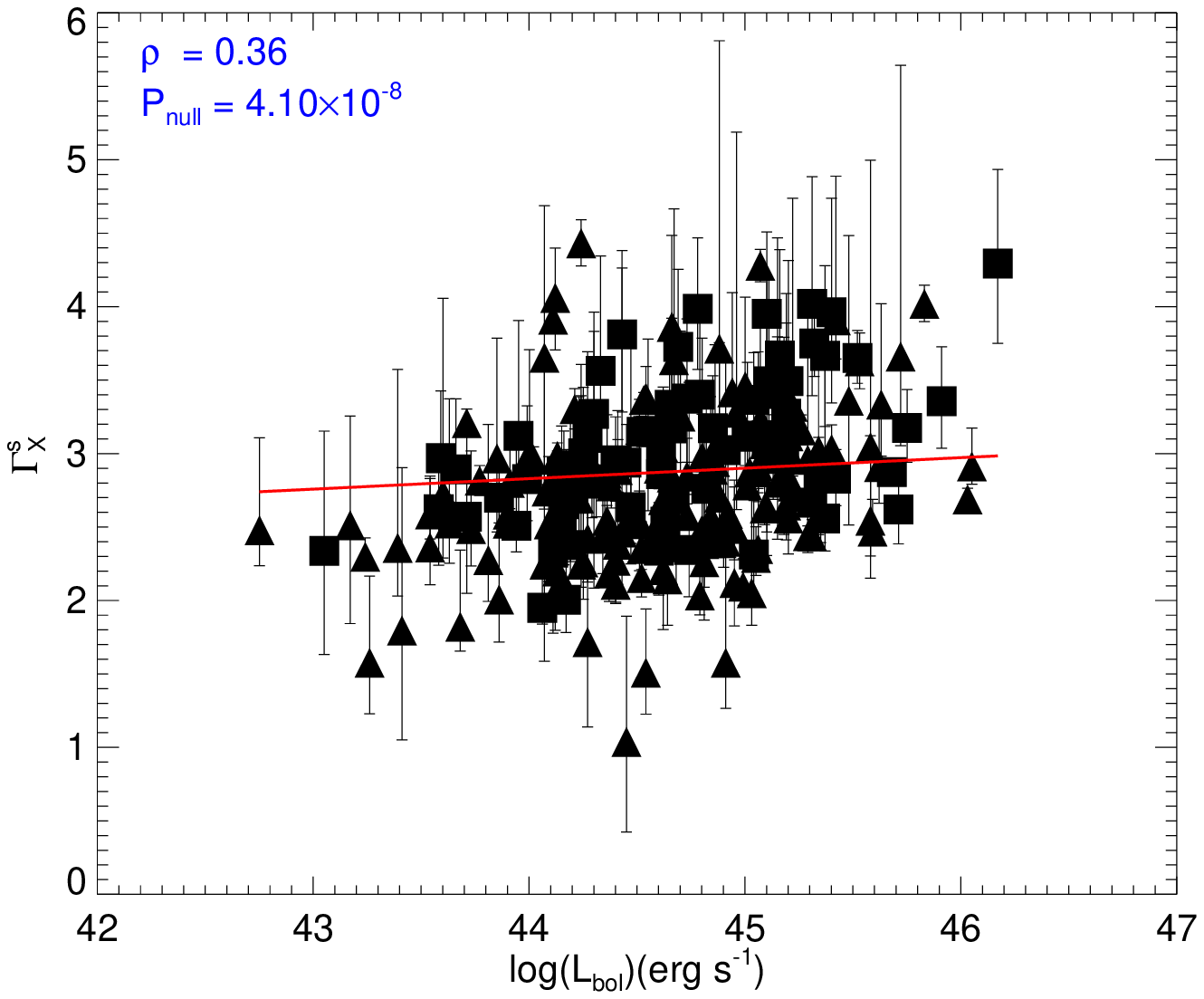}
     \hspace{-1.10cm}
 \includegraphics[width=0.37\textwidth,height=0.215\textheight,angle=00]{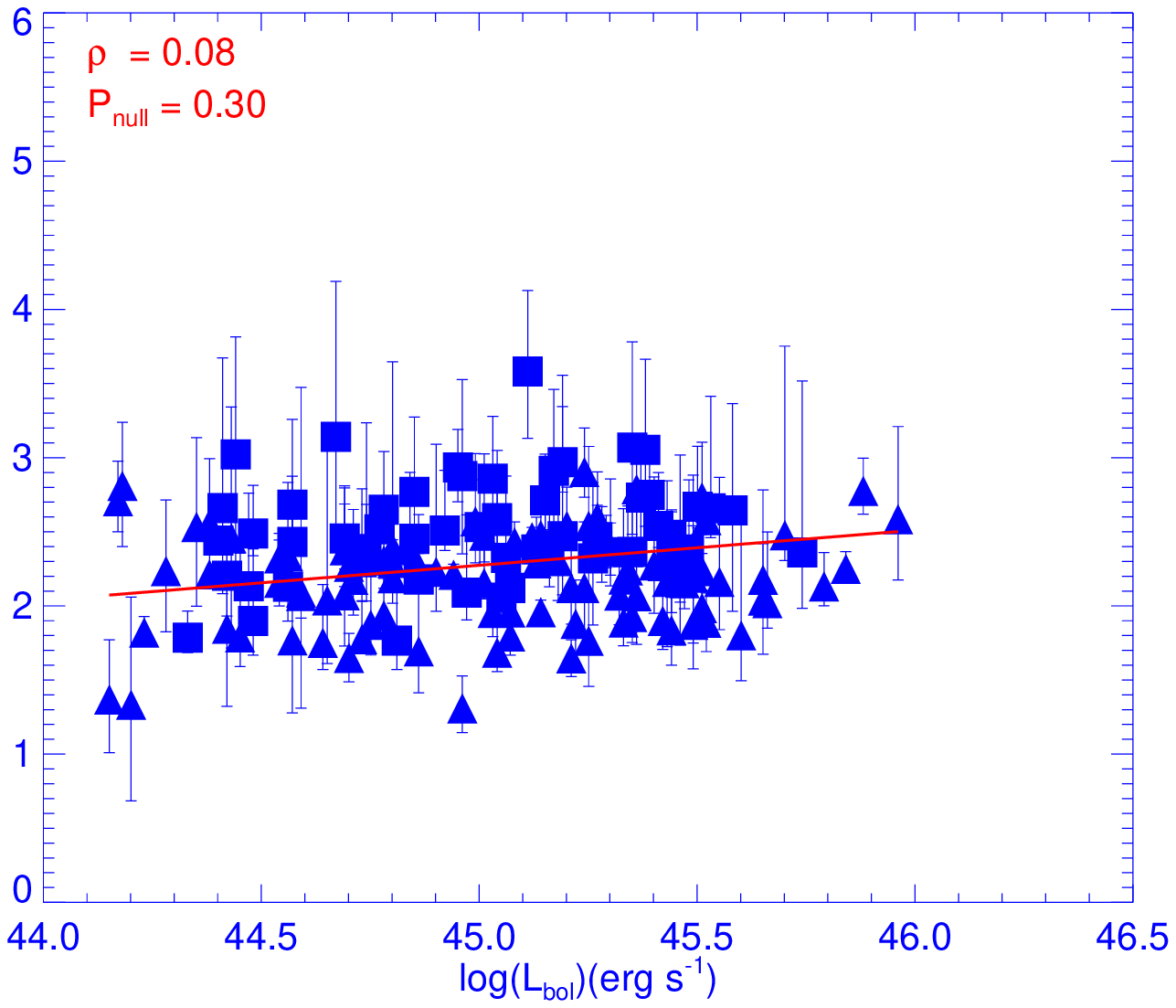}
      \hspace{-1.10cm}
 \includegraphics[width=0.37\textwidth,height=0.215\textheight,angle=00]{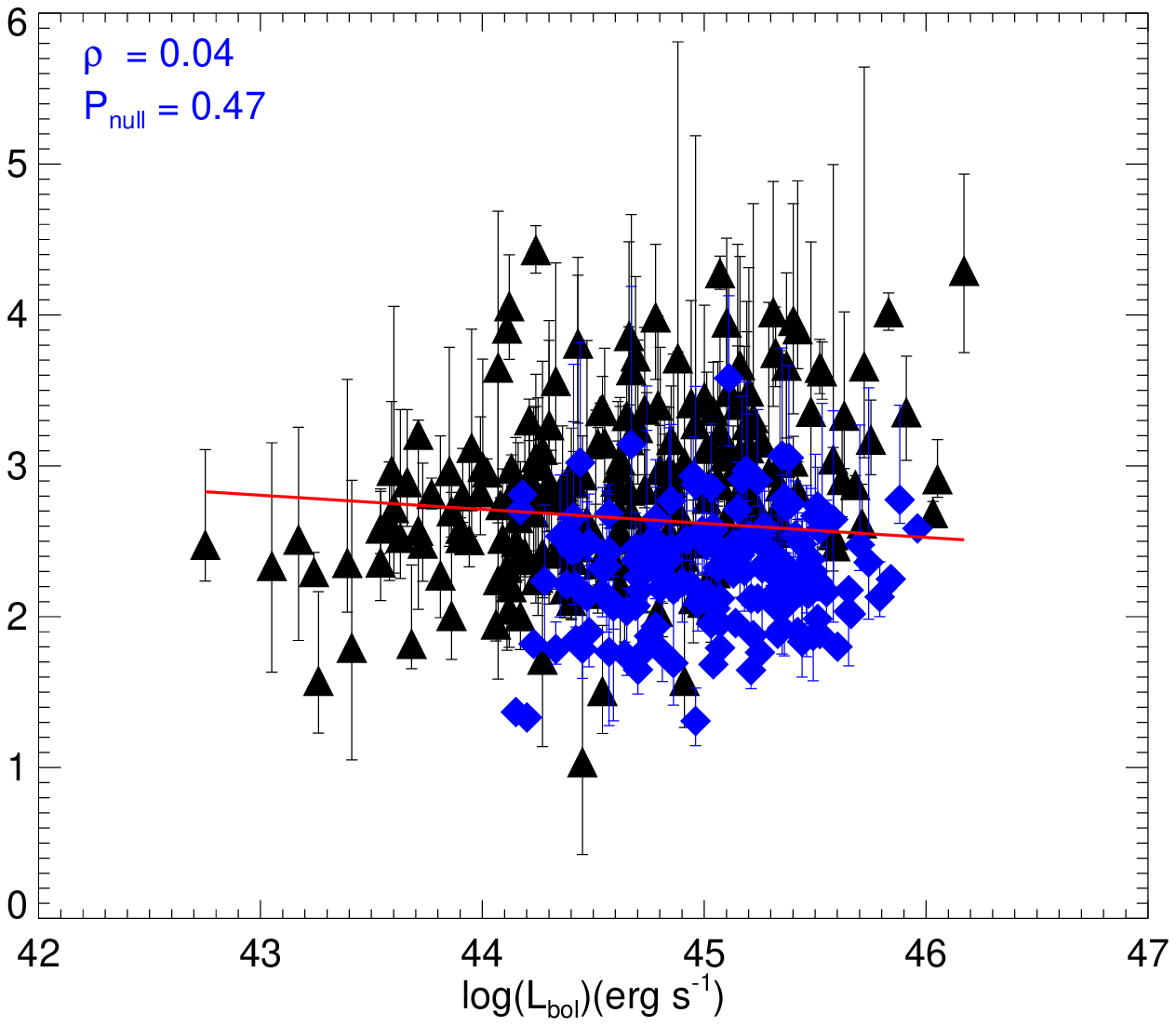}
 \includegraphics[width=0.37\textwidth,height=0.215\textheight,angle=00]{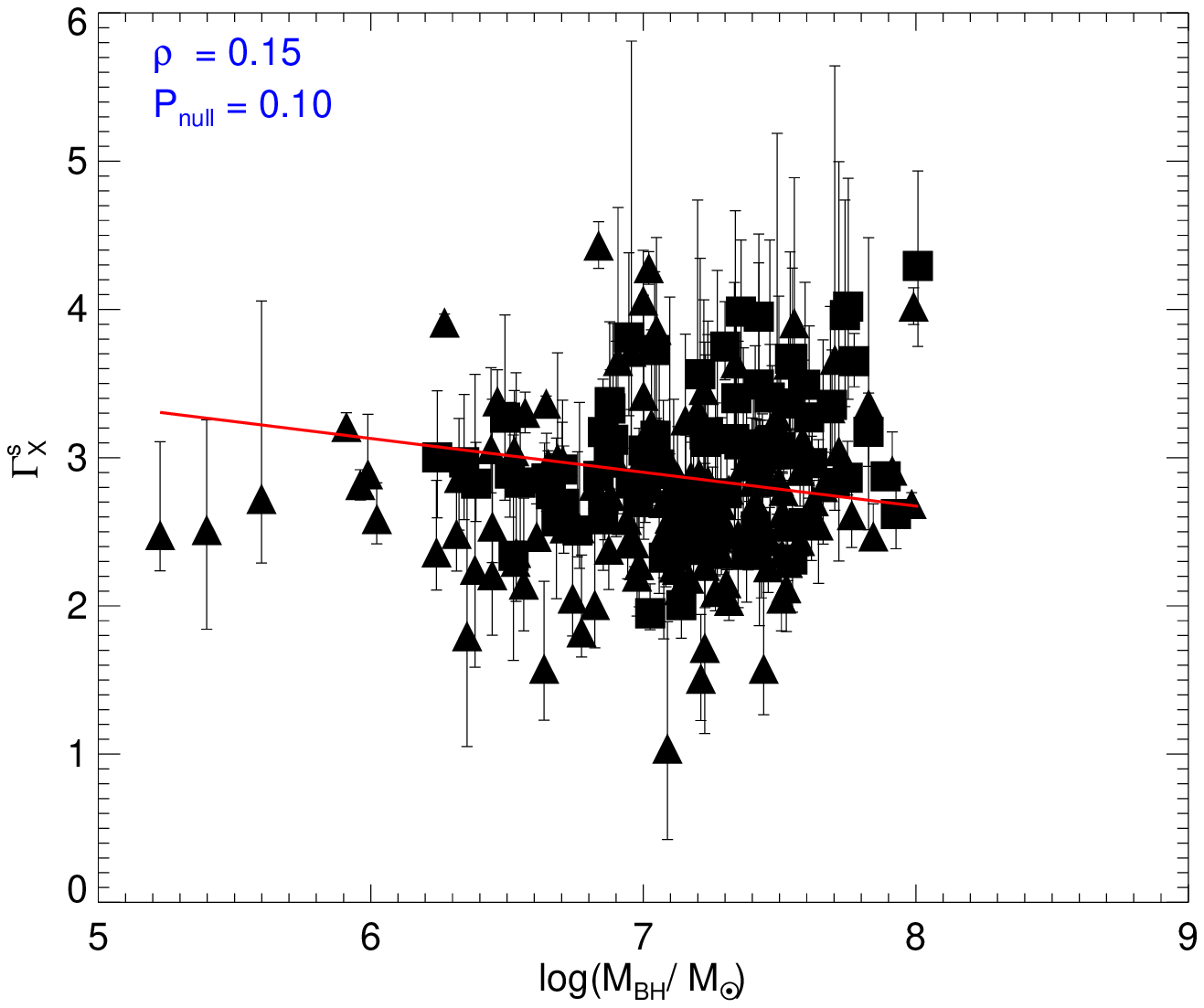}
      \hspace{-1.10cm}
 \includegraphics[width=0.37\textwidth,height=0.215\textheight,angle=00]{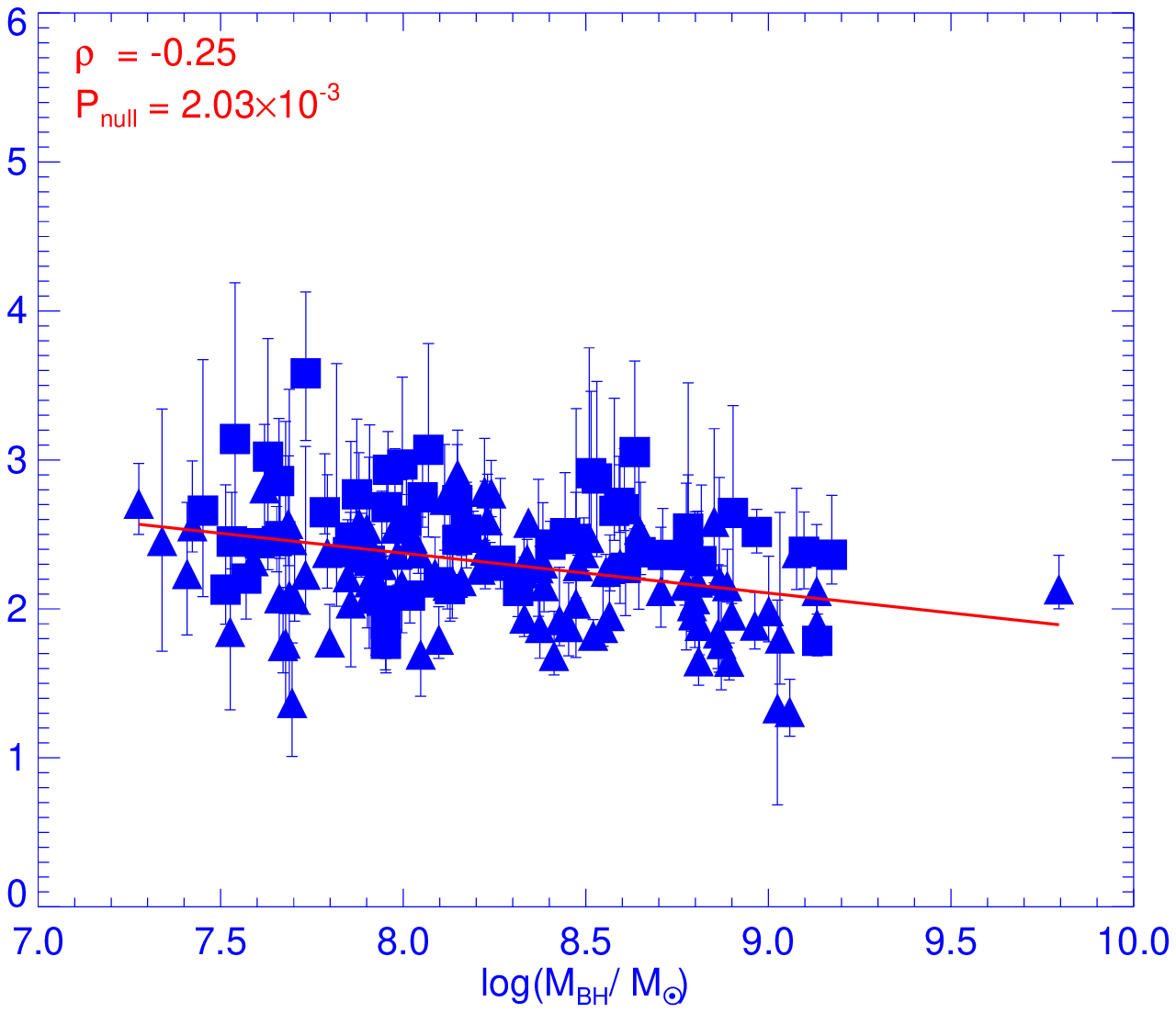}
      \hspace{-1.10cm}
 \includegraphics[width=0.37\textwidth,height=0.215\textheight,angle=00]{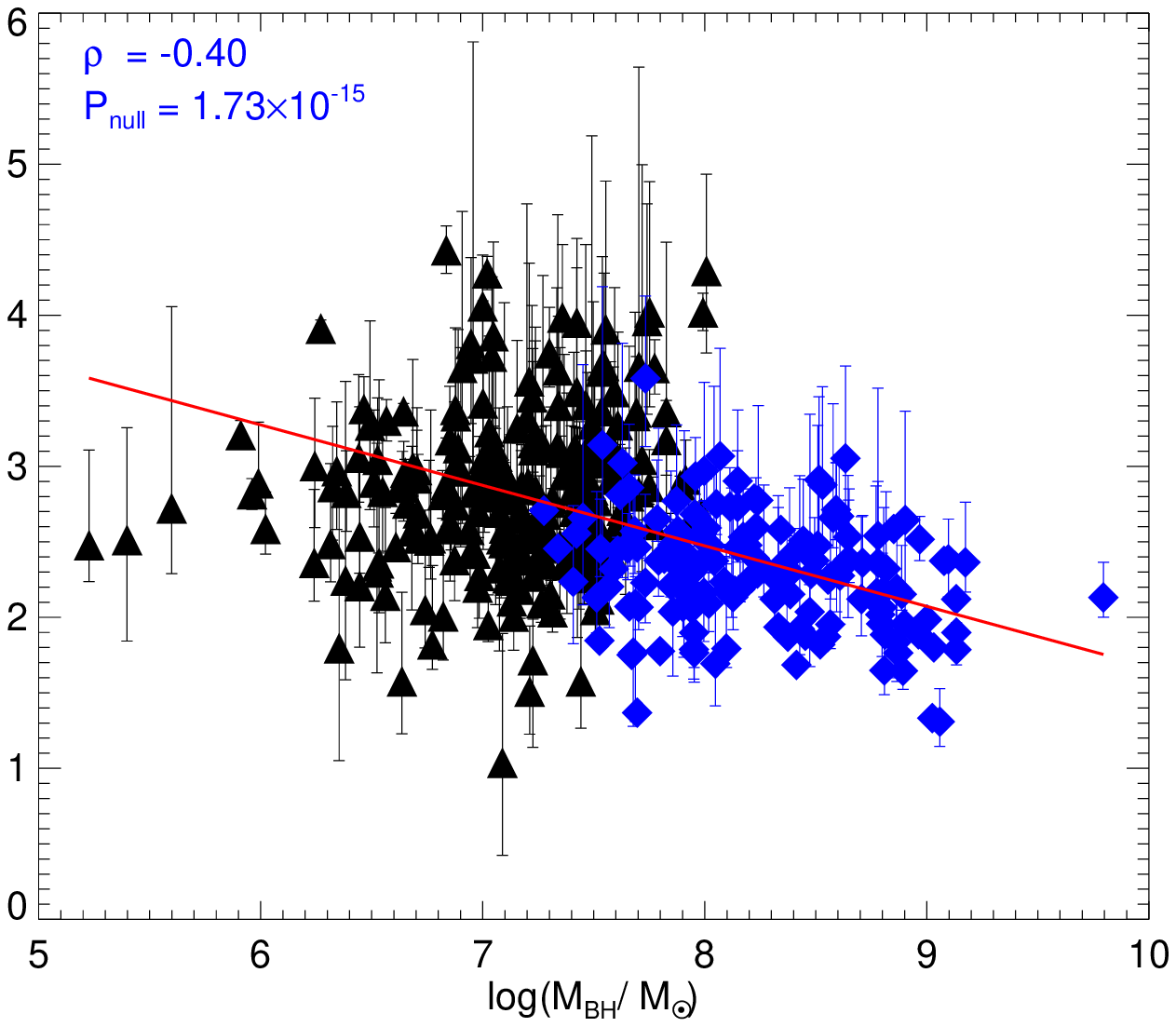}
 \includegraphics[width=0.37\textwidth,height=0.215\textheight,angle=00]{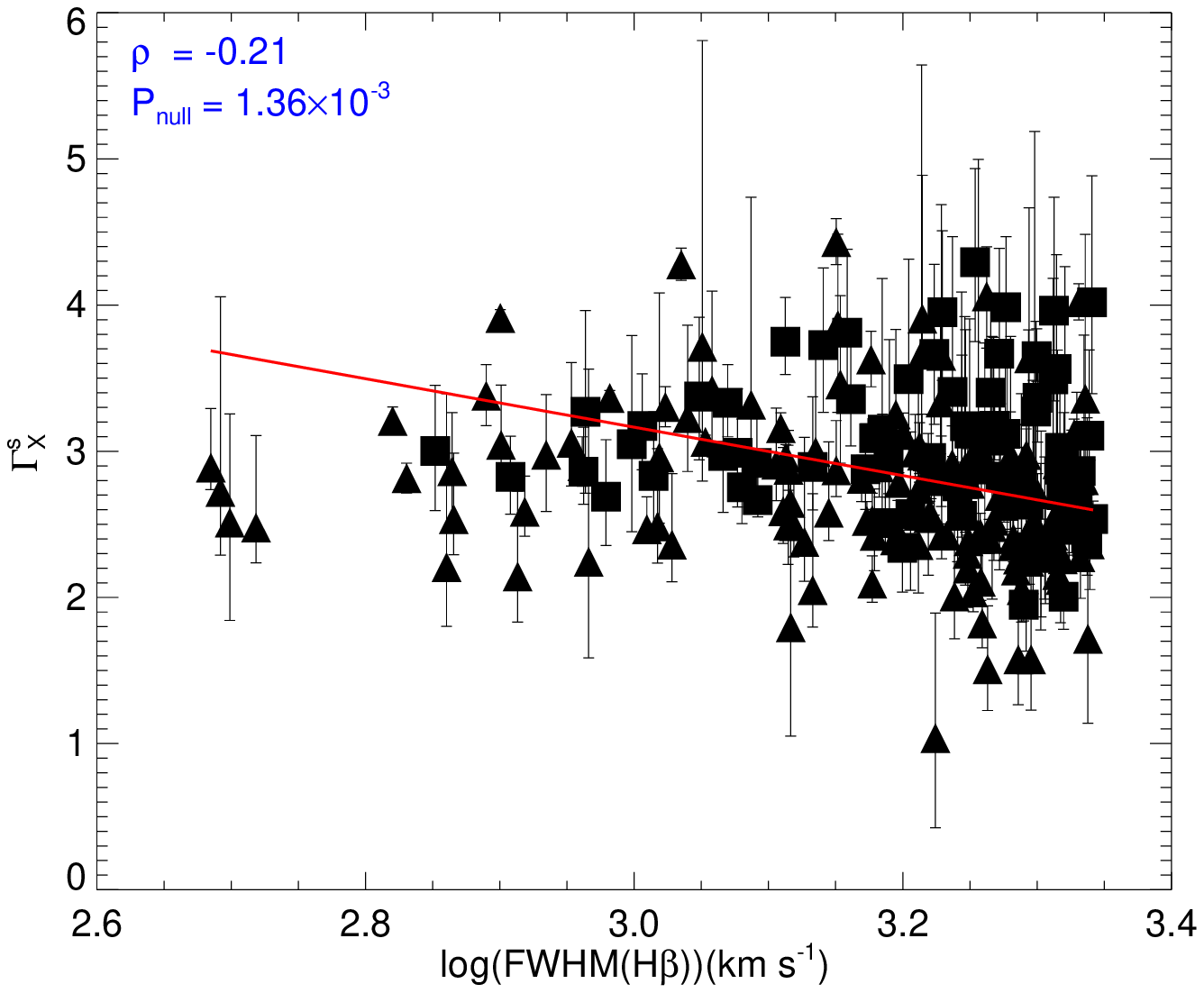}
       \hspace{-1.10cm}
 \includegraphics[width=0.37\textwidth,height=0.215\textheight,angle=00]{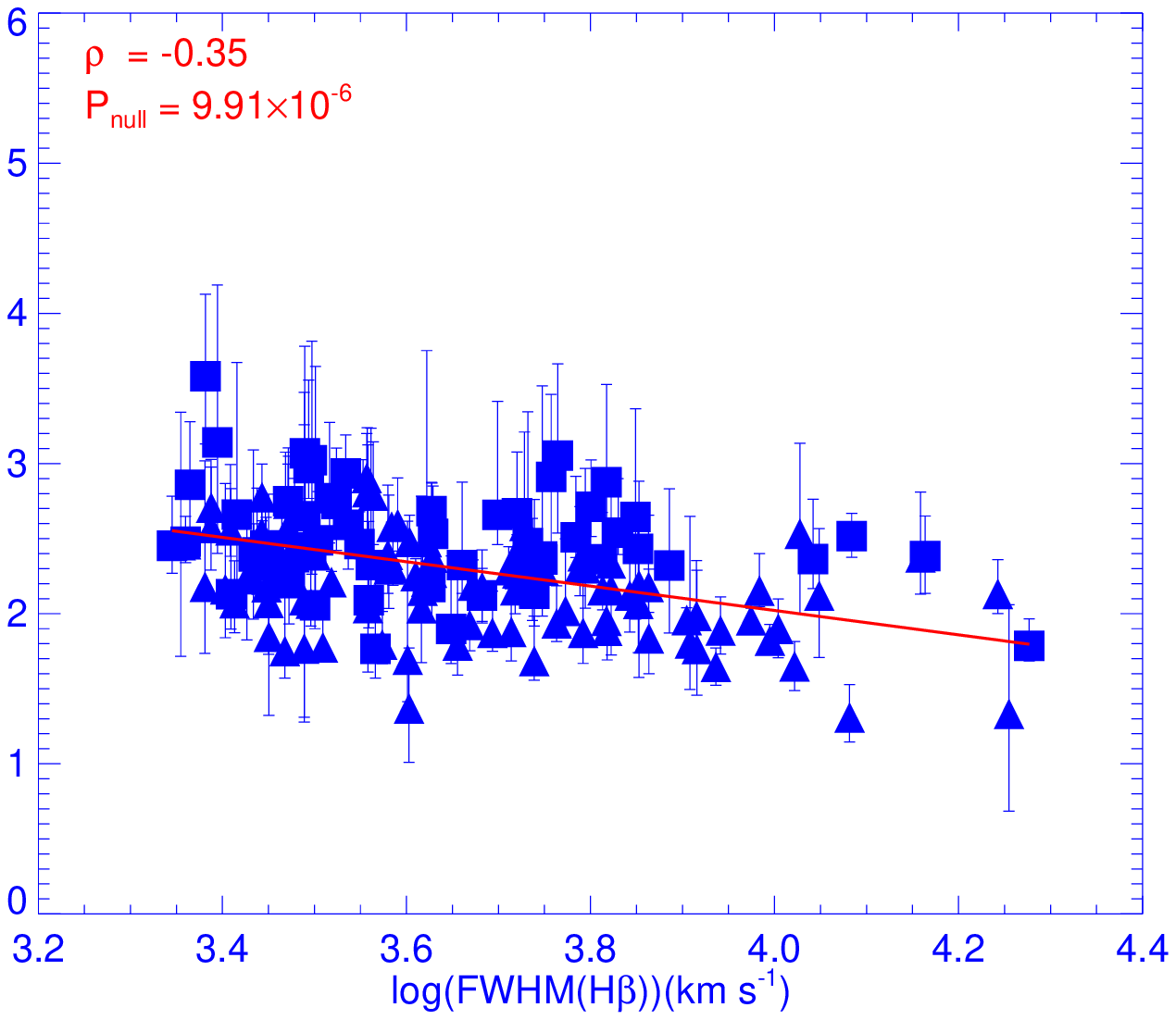} 
        \hspace{-1.10cm}
 \includegraphics[width=0.37\textwidth,height=0.215\textheight,angle=00]{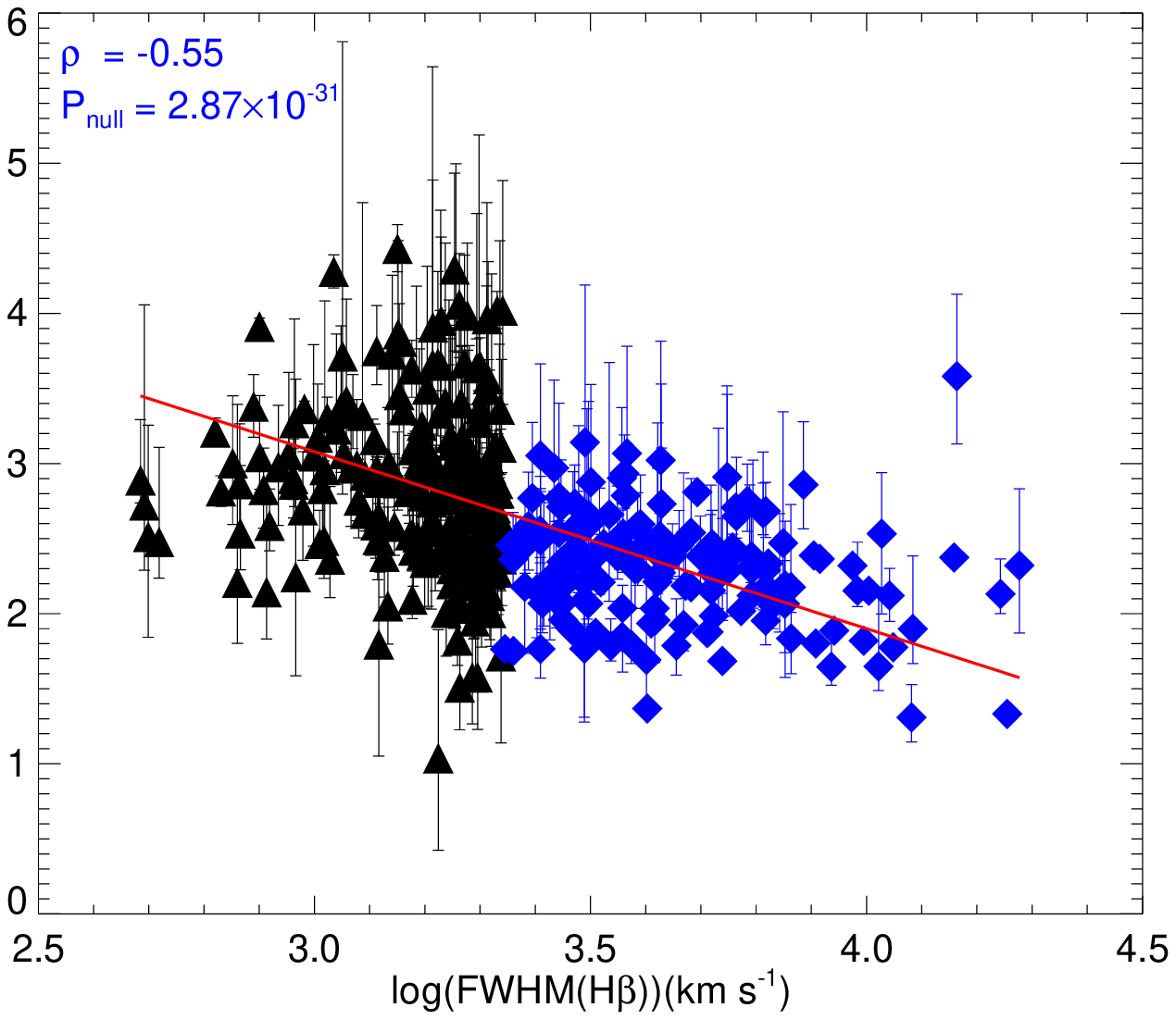}
  \end{minipage}
   \caption{{\scriptsize Correlations of the  221 NLSy1s (left) and  154 BLSy1s (middle)  either from the  {\it ROSAT} (filled squares) or from the {\it XMM-Newton} (filled triangles) for the soft (0.1-2.0 keV) X-ray photon indices ($\Gamma_{X}^{s}$) versus Eddington ratios, bolometric luminosities,  black hole masses and  FWHM of H$\beta$ lines, respectively from top to bottom panels, along with their error-weighted linear fit (red, solid line). The last column presents the correlations among the same, but for the joint sample of  221 NLSy1s (black, filled triangles) and  154 BLSy1s (blue, filled diamonds). The plots also give the Spearman's correlation coefficient ($\rho$) and the probability of null correlation (P$_{null}$) values (left corner of each panel).}}
 \label{fig:nlsy_blsy_corr_rosat_xmm_soft}
\end{figure*}

Results based on our correlations analysis for the samples of 
 221 NLSy1s,  154 BLSy1s and their combined sample (i.e.,  375 [NLSy1$+$BLSy1]) are shown in Fig.~\ref{fig:nlsy_blsy_corr_rosat_xmm_soft}.
This figure shows the plots of R$_{\mathrm Edd}$, L$_{\mathrm bol}$, M$_{\mathrm BH}$ and FWHM(H$\beta$) versus $\Gamma_{X}^{s}$
for NLSy1, BLSy1 and [NLSy1$+$BLSy1] galaxies in the left, middle and right panels, respectively. The
statistical quantifications of correlations of these parameters with $\Gamma_{X}^{s}$ are
summarised in Table~\ref{table_nlsy1_blsy1_corr}. As can be seen from
the top panels of Fig.~\ref{fig:nlsy_blsy_corr_rosat_xmm_soft} that the positive
correlations between $\Gamma_{X}^{s}$ and log(R$_{\mathrm Edd}$) are quite apparent
for the samples of NLSy1 (top left panel), BLSy1 (top middle panel), and [NLSy1$+$BLSy1] (top right panel) galaxies, respectively.
We have also quantified these correlations with the fitting
function of the form $y=mx+c$ by standard $\chi^{2}$ - minimization
method which yields the relations, for the sample of NLSy1s as

\begin{equation}
\Gamma_{X}^{s} = (0.67\pm0.04) \log(R_{\mathrm Edd}) + (3.27\pm0.02)
\end{equation} 
 
, for the sample of BLSy1s as

\begin{equation}
  \Gamma_{X}^{s} = (0.37\pm0.03) \log(R_{\mathrm Edd}) + (2.79\pm0.04)
\end{equation}

and for the joint sample of NLSy1 and BLSy1 galaxies as

\begin{equation}
  \Gamma_{X}^{s} = (0.62\pm0.02) \log(R_{\mathrm Edd}) + (3.19\pm0.02)
\end{equation}

as shown by the solid red line in the plots of $\Gamma_{X}^{s} -
\log(R_{\mathrm Edd}$) in the top panels of
Fig.~\ref{fig:nlsy_blsy_corr_rosat_xmm_soft}. This very good correlation is also supported based on their
Spearman's rank correlation coefficient ($\rho$) of  0.44, 0.41 and 0.62
with the probability of null correlation (P$_{null}$) of 1.28$\times10^{-11}$, 1.71$\times10^{-7}$, and 1.16$\times10^{-40}$
for the samples of NLSy1, BLSy1 and [NLSy1$+$BLSy1]  galaxies.
We note here that the correlation coefficients found for the samples of NLSy1 and BLSy1
galaxies are almost similar, however, the difference is significant in the slopes of their
$\Gamma_{X}^{s} - \log(R_{\mathrm Edd}$) linear fit, with  \emph{m=
  0.67$\pm$0.04} and  \emph{0.37$\pm$0.03}, respectively.  Our
above correlations between $\Gamma_{X}^{s}$ and log(R$_{\mathrm Edd}$) give
hint that $\Gamma_{X}^{s} \propto L_{\mathrm bol}/L_{\mathrm Edd}$ which implies that
$\Gamma_{X}^{s} \propto L_{\mathrm bol}$ and $\Gamma_{X}^{s} \propto
L_{\mathrm Edd}^{-1}$. As we know that L$_{\mathrm Edd} \propto M_{\mathrm BH}$ and M$_{\mathrm BH}
\propto (FWHM(H\beta))^{2}$,
therefore, a very good $\Gamma_{X}^{s} - \log(R_{\mathrm Edd}$) correlation can
be due to intrinsic $\Gamma_{X}^{s} - \log(L_{\mathrm bol}$) and
$\Gamma_{X}^{s} - \log(M_{\mathrm BH}^{-1}$) correlations or due to both these
correlations. We tested these possibilities in the samples of
NLSy1, BLSy1 and [NLSy1$+$BLSy1] galaxies. We find a significant $\Gamma_{X}^{s} -
\log(L_{\mathrm bol}$) correlation for the sample of NLSy1s with  $\rho = 0.36$,
P$_{null} = 4.10\times10^{-8}$. The corresponding correlations are found to be nonsignificant for the samples of
BLSy1 (with $\rho = 0.08$ \&  P$_{null}$ = 0.30) and [NLSy1$+$BLSy1] galaxies (with  $\rho = 0.04$ \& P$_{null}$ = 0.47). However,
for the $\Gamma_{X}^{s}-\log(M_{\mathrm BH}$), we found a good anticorrelation in the joint sample of NLSy1 and BLSy1 galaxies (with  $\rho = -0.40$ \& P$_{null} = 1.73\times10^{-15}$), though it was found to be nonsignificant when the samples of NLSy1 and BLSy1 galaxies considered separately.\par
On the other hand, the significant $\Gamma_{X}^{s}-\log(FWHM(H{\beta}$)) anticorrelations are found
for the samples of BLSy1  (with  $\rho = -0.35$  \&  P$_{null} = 9.91\times10^{-6}$)
and [NLSy1$+$BLSy1] galaxies (with  $\rho = -0.55$ \& P$_{null} = 2.87\times10^{-31}$).
  However, this correlation is found to be nominal for the sample of NLSy1s with
 $\rho = -0.21$ and P$_{null} = 1.36\times10^{-3}$. Summary of 
all the above correlations along with $\rho$ and P$_{null}$ values separately for the NLSy1s, BLSy1s, and their combined samples is
given in Table~\ref{table_nlsy1_blsy1_corr}.\par
\begin{figure*}[!t]
  \begin{minipage}[]{1.0\textwidth}
 \includegraphics[width=0.37\textwidth,height=0.215\textheight,angle=00]{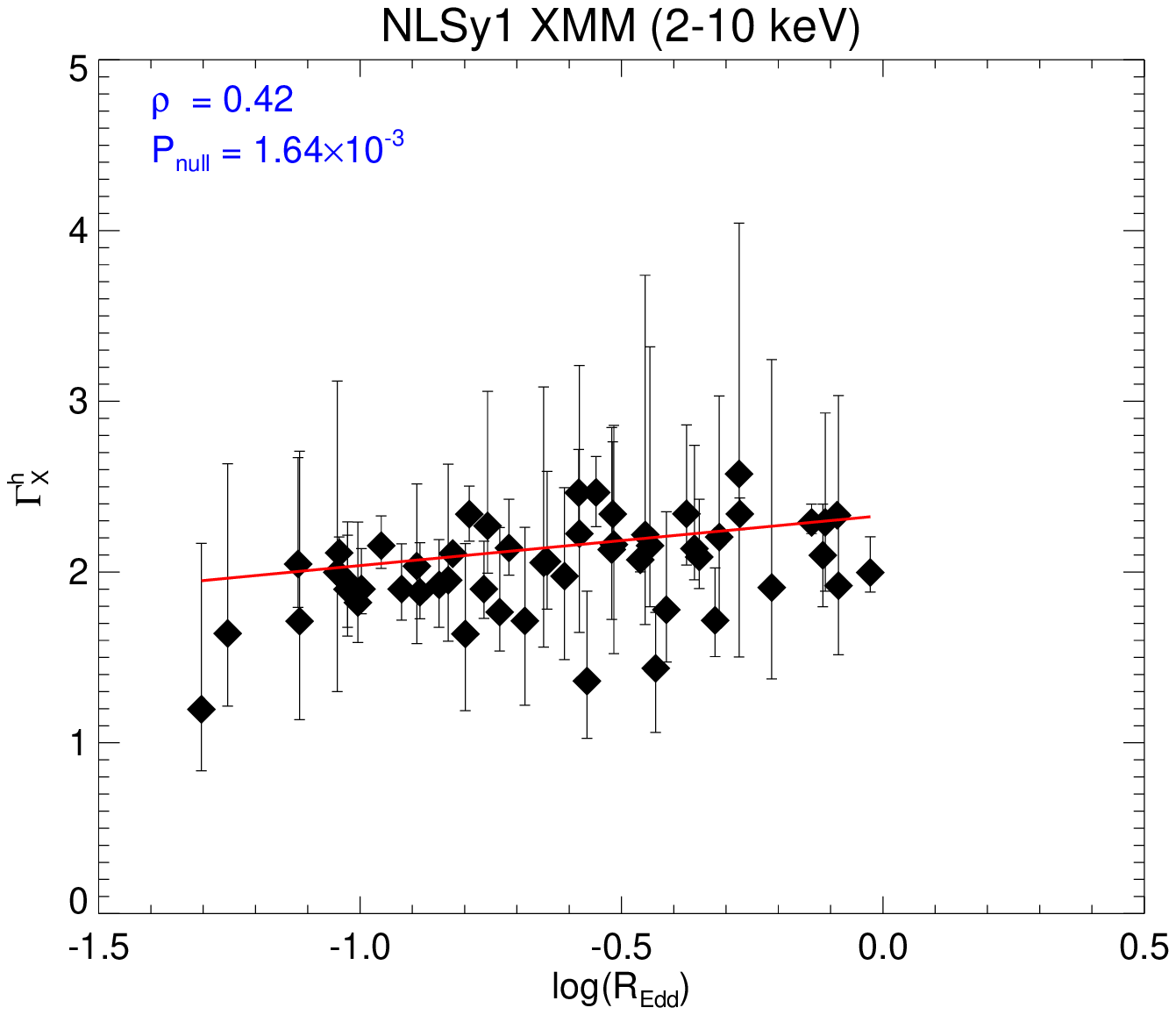}
     \hspace{-1.10cm}
 \includegraphics[width=0.37\textwidth,height=0.215\textheight,angle=00]{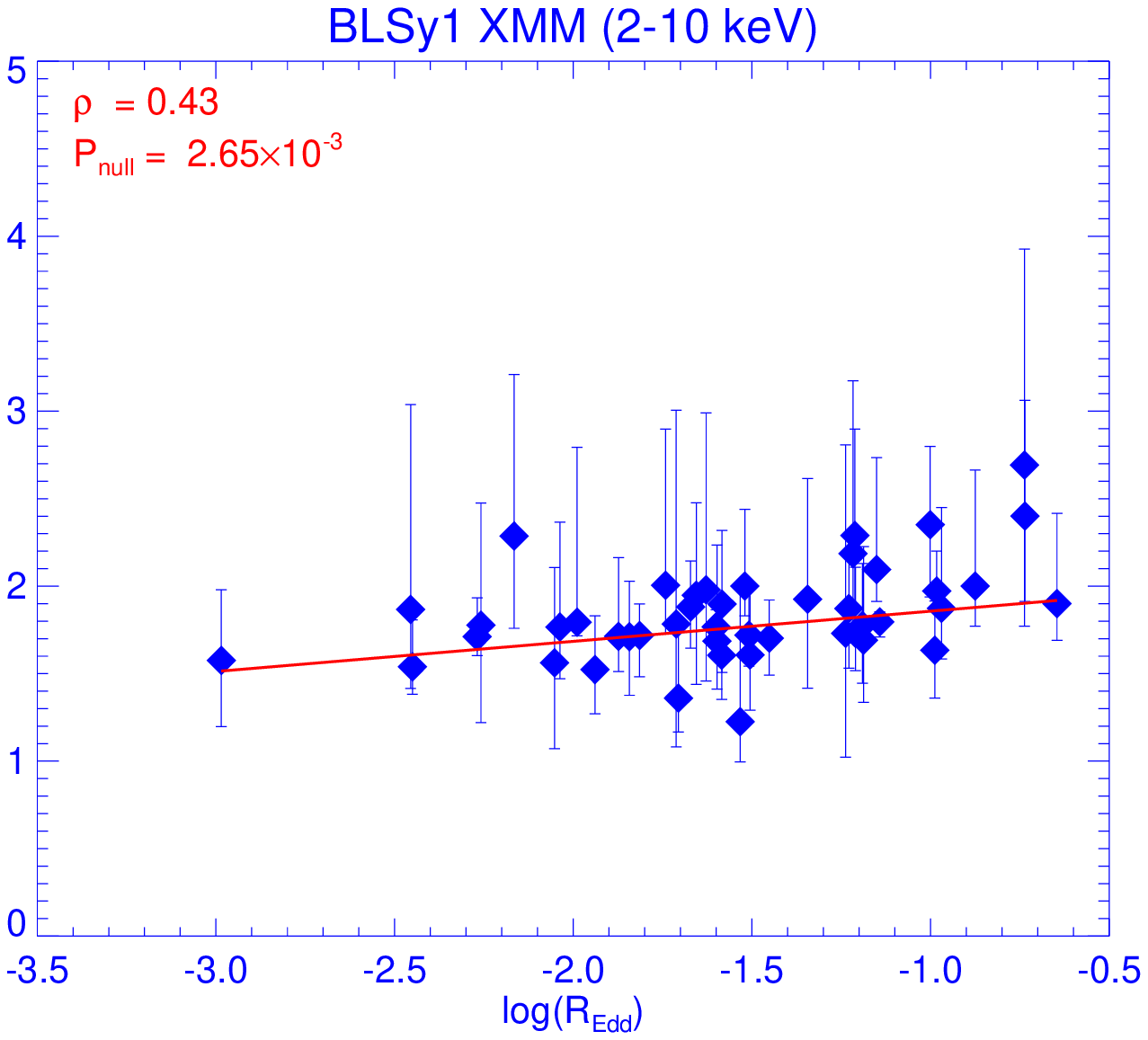}
    \hspace{-1.10cm}
 \includegraphics[width=0.37\textwidth,height=0.215\textheight,angle=00]{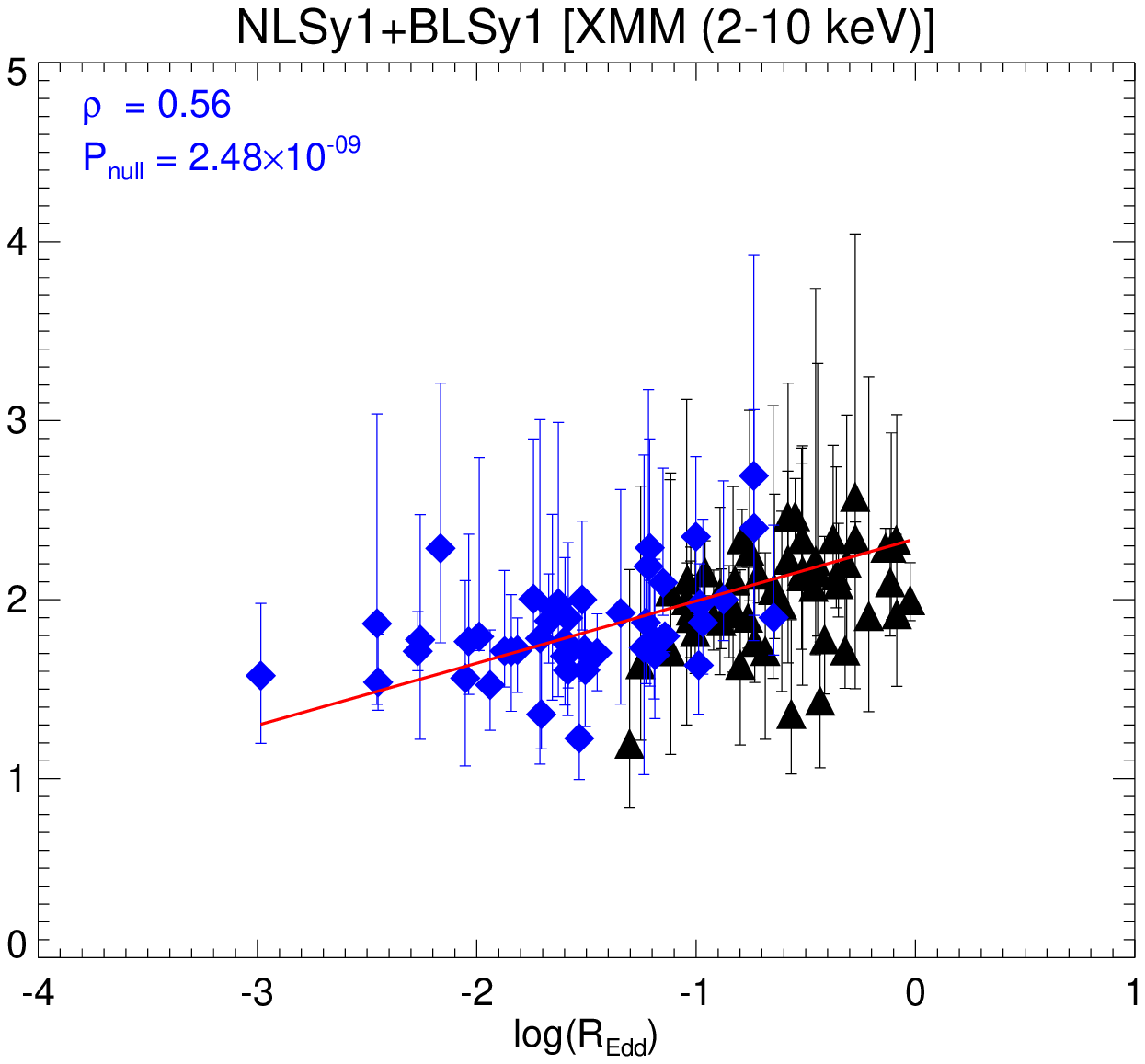}
 \includegraphics[width=0.37\textwidth,height=0.215\textheight,angle=00]{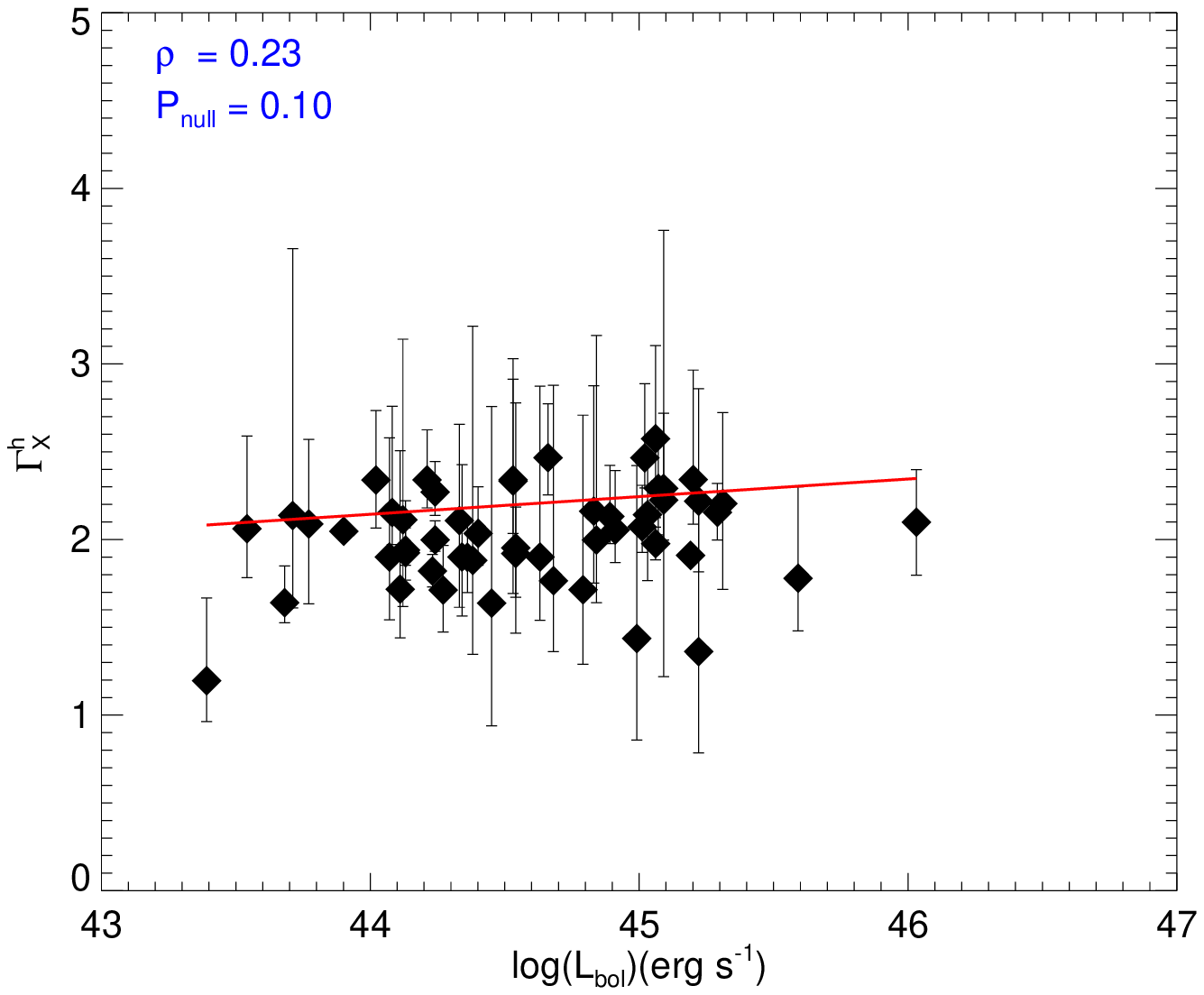}
    \hspace{-1.10cm}
 \includegraphics[width=0.37\textwidth,height=0.215\textheight,angle=00]{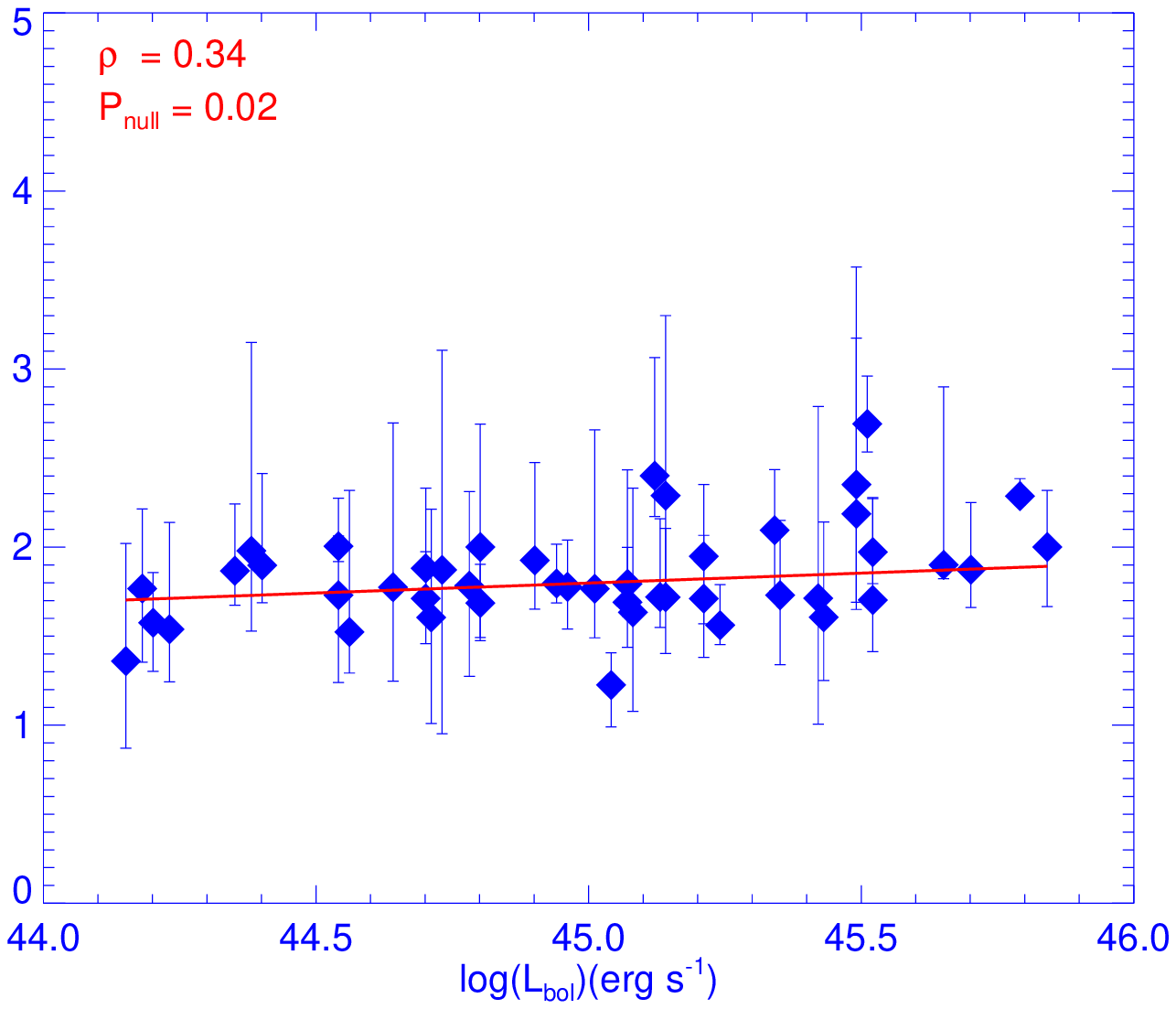}
     \hspace{-1.10cm}
 \includegraphics[width=0.37\textwidth,height=0.215\textheight,angle=00]{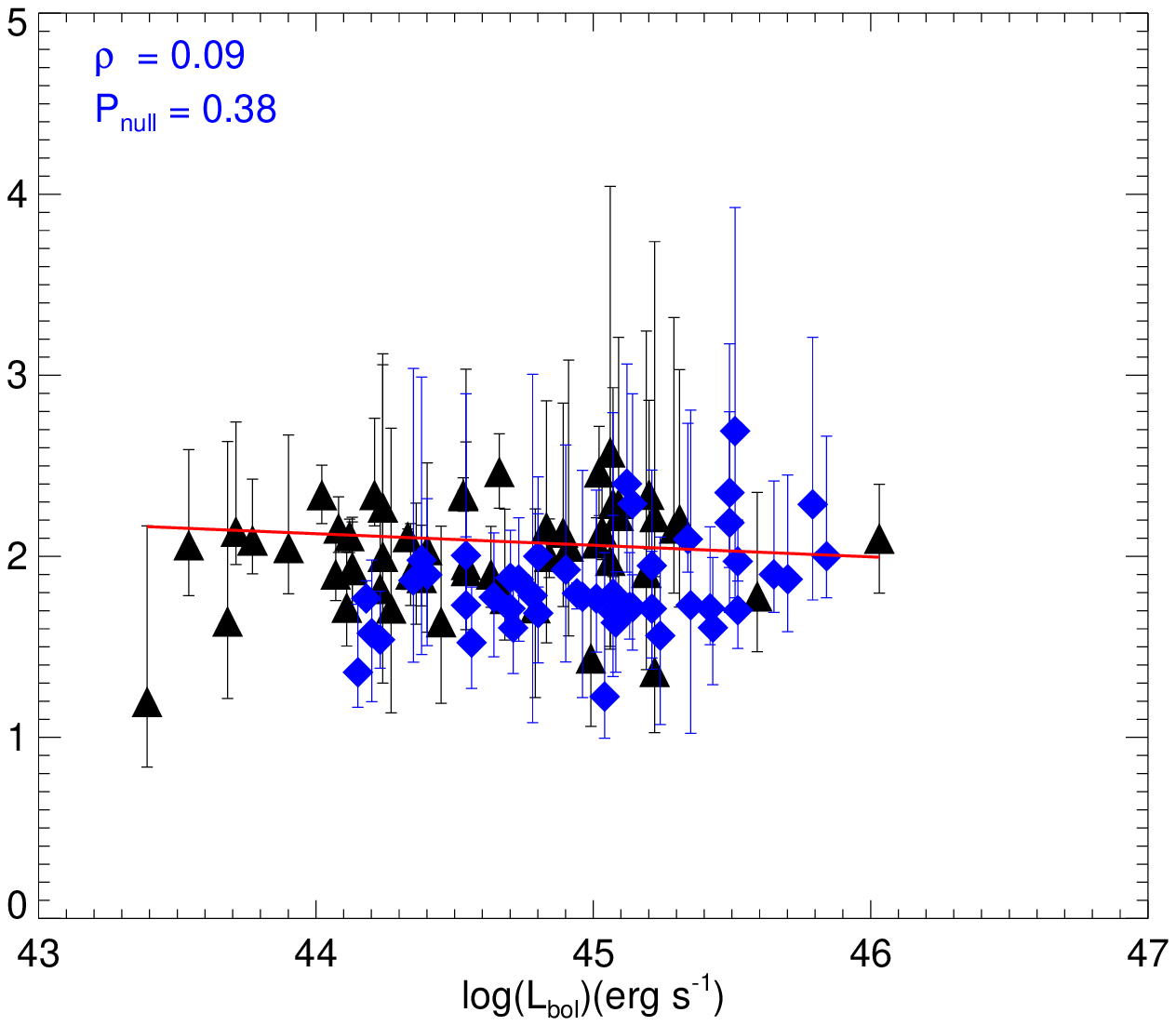}
 \includegraphics[width=0.37\textwidth,height=0.215\textheight,angle=00]{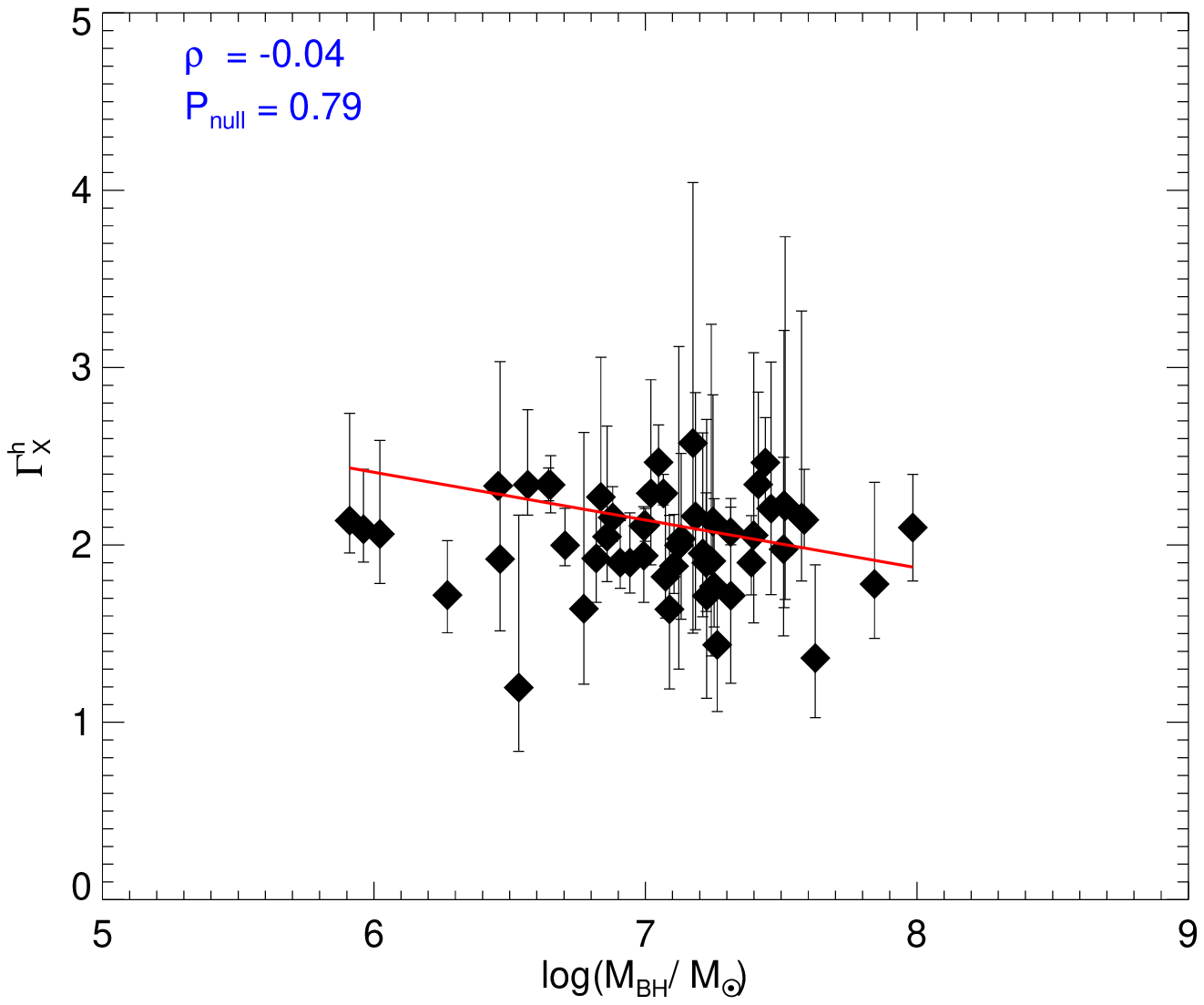}
     \hspace{-1.10cm}
 \includegraphics[width=0.37\textwidth,height=0.215\textheight,angle=00]{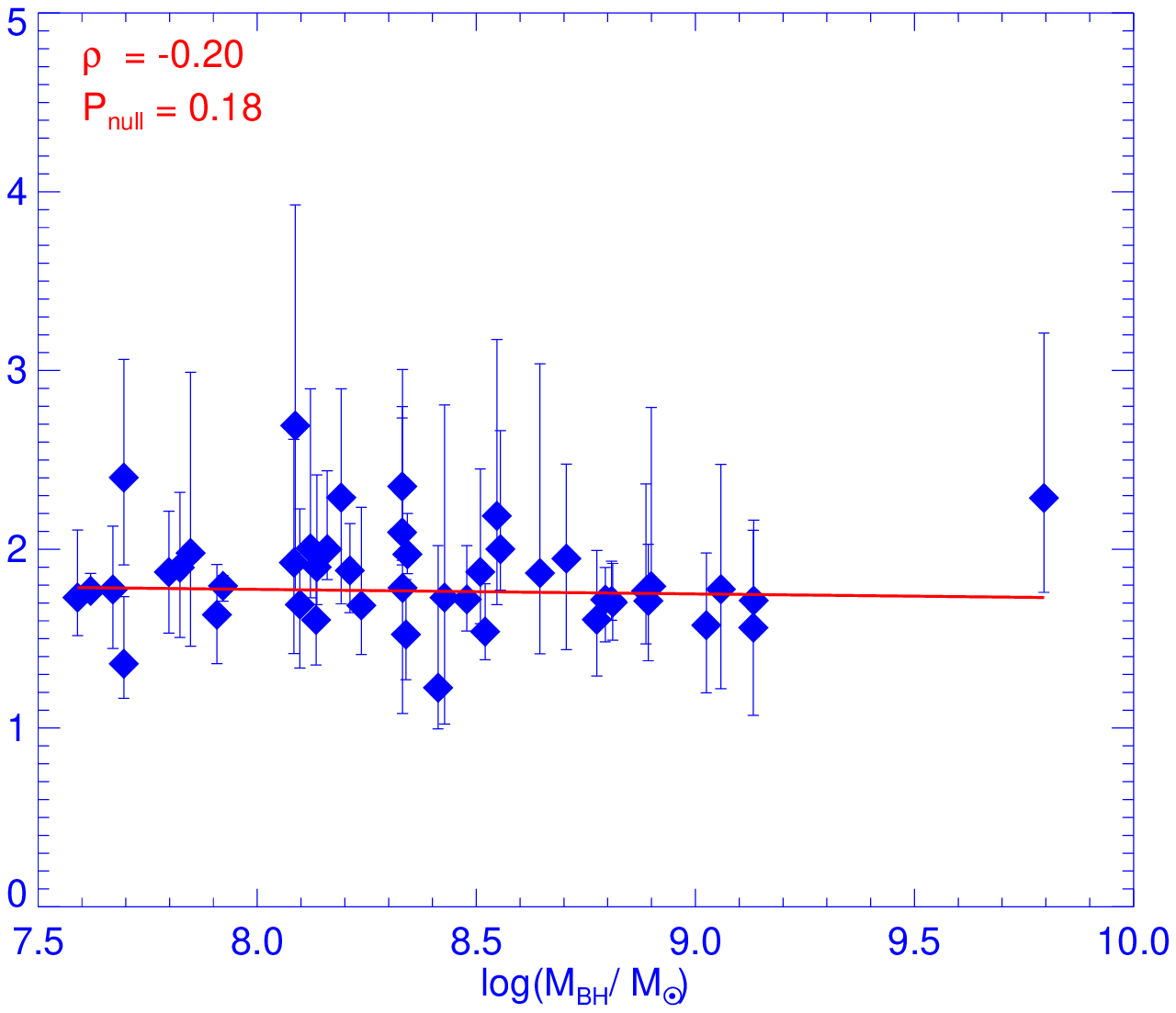}
     \hspace{-1.10cm}
 \includegraphics[width=0.37\textwidth,height=0.215\textheight,angle=00]{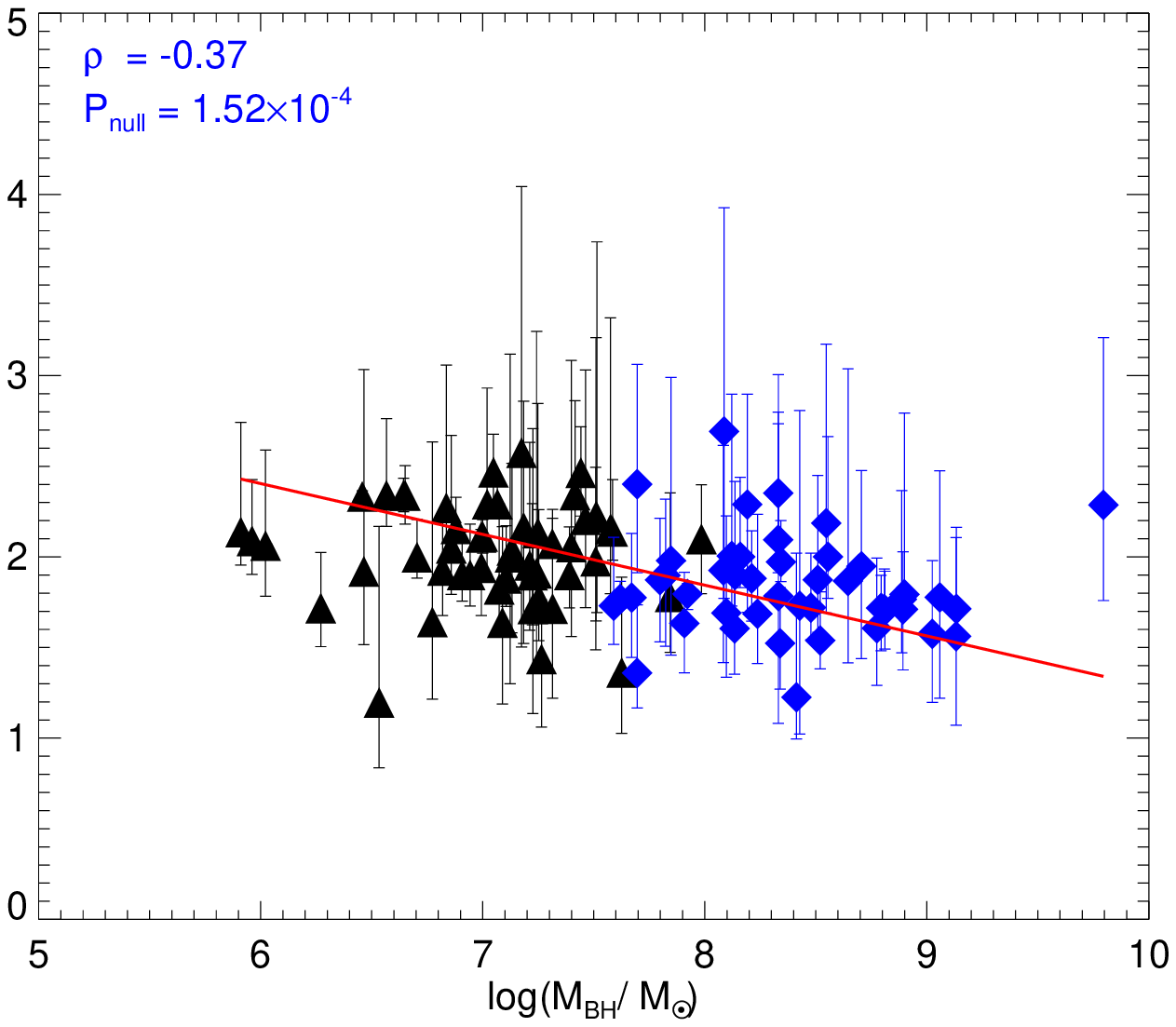}
 \includegraphics[width=0.37\textwidth,height=0.215\textheight,angle=00]{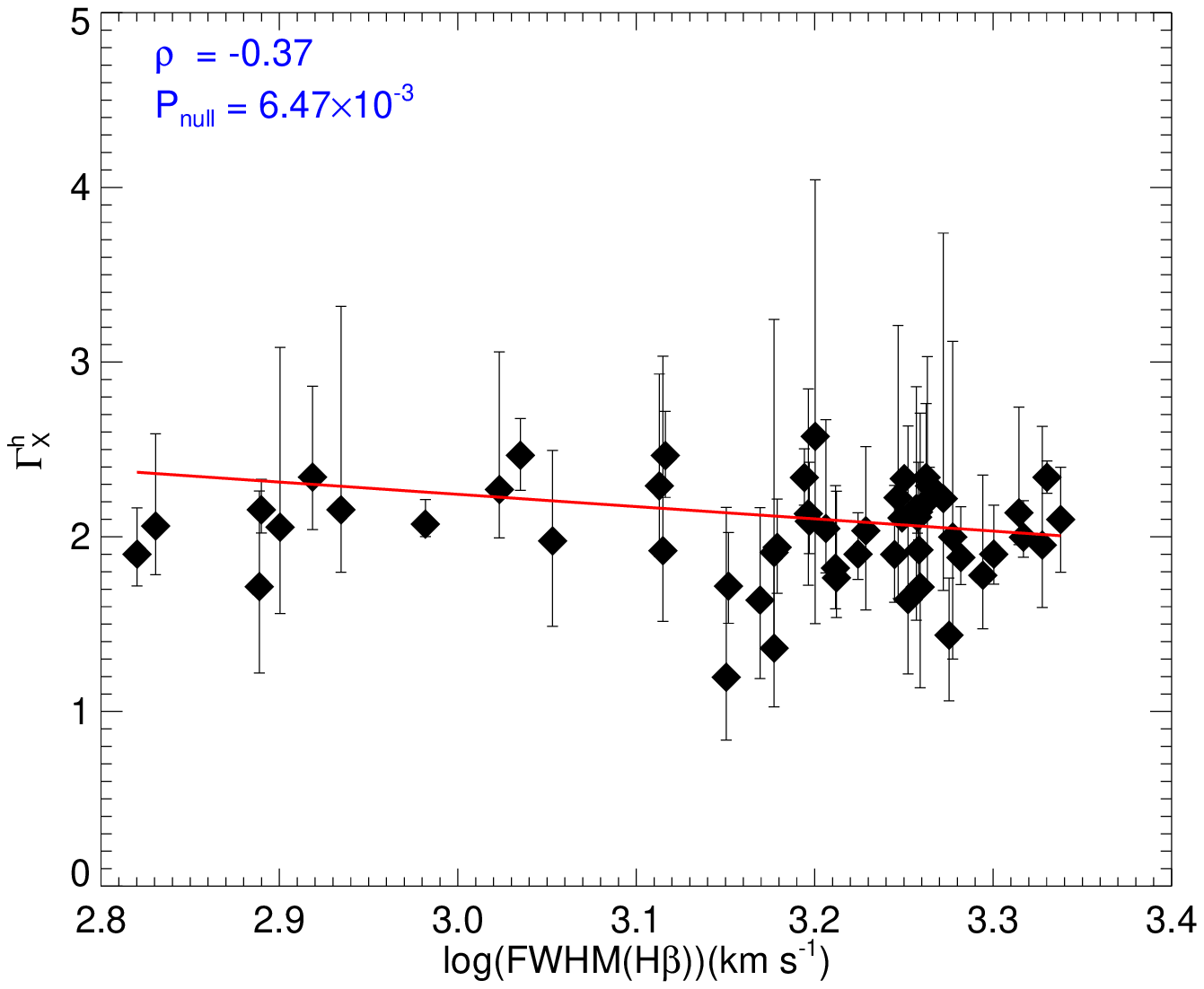}
      \hspace{-1.10cm}
 \includegraphics[width=0.37\textwidth,height=0.215\textheight,angle=00]{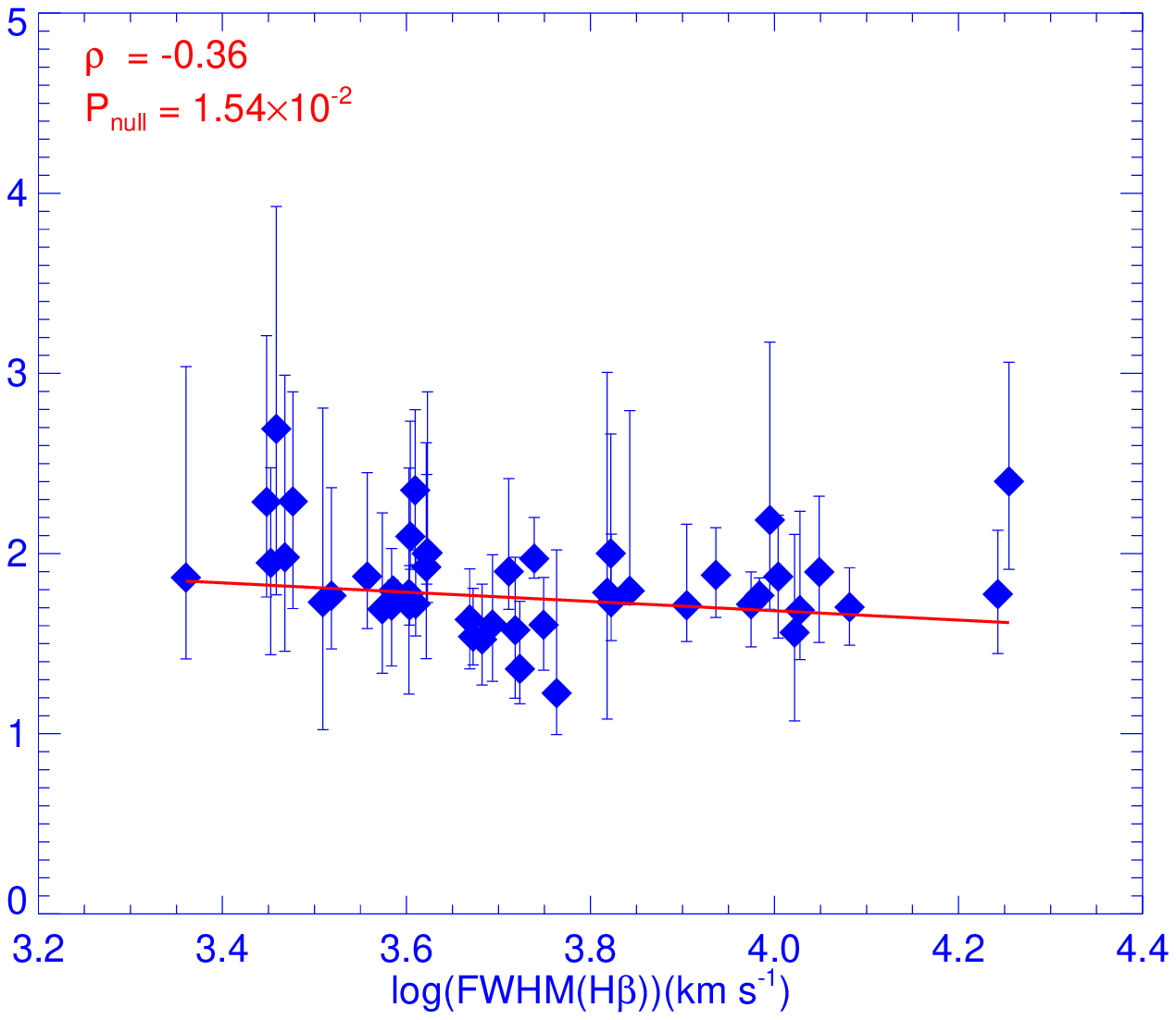}
      \hspace{-1.10cm}
 \includegraphics[width=0.37\textwidth,height=0.215\textheight,angle=00]{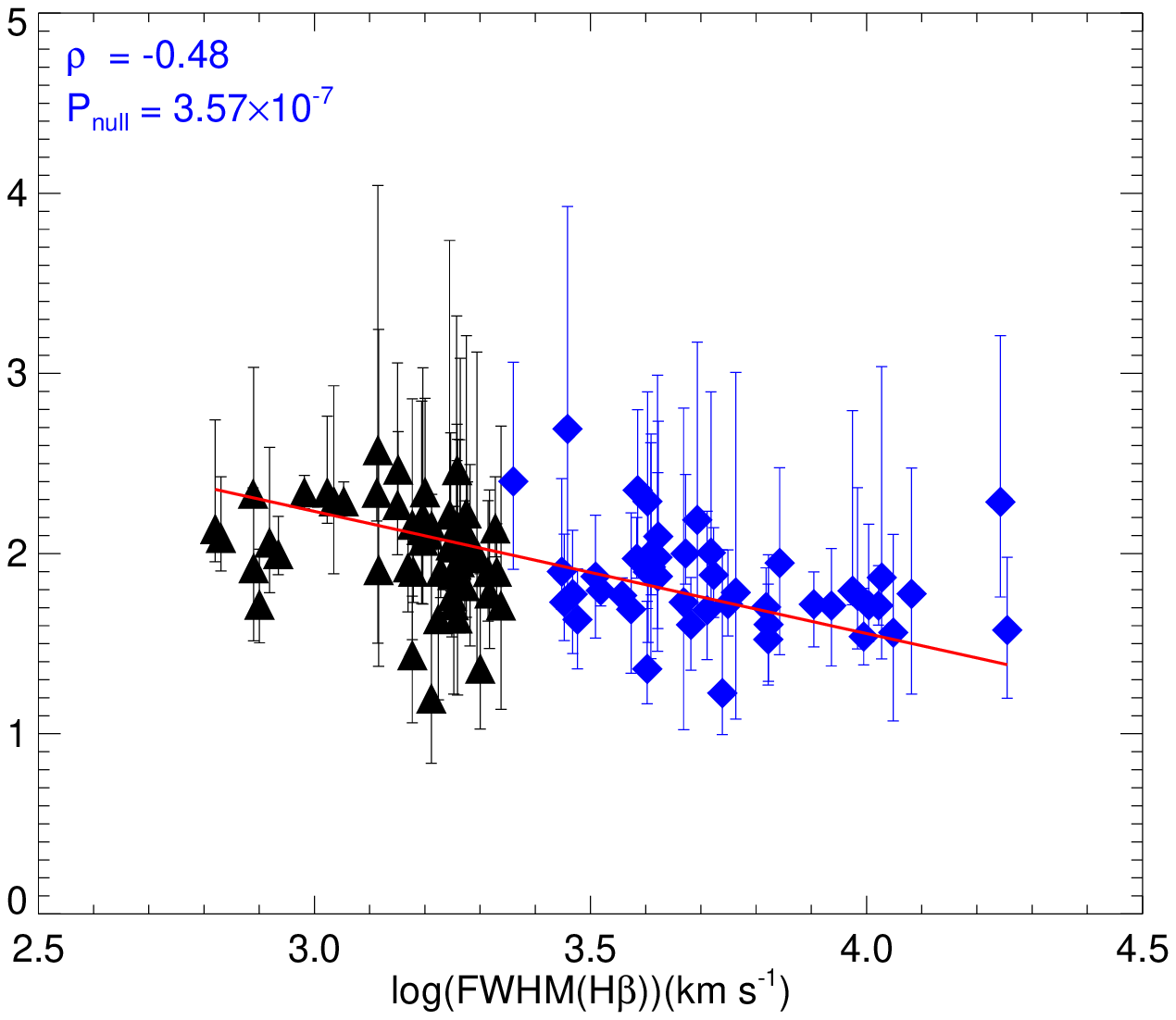}
  \end{minipage}
  \caption{{\scriptsize Same as Fig.~\ref{fig:nlsy_blsy_corr_rosat_xmm_soft}, but using the  53 NLSy1 and  46 BLSy1 galaxies for the {\it XMM-Newton} hard (2-10 keV) X-ray photon indices ($\Gamma_{X}^{h}$).}}
  \label{fig:nlsy_blsy_corr_xmm_hard}
\end{figure*}

It may be noted here  that the effect of soft X-ray excess, cold absorbers, warm absorbers, 
and other low energy spectral complexities are generally
prominent below 2 keV~\citep{Brandt1997MNRAS.285L..25B} in the NLSy1 and BLSy1
galaxies. So, to confirm the aforementioned correlations, found for the samples of
NLSy1 and BLSy1 galaxies between $\Gamma_{X}^{s}$ and log(R$_{\mathrm Edd}$), we analyzed
$\Gamma_{X}^{h}-\log(R_{\mathrm Edd}$) correlation in hard energy (2-10 keV) band which is thought to be
probably less affected by soft X-ray excess.
For this, we analyzed  53 NLSy1 and  46 BLSy1 galaxies to whom $\Gamma_{X}^{h}$ had been obtained (e.g. Sect.~\ref{section 4.0}).
The analysis of these two subsamples
along with  their joint subsample (i.e., [NLSy1$+$BLSy1]) resulted
in a good positive correlation between $\Gamma_{X}^{h}$ and log(R$_{\mathrm Edd}$)
for  53 NLSy1,  46 BLSy1 and  99 [NLSy1$+$BLSy1] galaxies with  $\rho$ = 0.42, 0.43, and 0.56, respectively.  This can be seen in the top panels of Fig.~\ref{fig:nlsy_blsy_corr_xmm_hard}, and also from the  middle part of 
Table~\ref{table_nlsy1_blsy1_corr}.  Mild anticorrelations are
found between $\Gamma_{X}^{h}$ and log(FWHM($H{\beta}$)) for the
subsamples of NLSy1, BLSy1 and [NLSy1$+$BLSy1] galaxies with $\rho$ =  $-$0.37, $-$0.36, and
$-$0.48, respectively. However, no significant correlations are found
for $\Gamma_{X}^{h}-\log(L_{\mathrm bol}$) and $\Gamma_{X}^{h}-\log(M_{\mathrm BH}$) in these
subsamples (e.g., see Table~\ref{table_nlsy1_blsy1_corr} and
Fig.~\ref{fig:nlsy_blsy_corr_xmm_hard}). The $\chi^{2}$-minimization
using the functional form of $y=mx+c$ (see above) with  $y=\Gamma_{X}^{h}$ and X either
$\log(R_{\mathrm Edd}$) or $\log(FWHM(H{\beta})$), yielded for the subsample of NLSy1 galaxies as

\begin{equation}
  \begin{split}
  \Gamma_{X}^{h} = (0.29\pm0.06) \log(R_{\mathrm Edd}) + (2.33\pm0.03)\\
  \Gamma_{X}^{h} = (-0.70\pm0.13) \log(FWHM(H\beta)) + (4.35\pm0.41)
  \end{split}
\end{equation}

, for the subsample of BLSy1 galaxies as

\begin{equation}
  \begin{split}
  \Gamma_{X}^{h} = (0.17\pm0.09)\log(R_{\mathrm Edd}) + (2.03\pm0.14)\\
  \Gamma_{X}^{h} = (-0.26\pm0.20)\log(FWHM(H\beta)) + (2.71\pm0.73)
  \end{split}
\end{equation}

and for the joint subsample of [NLSy1$+$BLSy1] galaxies as

\begin{equation}
  \begin{split}
  \Gamma_{X}^{h} = (0.35\pm0.03) \log(R_{\mathrm Edd}) + (2.34\pm0.03)\\
  \Gamma_{X}^{h} = (-0.68\pm0.06) \log(FWHM(H\beta)) + (4.27\pm0.19)
  \end{split}
\end{equation}

as shown by solid red lines in the plots of Fig.~\ref{fig:nlsy_blsy_corr_xmm_hard}.\par

\begin{figure*}[!t]
  \begin{minipage}[]{1.0\textwidth}
 \includegraphics[width=0.37\textwidth,height=0.215\textheight,angle=00]{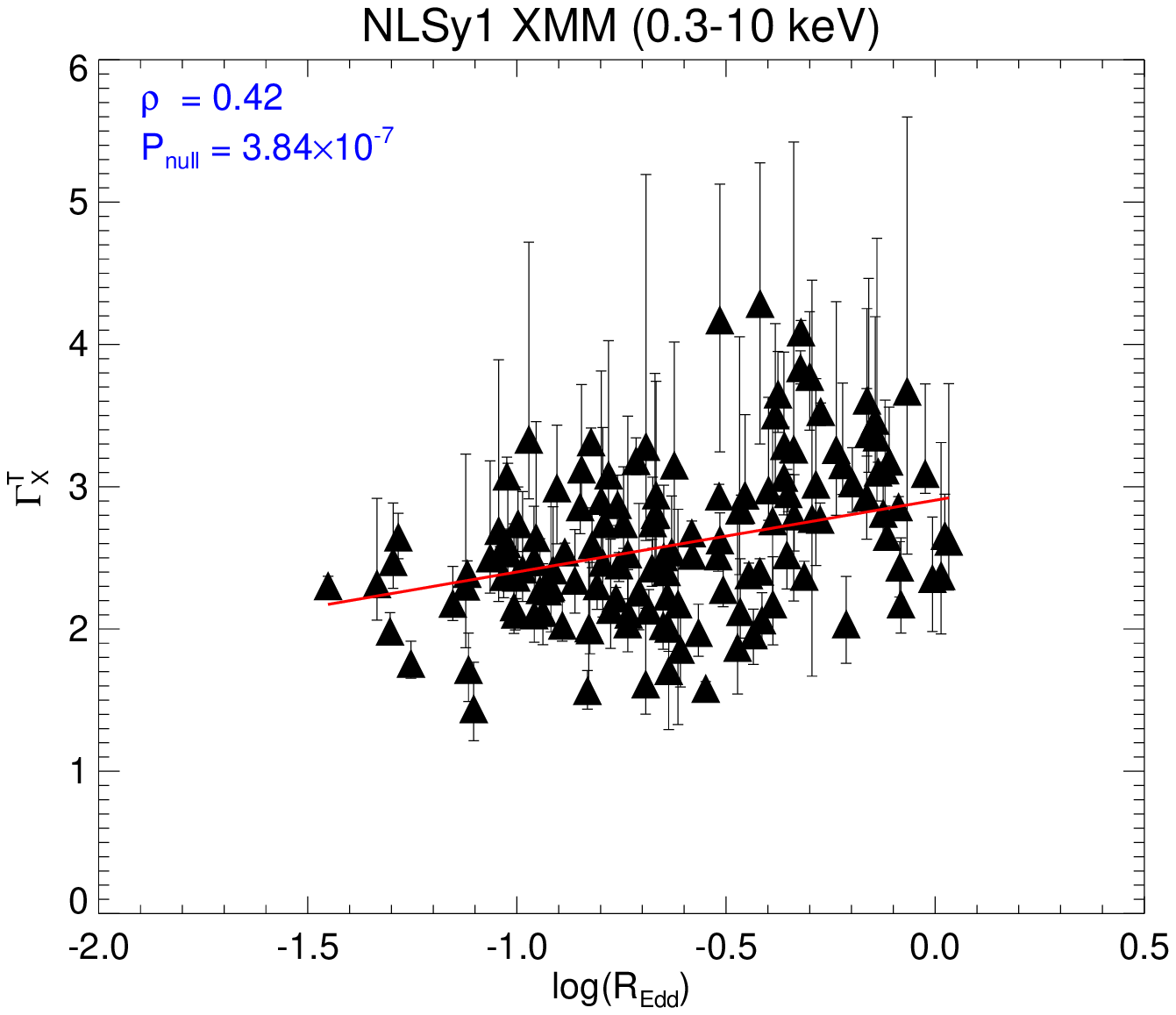}
     \hspace{-1.10cm}
 \includegraphics[width=0.37\textwidth,height=0.215\textheight,angle=00]{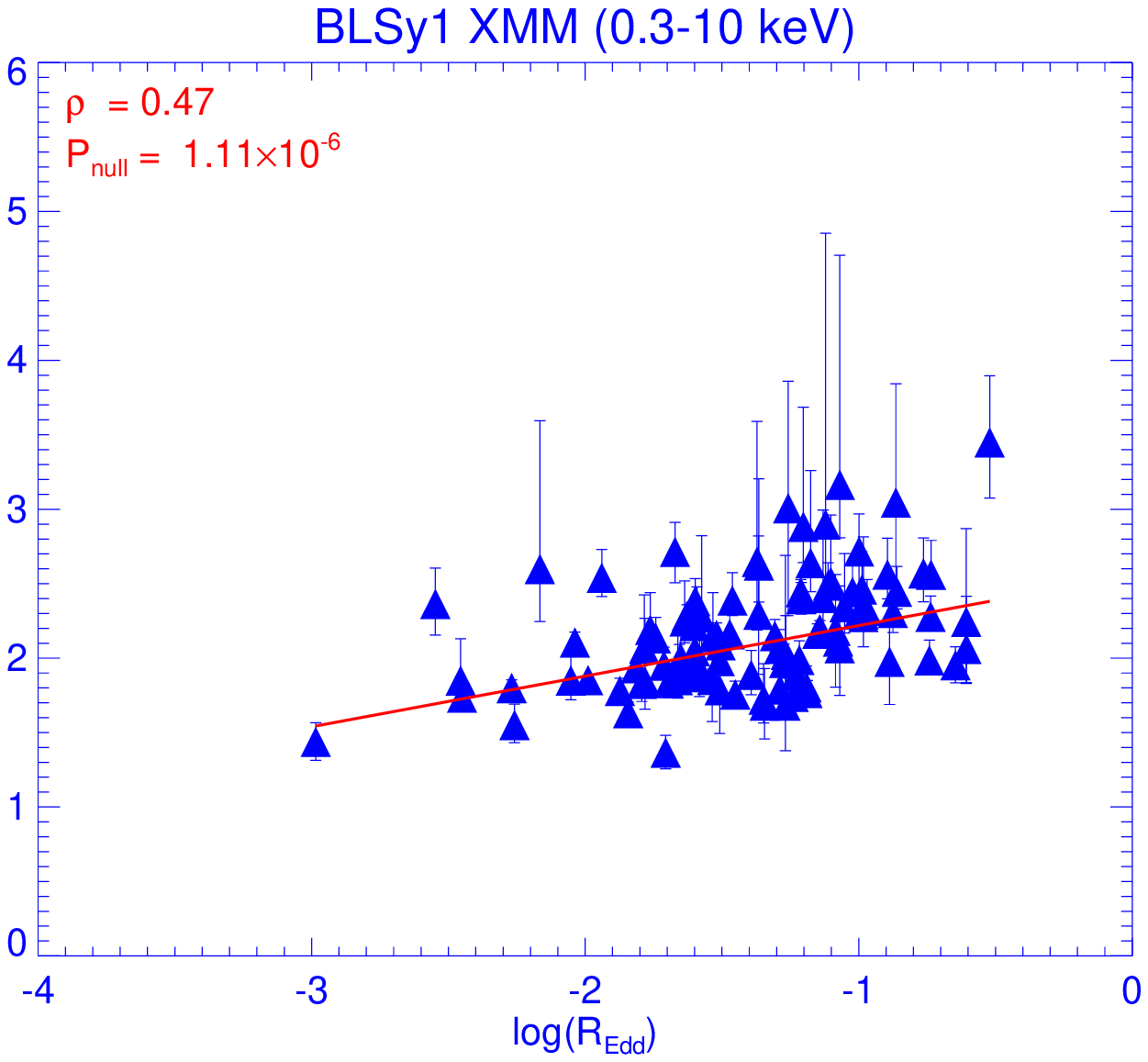}
    \hspace{-1.10cm}
 \includegraphics[width=0.37\textwidth,height=0.215\textheight,angle=00]{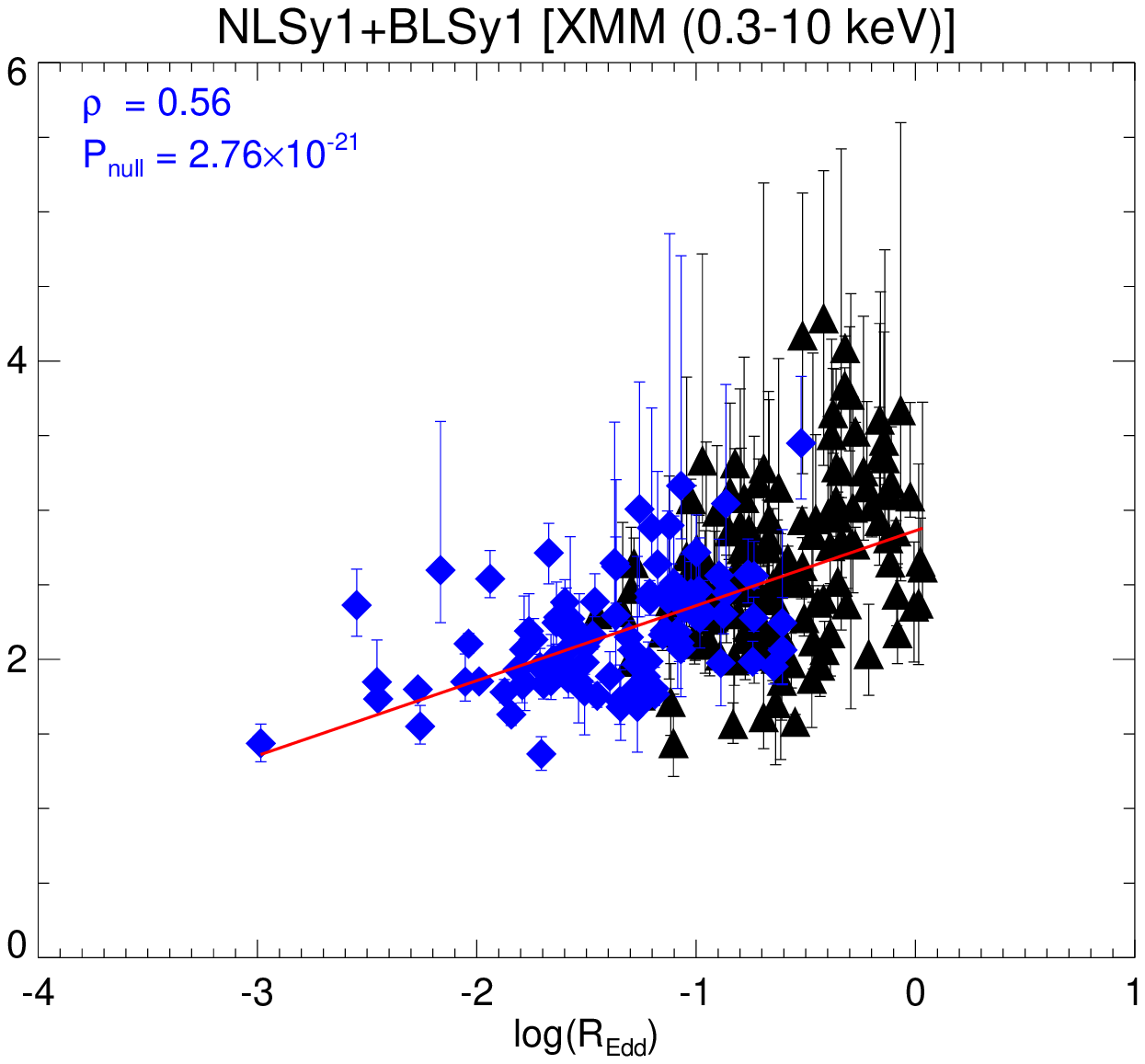}
 \includegraphics[width=0.37\textwidth,height=0.215\textheight,angle=00]{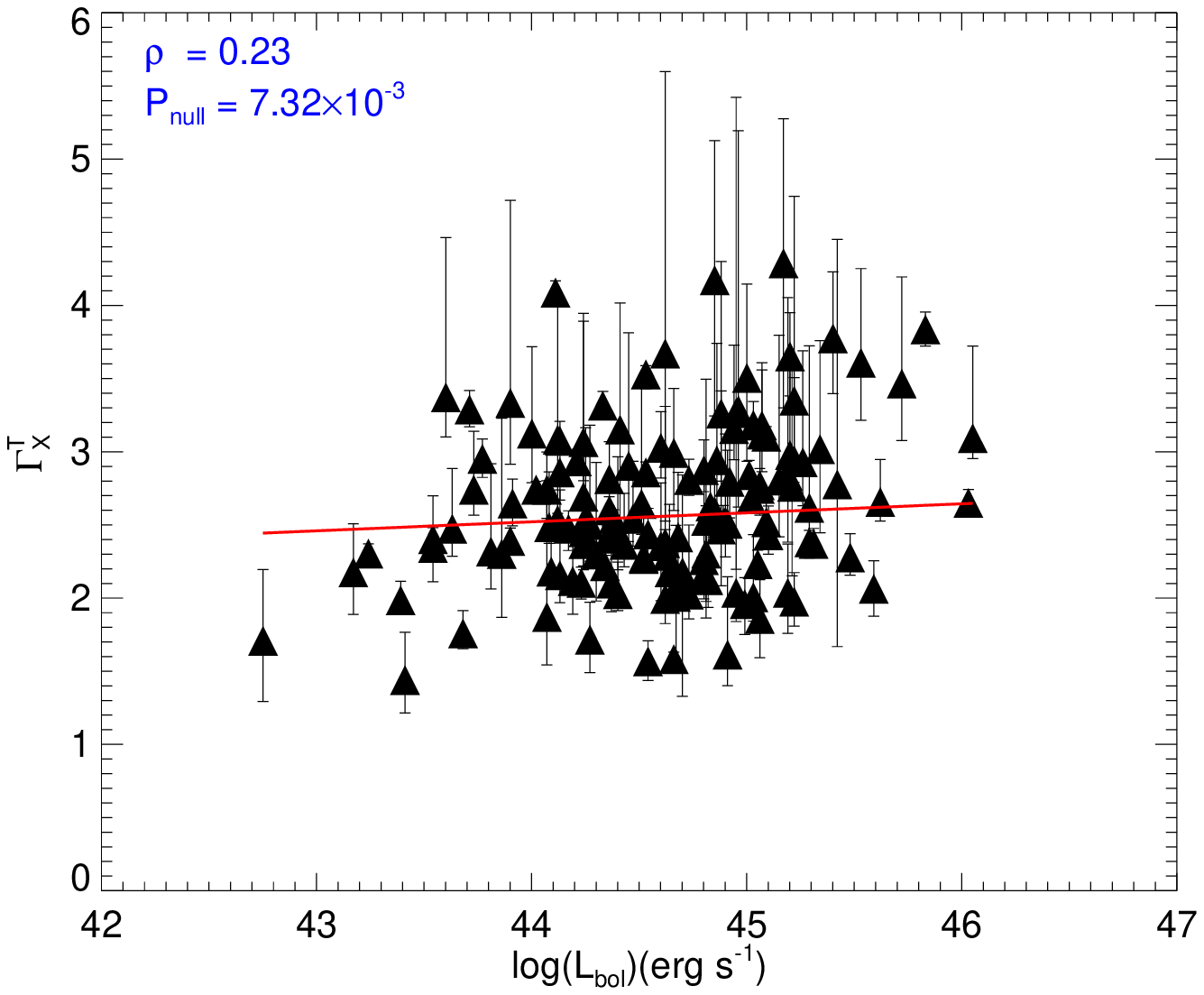}
    \hspace{-1.10cm}
 \includegraphics[width=0.37\textwidth,height=0.215\textheight,angle=00]{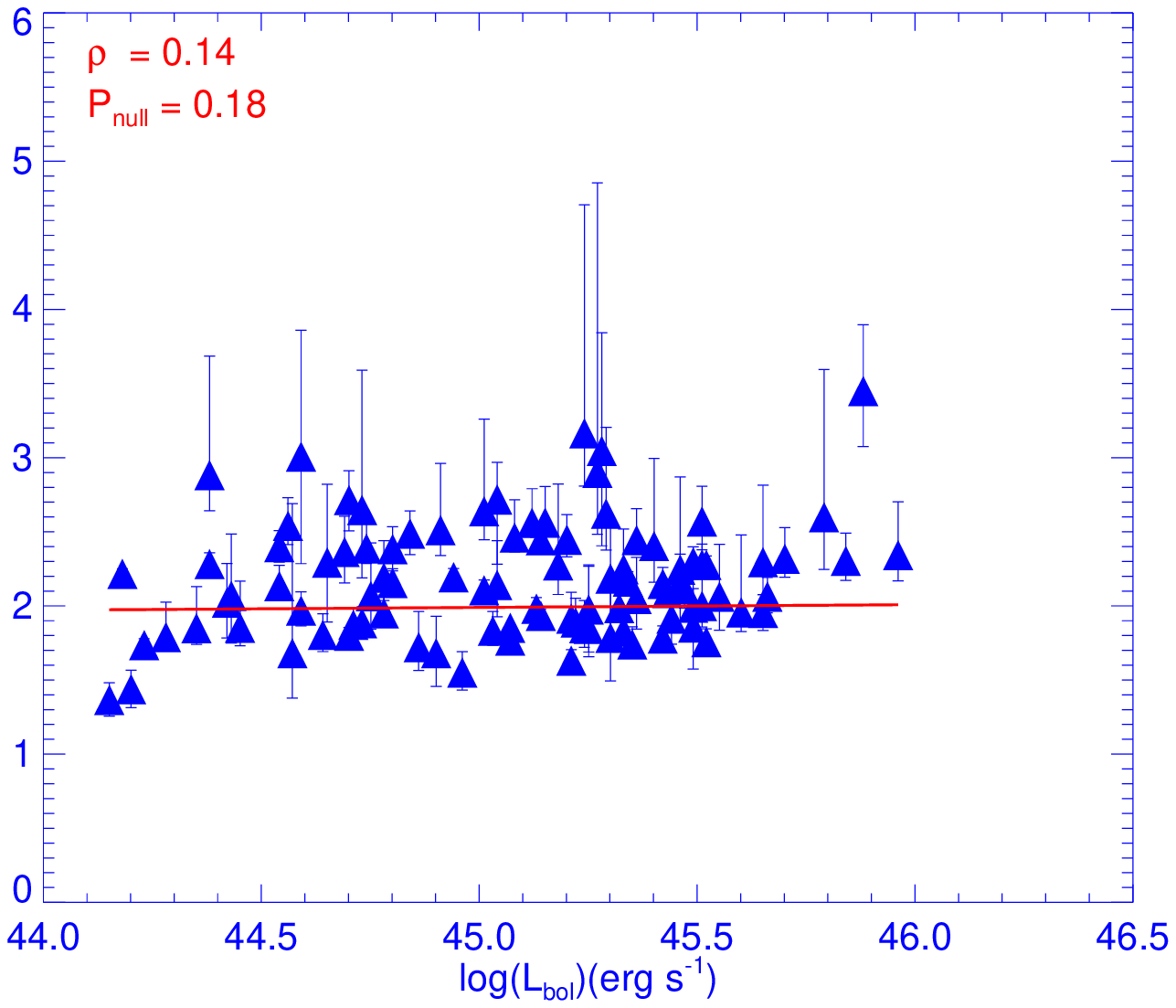}
     \hspace{-1.10cm}
 \includegraphics[width=0.37\textwidth,height=0.215\textheight,angle=00]{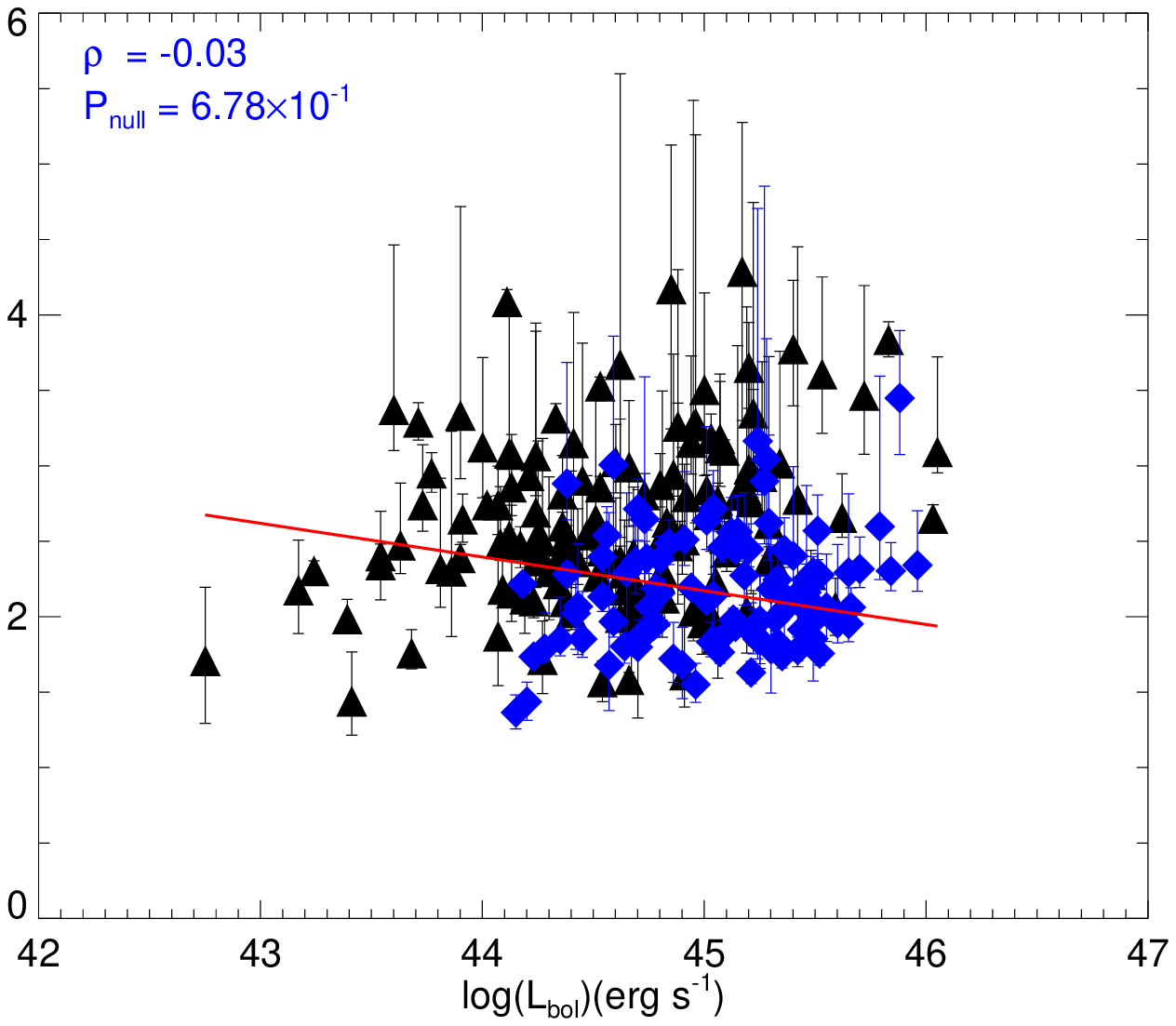}
 \includegraphics[width=0.37\textwidth,height=0.215\textheight,angle=00]{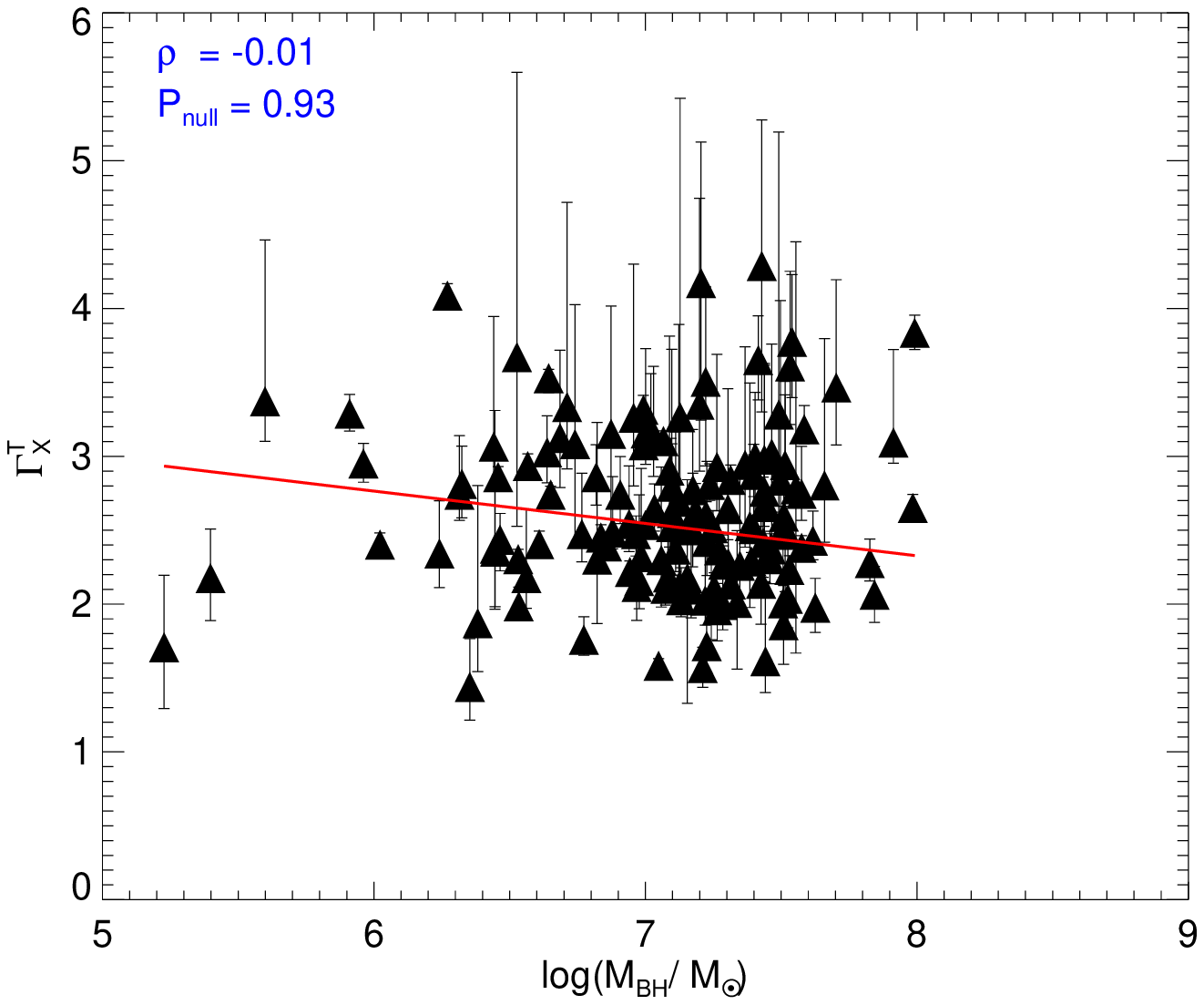}
     \hspace{-1.10cm}
 \includegraphics[width=0.37\textwidth,height=0.215\textheight,angle=00]{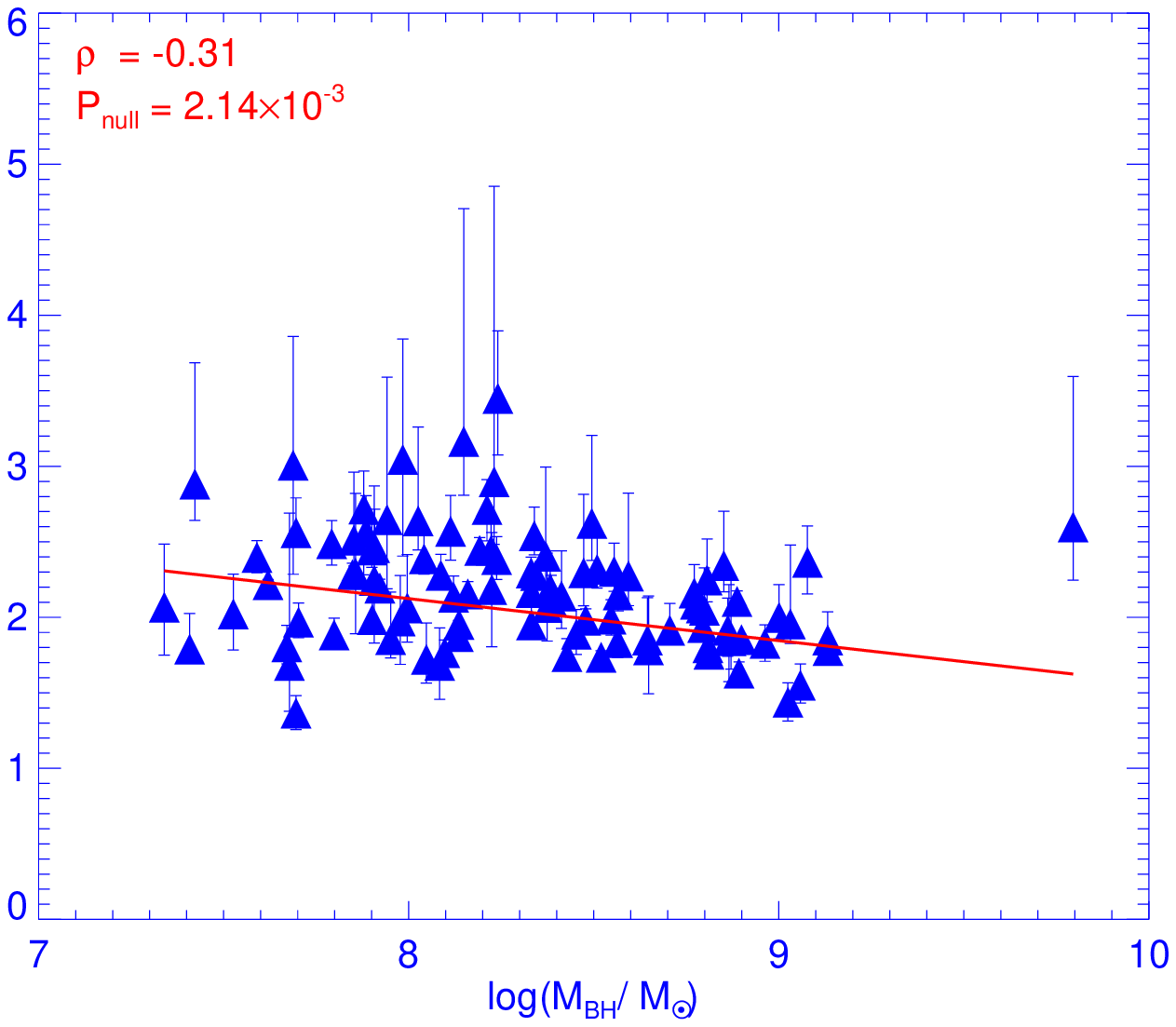}
     \hspace{-1.10cm}
 \includegraphics[width=0.37\textwidth,height=0.215\textheight,angle=00]{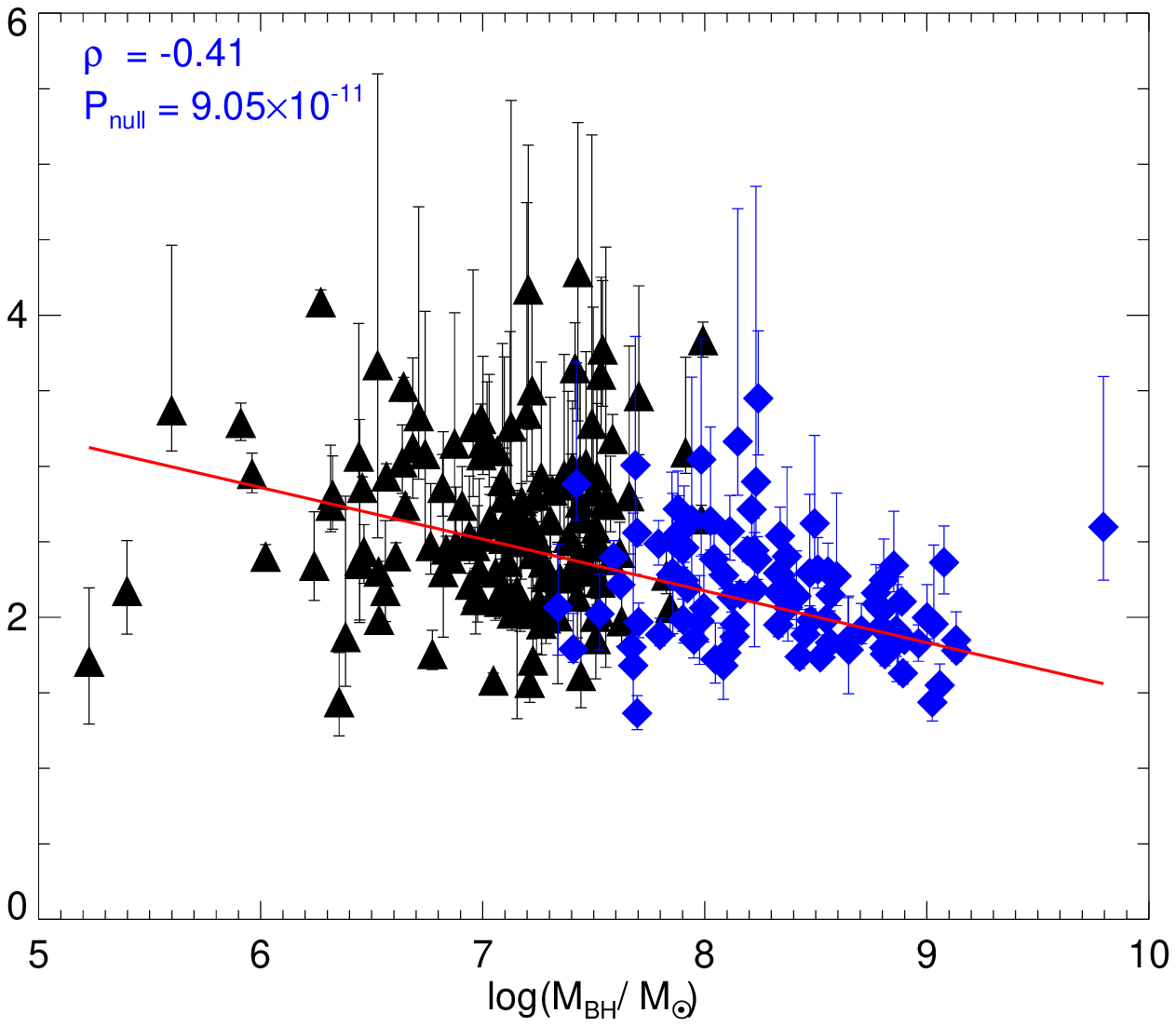}
 \includegraphics[width=0.37\textwidth,height=0.215\textheight,angle=00]{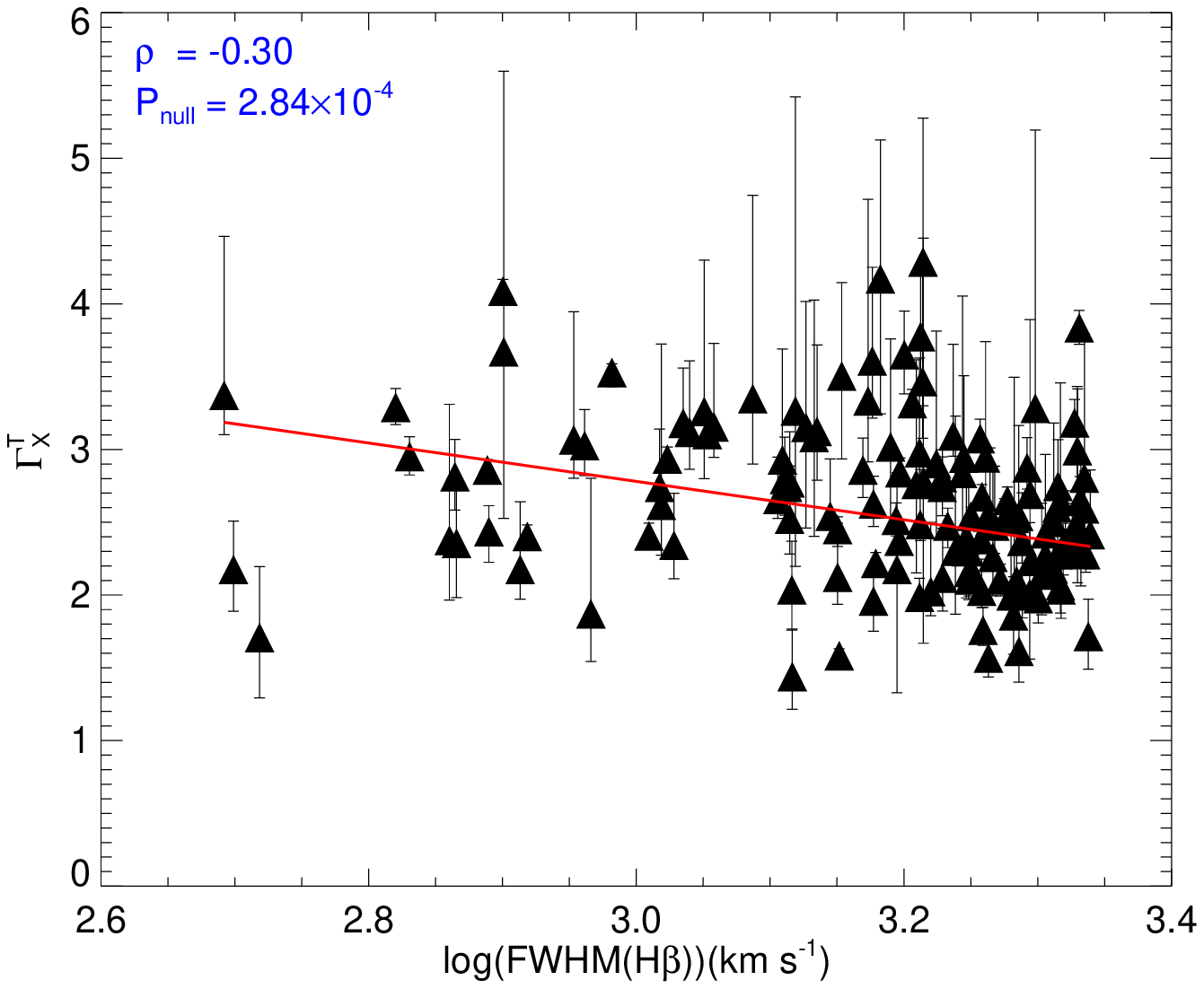}
      \hspace{-1.10cm}
 \includegraphics[width=0.37\textwidth,height=0.215\textheight,angle=00]{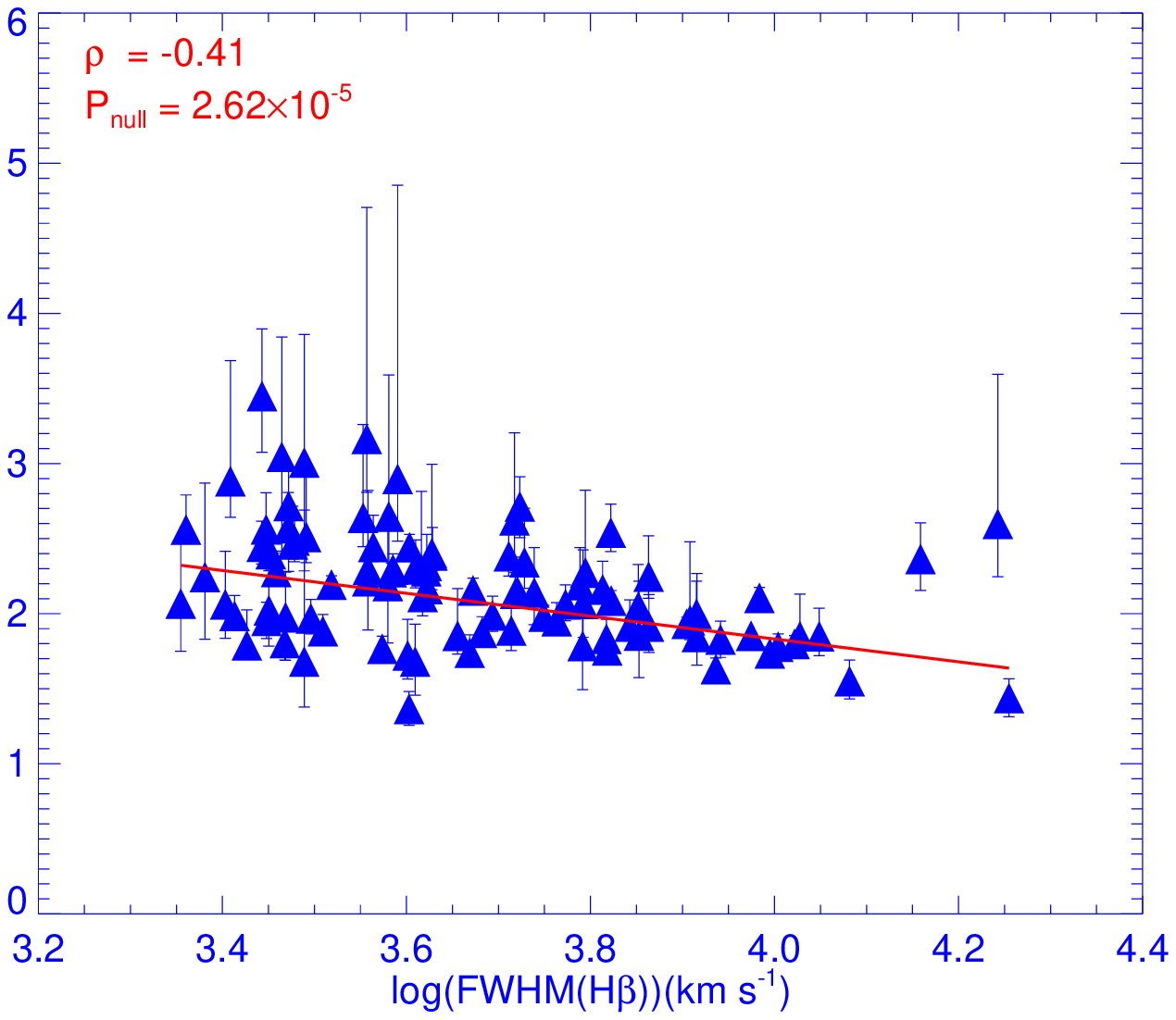}
      \hspace{-1.10cm}
 \includegraphics[width=0.37\textwidth,height=0.215\textheight,angle=00]{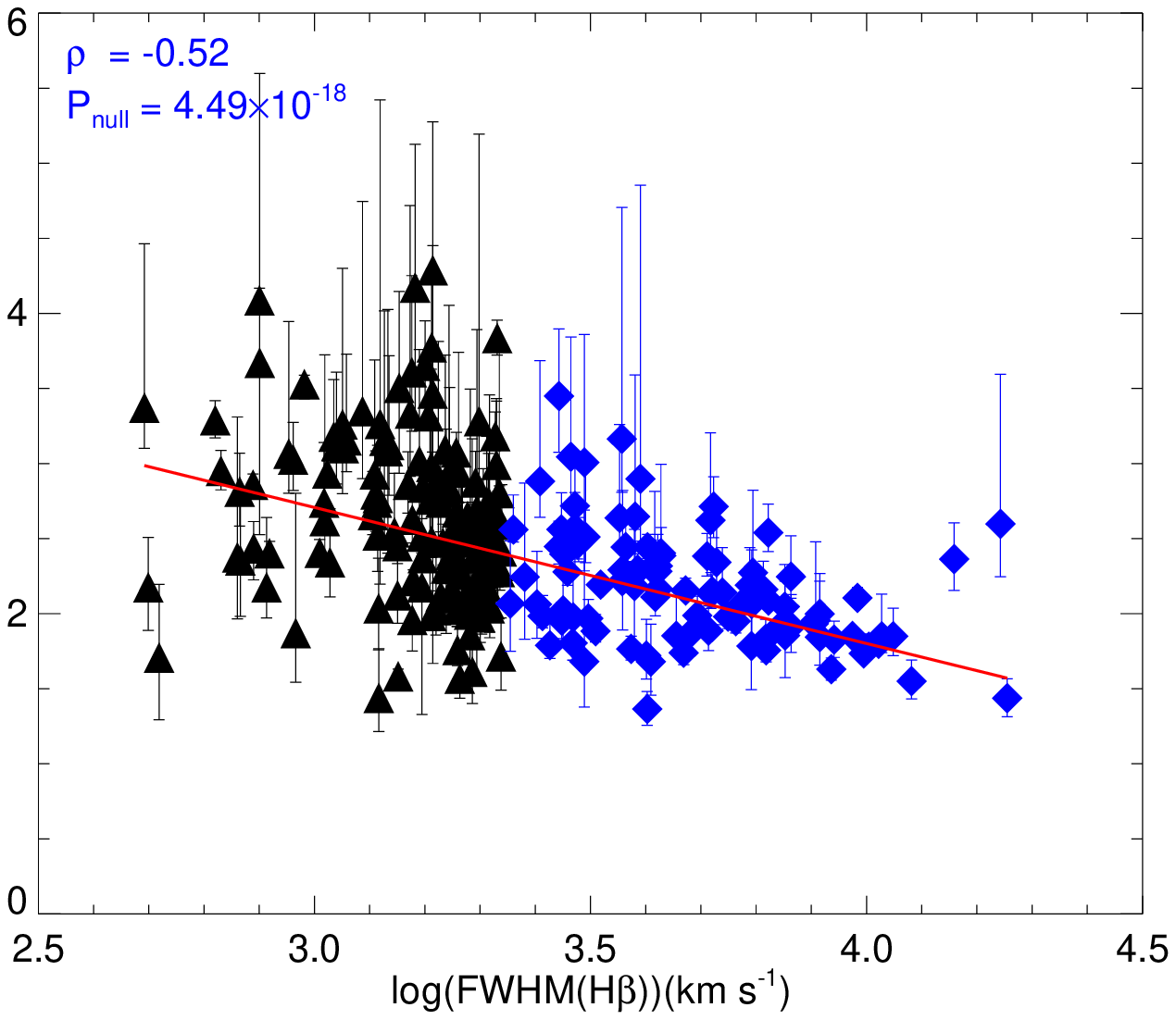}
  \end{minipage}
  \caption{\scriptsize  Same as Fig.~\ref{fig:nlsy_blsy_corr_rosat_xmm_soft}, but using the 139 NLSy1 and 97 BLSy1 galaxies for the {\it XMM-Newton} total (0.3-10 keV) X-ray photon indices ($\Gamma_{X}^{T}$).}
  \label{fig:nlsy_blsy_corr_xmm_total_energy}
\end{figure*}

Additionally, the soft (0.1-2.0 keV) X-ray photon indices of most of the NLSy1s are affected by soft excess and many of them also by absorption features due to ``warm absorbers''~\citep{Vaughan1999MNRAS.309..113V}. Therefore, a detailed X-ray spectral analysis in the 0.3-10 keV energy band is also carried out  to confirm the aforementioned correlations found for the samples of NLSy1 and BLSy1 galaxies between $\Gamma_{X}^{s}$ and log(R$_{\mathrm Edd}$). For this, we analyzed 139 NLSy1 and 97 BLSy1 galaxies in 0.3-10 keV energy band (e.g. Sect.~\ref{section 2.0}) of {\it XMM-Newton} Telescope.
The analysis of these two subsamples
along with their joint subsample (i.e., [NLSy1$+$BLSy1]) resulted
in a good positive correlation between $\Gamma_{X}^{T}$ and log(R$_{\mathrm Edd}$)
for NLSy1, BLSy1 and [NLSy1$+$BLSy1] galaxies subsamples with  $\rho$ = 0.42, 0.47, and 0.56, respectively. This can be seen in the top panels
of Fig.~\ref{fig:nlsy_blsy_corr_xmm_total_energy}, and also from the bottom part of 
Table~\ref{table_nlsy1_blsy1_corr}. Mild anticorrelations are
found between $\Gamma_{X}^{T}$ and log(FWHM($H{\beta}$)) for the
subsamples of NLSy1s, BLSy1s with $\rho$ = $-$0.30, $-$0.41, respectively. This is found to
be stronger with $\rho$ of $-$0.52 when both subsamples are combined together. However, no significant correlations are found
for $\Gamma_{X}^{T}-\log(L_{\mathrm bol}$) in these
subsamples except a mild anticorrelations found for $\Gamma_{X}^{T}-\log(M_{\mathrm BH}$) with $\rho$ of  $-$0.30 and  $-$0.41, in the case of 97 BLSy1 and 236 [NLSy1$+$BLSy1] galaxies respectively,  (e.g., see Table~\ref{table_nlsy1_blsy1_corr} and
Fig.~\ref{fig:nlsy_blsy_corr_xmm_total_energy}). The $\chi^{2}$-minimization
using the functional form of $y=mx+c$ (see above) with  $y=\Gamma_{X}^{T}$ and X either
$\log(R_{\mathrm Edd}$) or $\log(FWHM(H{\beta})$), yielded for the subsample of  NLSy1 galaxies as

\begin{equation}
  \begin{split}
  \Gamma_{X}^{T} = (0.51\pm0.03) \log(R_{\mathrm Edd}) + (2.91\pm0.03)\\
  \Gamma_{X}^{T} = (-1.32\pm0.08) \log(FWHM(H\beta)) + (6.73\pm0.25)
  \end{split}
\end{equation}

, for the subsample of  BLSy1 galaxies as

\begin{equation}
  \begin{split}
  \Gamma_{X}^{T} = (0.34\pm0.02)\log(R_{\mathrm Edd}) + (2.56\pm0.04)\\
  \Gamma_{X}^{T} = (-0.76\pm0.05)\log(FWHM(H\beta)) + (4.87\pm0.17)
  \end{split}
\end{equation}

and for the joint subsample of  [NLSy1$+$BLSy1] galaxies as

\begin{equation}
  \begin{split}
  \Gamma_{X}^{T} = (0.50\pm0.01) \log(R_{\mathrm Edd}) + (2.86\pm0.02)\\
  \Gamma_{X}^{T} = (-0.91\pm0.02) \log(FWHM(H\beta)) + (5.42\pm0.07)
  \end{split}
\end{equation}

as shown by solid red lines in the plots of Fig.~\ref{fig:nlsy_blsy_corr_xmm_total_energy}.

Additionally, in view of the steeper $\Gamma_{X}^{s}$ for the sample
of NLSy1s and recalling that NLSy1 galaxies do have smaller FWHM of
the emission lines as compared to BLSy1 galaxies. It will be worth to
explore the possible correlation between FWHM of emission lines and X-ray spectral indices as well. Therefore in the bottom panels of
  Figs.~\ref{fig:nlsy_blsy_corr_rosat_xmm_soft},
  ~\ref{fig:nlsy_blsy_corr_xmm_hard}~\&~\ref{fig:nlsy_blsy_corr_xmm_total_energy},
  we have plotted $\Gamma^{s}_{X}$, $\Gamma^{h}_{X}$ \&
  $\Gamma^{T}_{X}$ versus log(FWHM(H$\beta$)) in the soft, hard and
  total 0.3-10 keV  energy bands, respectively. As can be seen from these
  figures (bottom right panel) that the anticorrelation in
  $\Gamma_{X}^{s}$, $\Gamma^{h}_{X}$ \&  $\Gamma^{T}_{X}$ versus
  log(FWHM(H$\beta$)) plots, based on the joint sample of NLSy1 and
  BLSy1 galaxies is significant with $\rho$= $-$0.55, $-$0.48 and
  $-$0.52, with their  corresponding  P$_{null}$ of
   2.87$\times$10$^{-31}$, 3.57$\times$10$^{-7}$ and
  4.49$\times$10$^{-18}$, respectively (see, Table~\ref{table_nlsy1_blsy1_corr}).

\subsection{Comparison of the $\Gamma_{X}^\mathrm{hot}$ correlations with AGN parameters among the subsamples NLSy1 and BLSy1 galaxies}

\begin{figure*}[!t]
  \begin{minipage}[]{1.0\textwidth}
  \includegraphics[width=0.5\textwidth,height=0.215\textheight,angle=00]{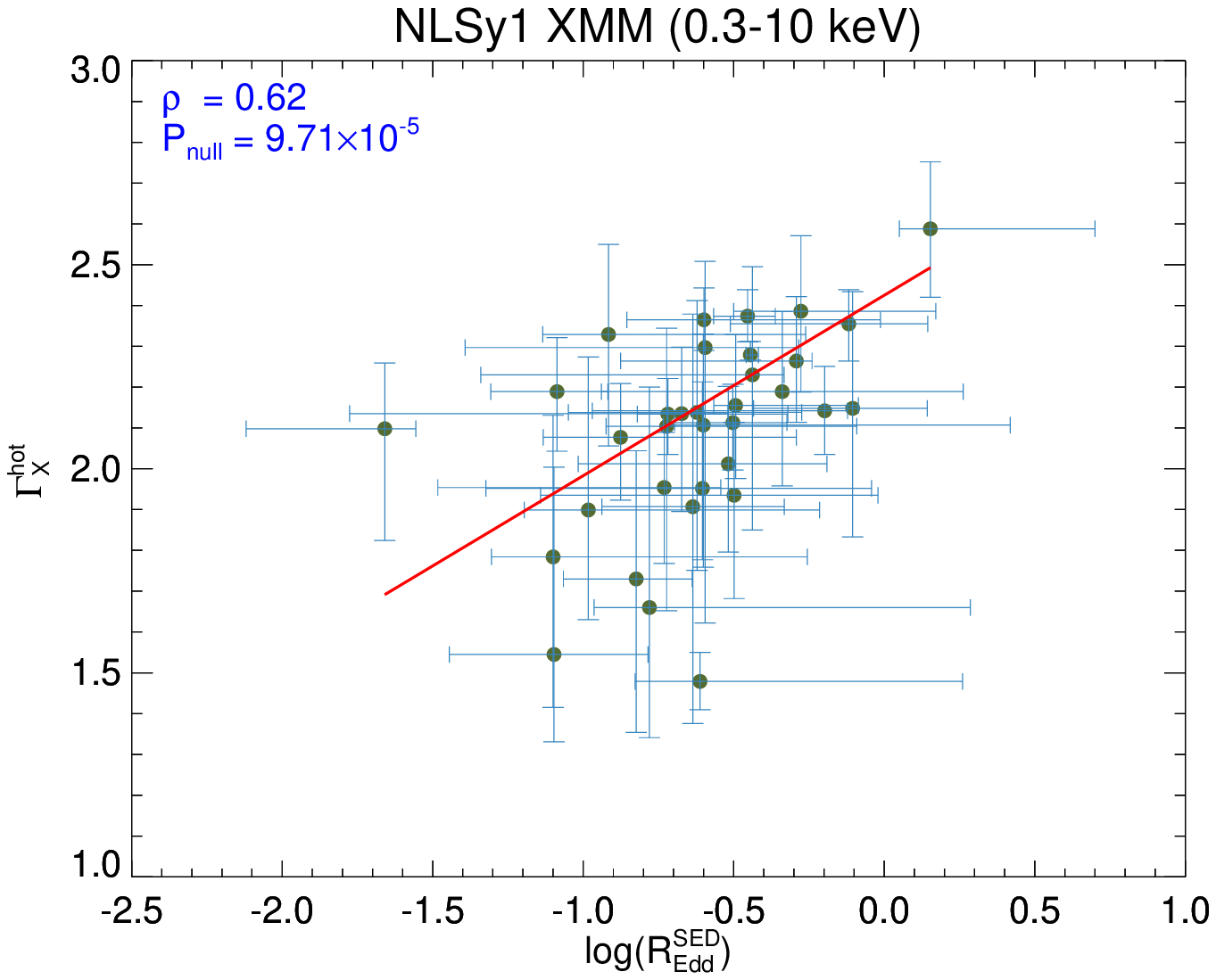}
  \includegraphics[width=0.5\textwidth,height=0.215\textheight,angle=00]{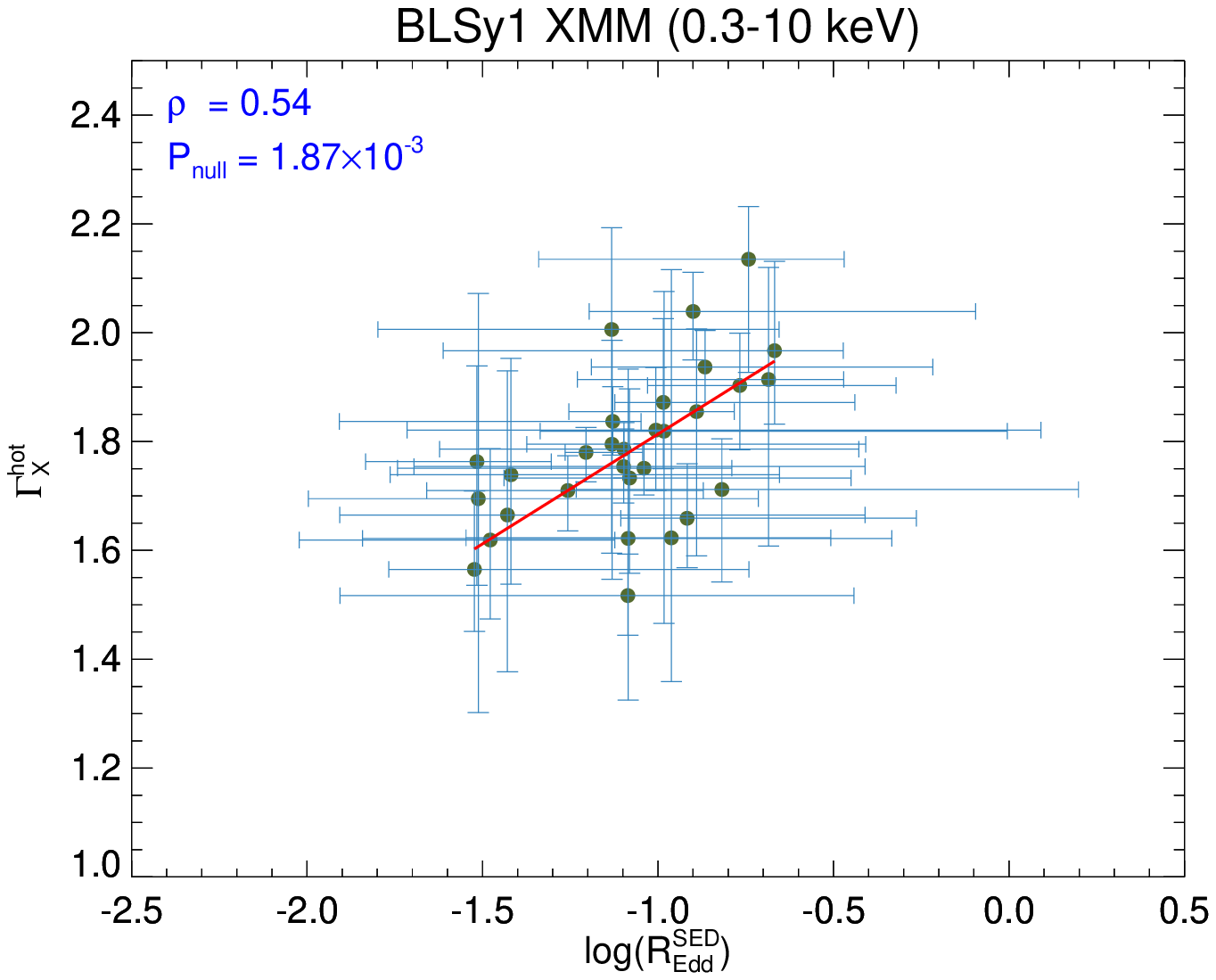} 
  \includegraphics[width=0.5\textwidth,height=0.215\textheight,angle=00]{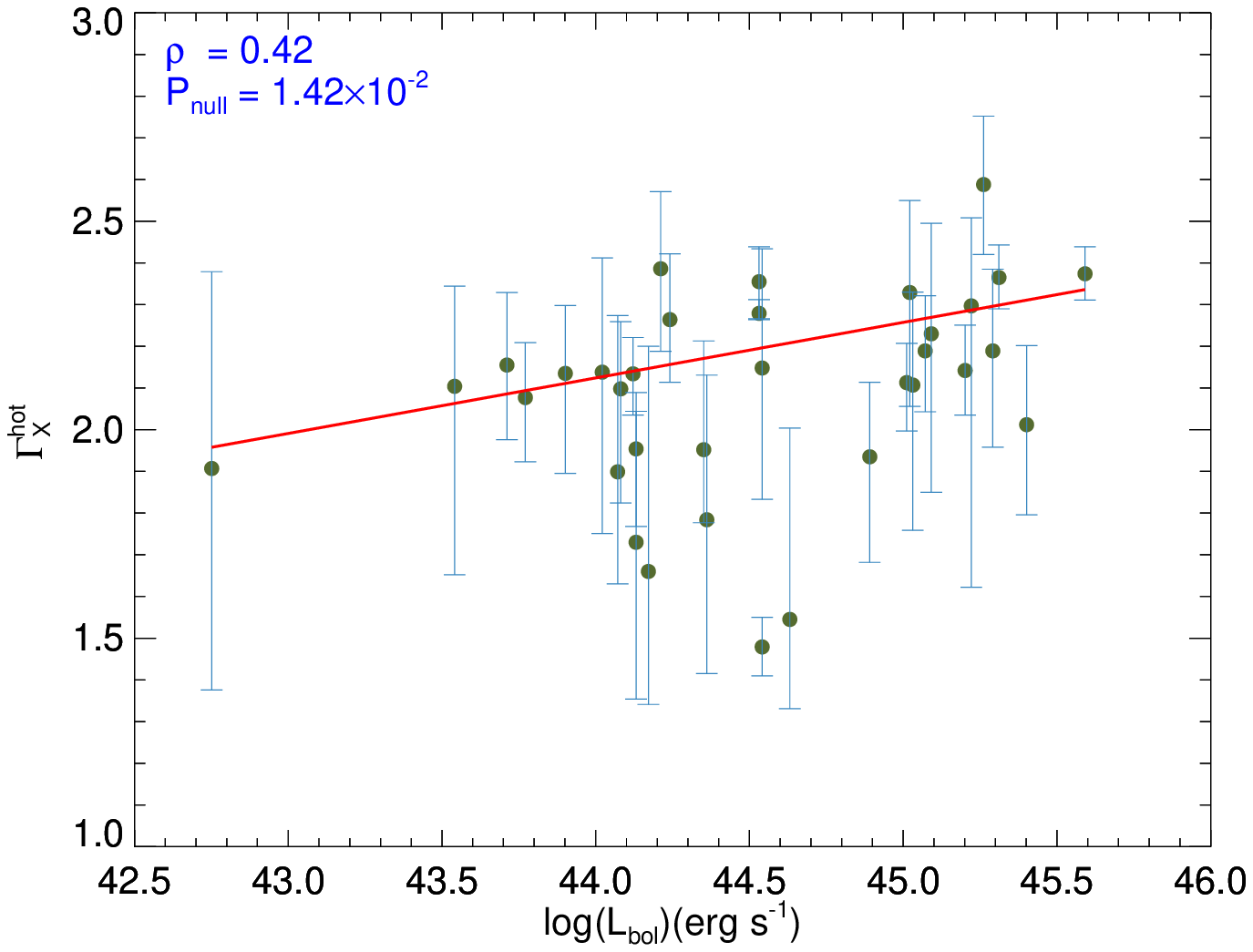} 
  \includegraphics[width=0.5\textwidth,height=0.215\textheight,angle=00]{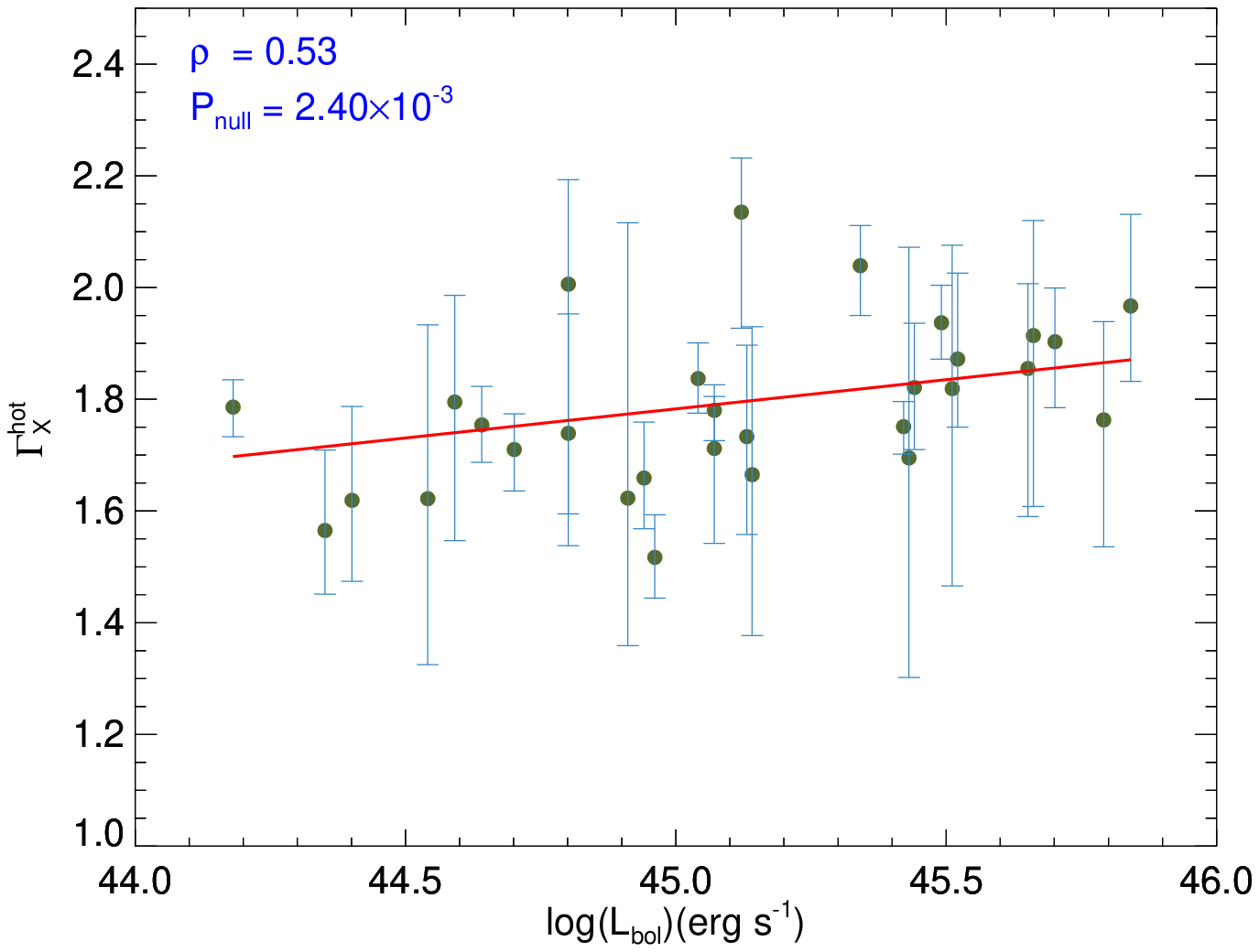} 
  \includegraphics[width=0.5\textwidth,height=0.215\textheight,angle=00]{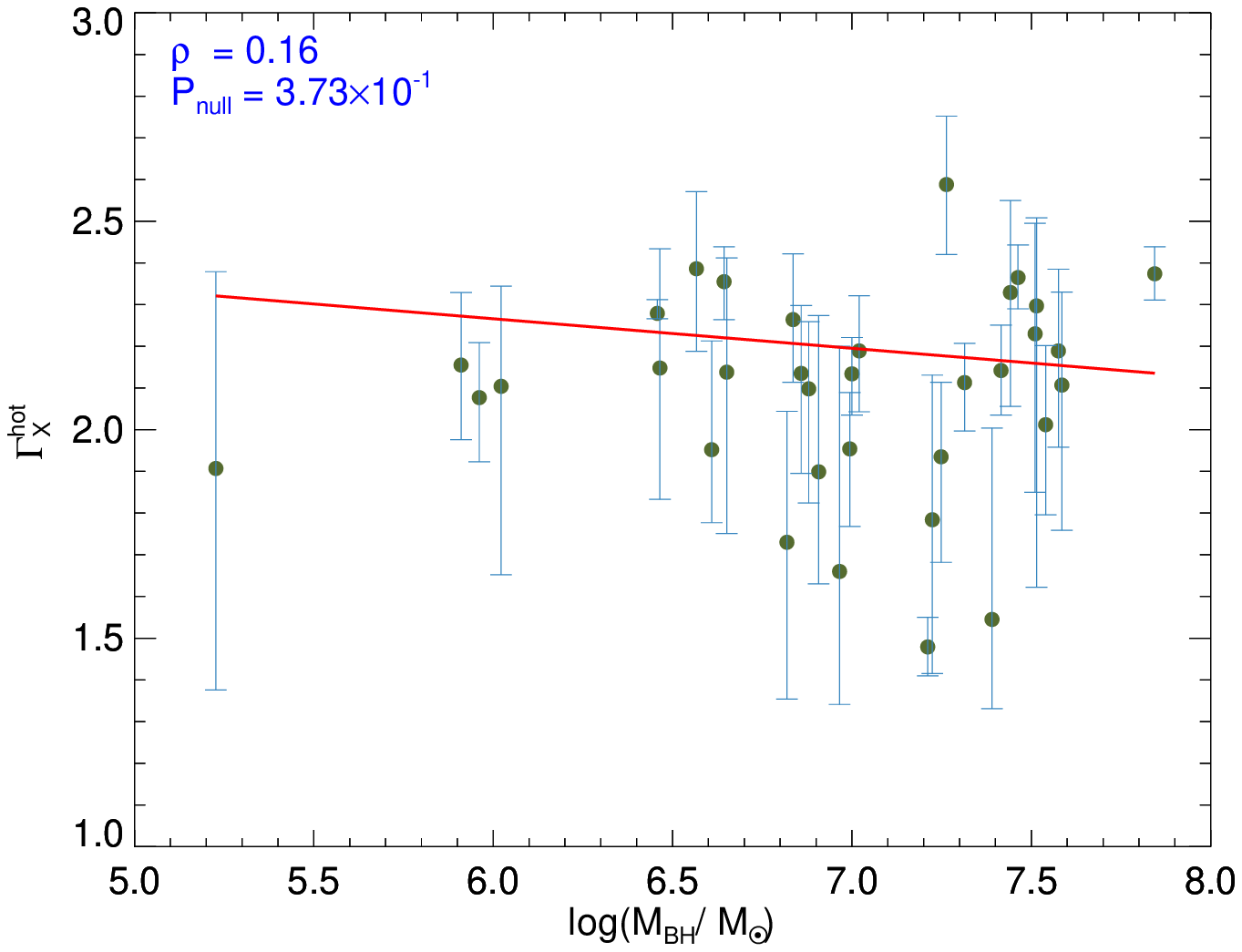}  
  \includegraphics[width=0.5\textwidth,height=0.215\textheight,angle=00]{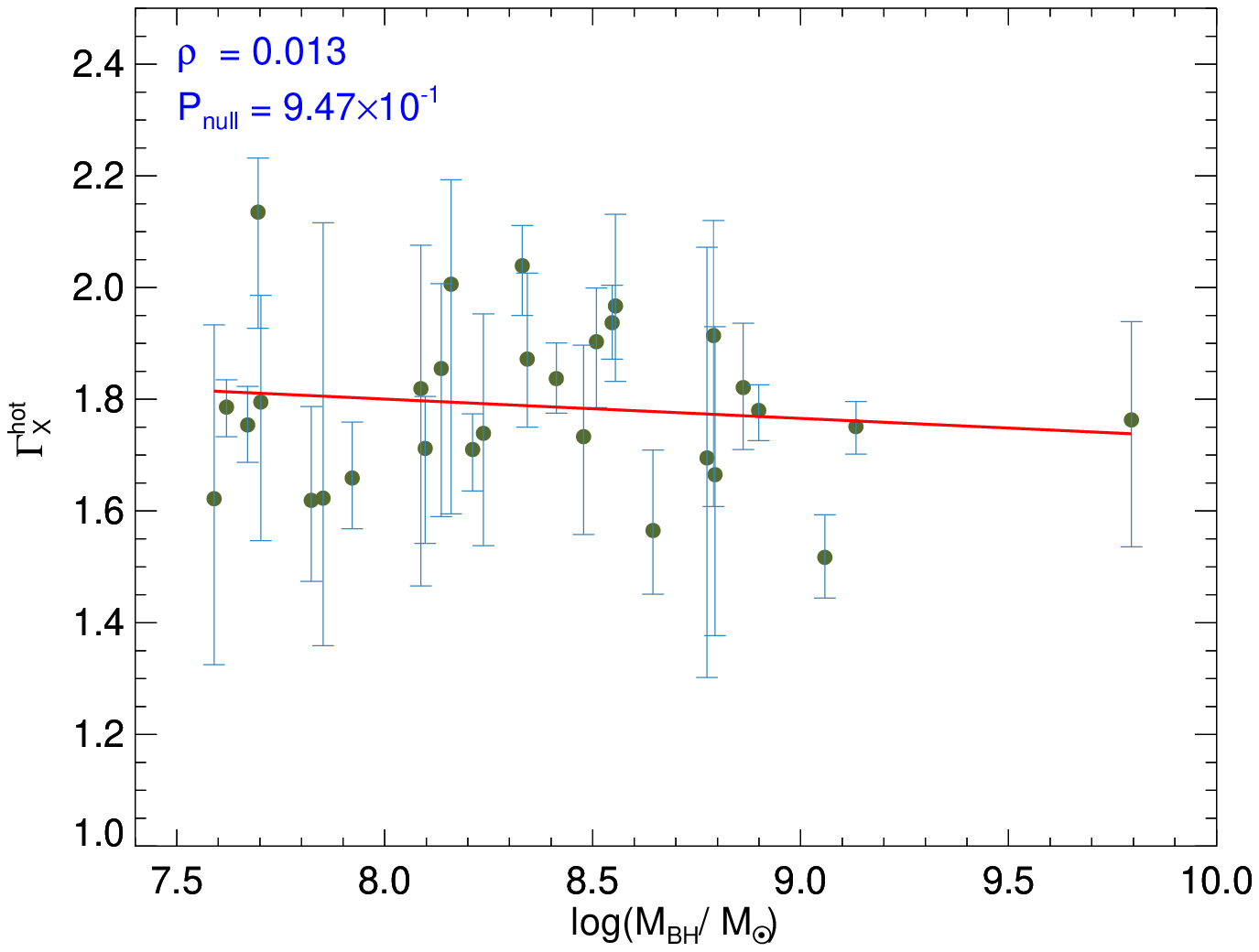}  
  \includegraphics[width=0.5\textwidth,height=0.215\textheight,angle=00]{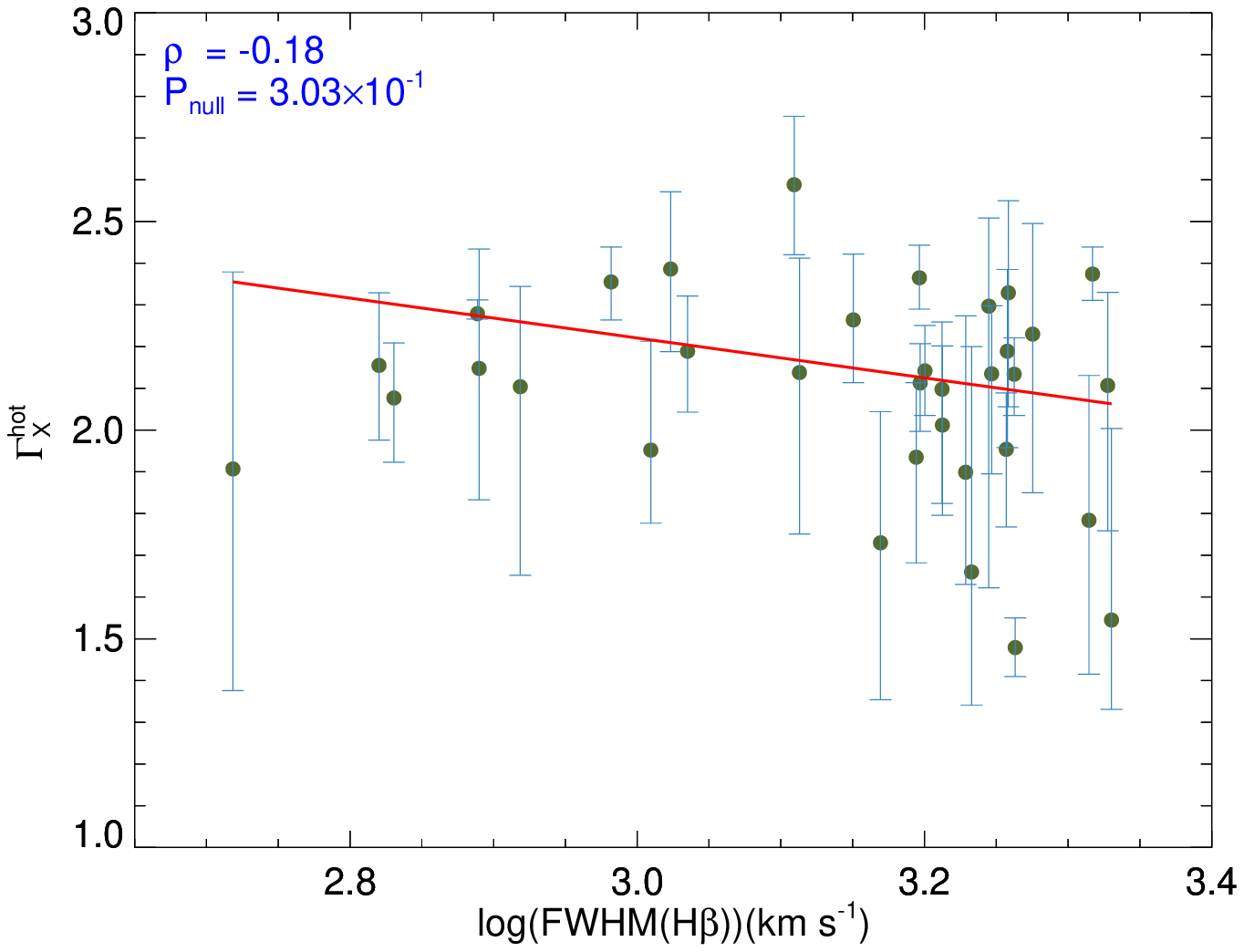} 
  \includegraphics[width=0.5\textwidth,height=0.215\textheight,angle=00]{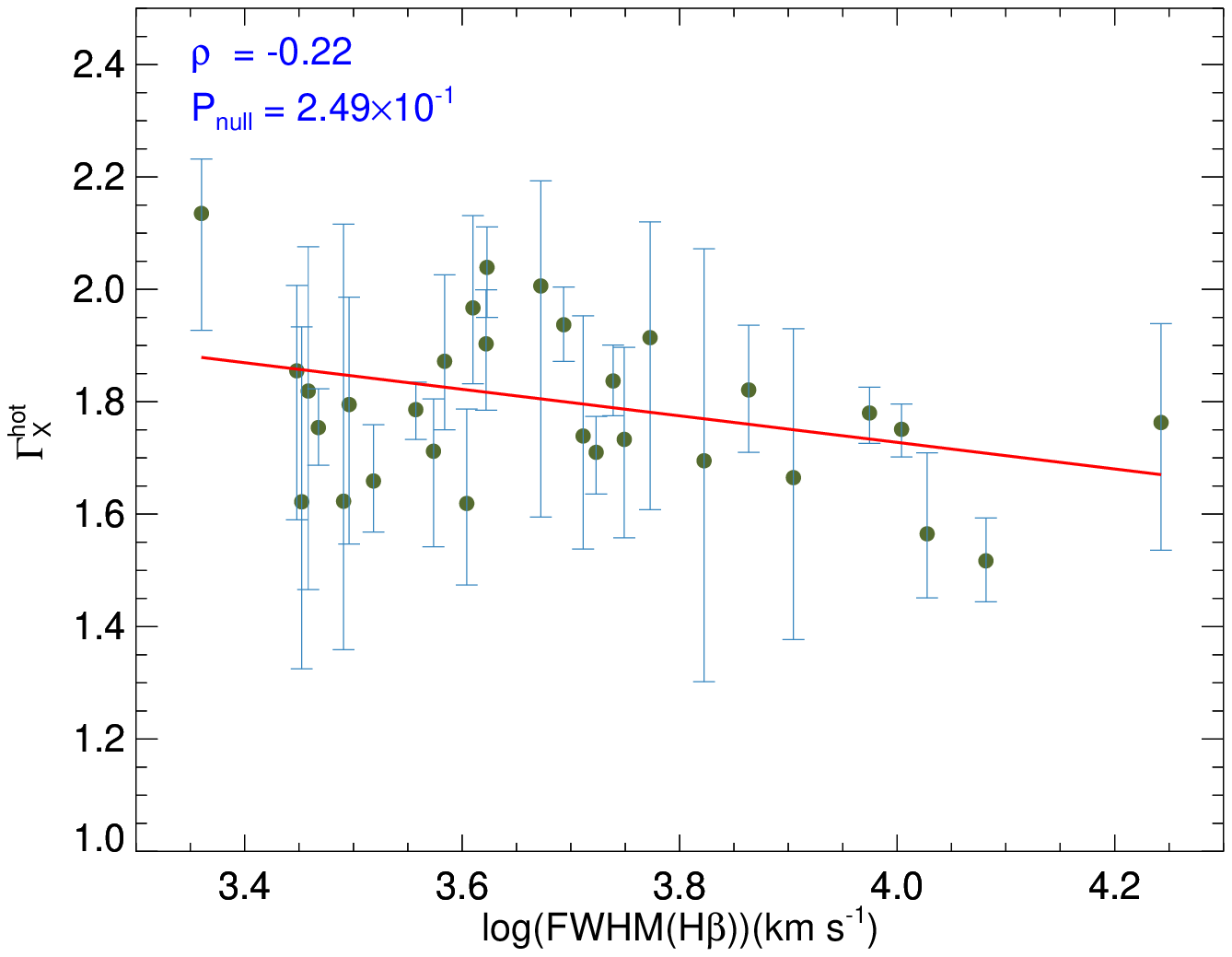} 
  \end{minipage}
  \caption{{\scriptsize Same as Fig.~\ref{fig:nlsy_blsy_corr_rosat_xmm_soft}, but using the spectral fit of  34 NLSy1 and  30 BLSy1 galaxies based on the {\it AGNSED} model for the X-ray hot photon indices ($\Gamma_{X}^\mathrm{hot}$) in the 0.3-10 keV energy band of the {\it XMM-Newton} data.}}
  
 \label{fig:nlsy_blsy_corr_xmm_hard_new_model_gamaa_hot_agnsed}
\end{figure*}

As detailed in Sect.~\ref{section 4.3} that  we could achieve spectral fit in 0.3-10 keV energy band
for 34 NLSy1 and 30 BLSy1 galaxies, based on the {\it `AGNSED'} model. This allowed us to estimate the hot spectral indices ($\Gamma_{X}^\mathrm{hot}$) for these sources along with its correlations with
R$_{\mathrm Edd}^{\mathrm SED}$, L$_{\mathrm bol}$, M$_{\mathrm BH}$ and FWHM(H$\beta$).
 The plots  of these correlations analysis
 are shown in Fig.~\ref{fig:nlsy_blsy_corr_xmm_hard_new_model_gamaa_hot_agnsed} and the results are listed 
in the upper part of Table~\ref{table_nlsy1_blsy1_corr_Gamma_hot_warm}.
From this table and the figure, it is clear that the correlations of 
$\Gamma_{X}^\mathrm{hot}$ with other parameters of nuclear activities are almost similar
to that of correlations obtained in the 0.1-2.0 keV energy band for the
NLSy1 (221 sources), BLSy1 (154 sources) and [NLSy1$+$BLSy1] (375 sources) galaxies.\par

\begin{figure*}[!t]
  \begin{minipage}[]{1.0\textwidth}
  
  \includegraphics[width=0.5\textwidth,height=0.275\textheight,angle=00]{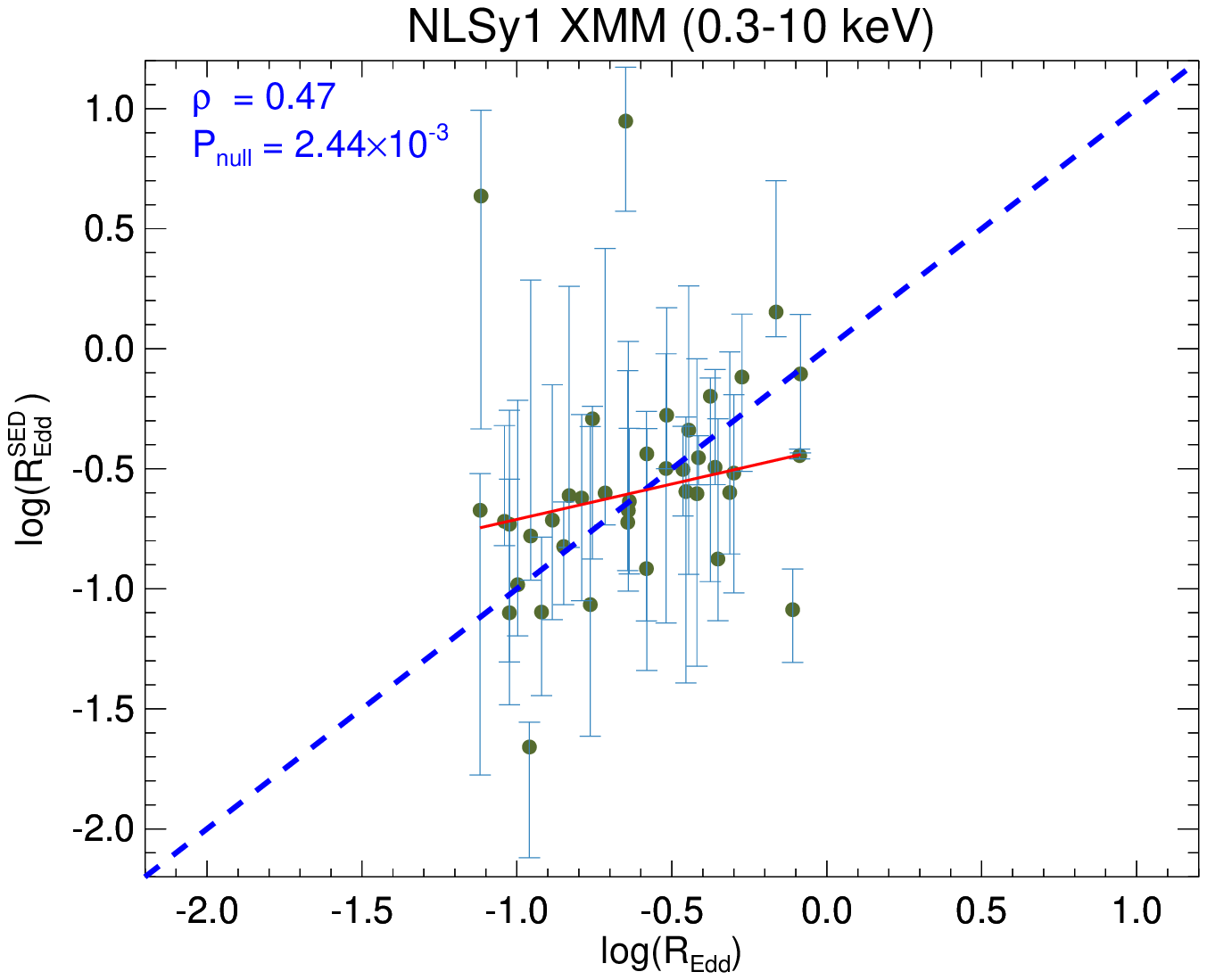} 
  \includegraphics[width=0.5\textwidth,height=0.275\textheight,angle=00]{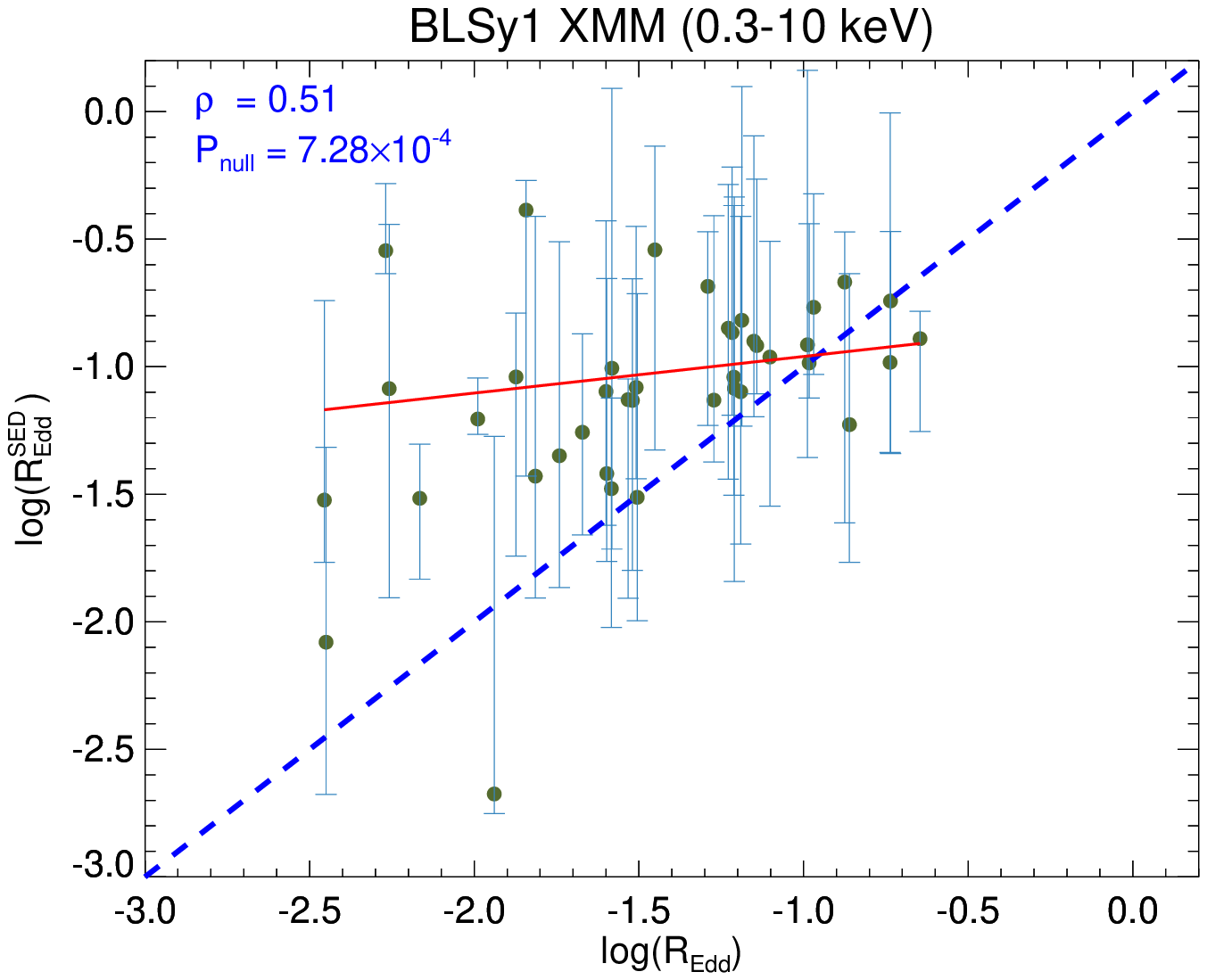} 
    
  \end{minipage}
  \caption{{\scriptsize Correlations of  34 NLSy1s (left) and  30 BLSy1s (right) for the X-ray Eddington ratio (log(R$_{\mathrm Edd}^{\mathrm SED}$)) and optical Eddington ratio (log(R$_{\mathrm Edd}$)) obtained from X-ray fitting in 0.3-10 keV energy band using the {\it AGNSED} model and optical scaling relationship (see, Sect.~\ref{section 4.1}), respectively. The dotted blue line is plotted in each panel following the equation $y=x$.}}  
    
  \label{fig:nlsy_blsy_corr_redd_opt_xray_xmm_hard_new_model_gamaa_hot_agnsed}
\end{figure*}

\begin{figure}
   \begin{minipage}[]{1.0\textwidth}
 
 \includegraphics[width=0.53\textwidth,height=0.30\textheight,angle=00]{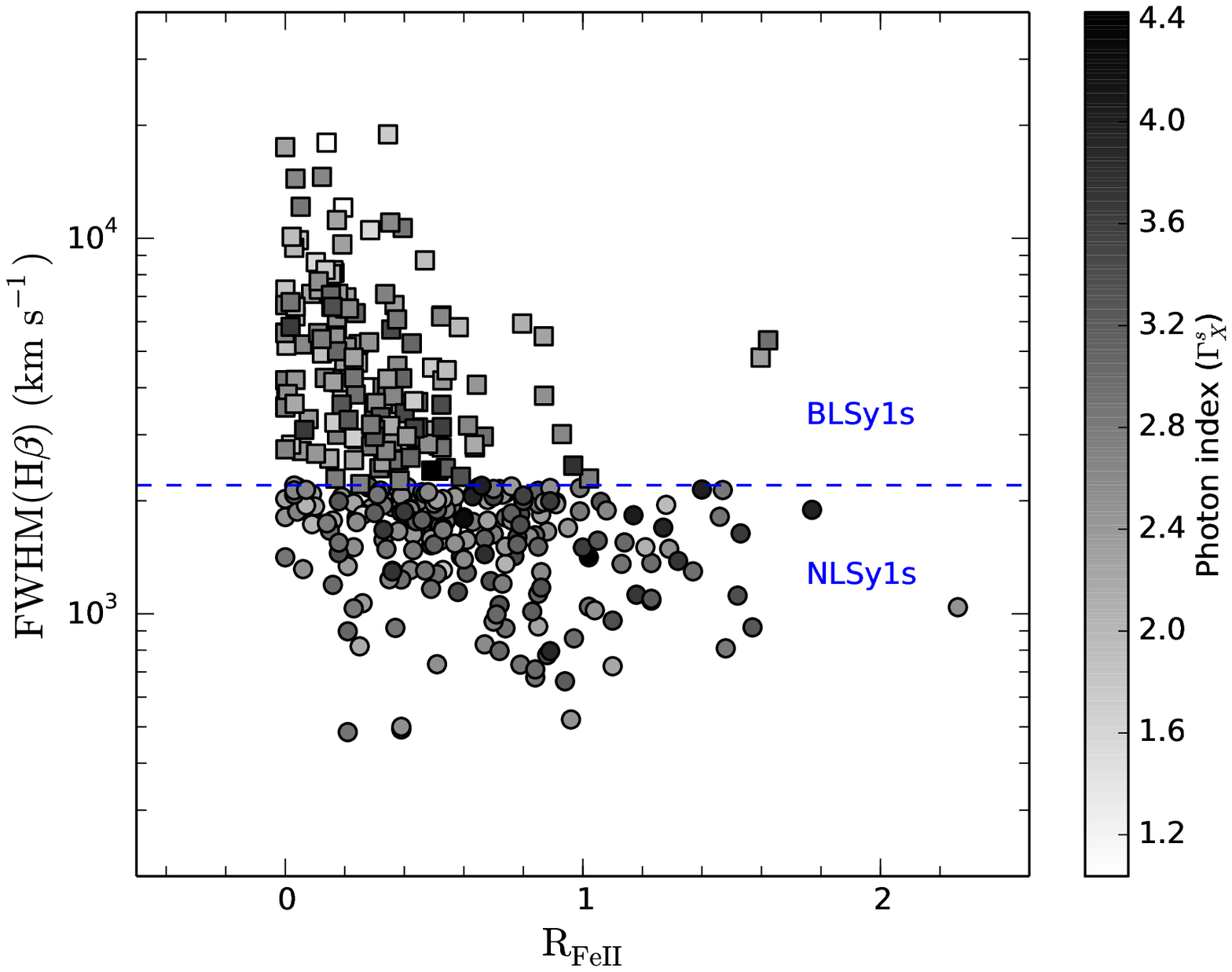}
   \end{minipage}
   \caption{{\scriptsize A plot of Eigenvector 1 EV1 (R$_{\mathrm FeII}$ vs FWHM(H($\beta$))) with space gray codded by the soft (0.1-2.0 keV) X-ray photon indices ($\Gamma_{X}^{s}$) for the  221 NLSy1 (gray filled circle) and  154 BLSy1 (gray filled square) galaxies.}}
 \label{fig:nlsy_blsy_corr_xmm_rosat_color_codded_FeII}
\end{figure}

\section{Discussion and conclusions}
\label{section 6.0}
In order to probe the X-ray emission mechanisms in NLSy1 and BLSy1
galaxies, a correlation study among X-ray spectral indices and parameters
of nuclear activity such as  R$_{\mathrm Edd}$, L$_{\mathrm bol}$, M$_{\mathrm BH}$, and FWHM(H$\beta$)
would be very important. 
For instance,~\citet{Brandt1997MNRAS.285L..25B} analyzed an Advanced Satellite for Cosmology and Astrophysics (ASCA)
sample of 15 NLSy1 and 19 BLSy1 galaxies, for the comparison of their
hard X-ray (2-10 keV) photon indices and found that NLSy1s have
steeper intrinsic hard X-ray photon indices than the BLSy1s. Here, we
have extended similar work by studying the soft (0.1-2.0 keV),
hard (2-10 keV) as well as  total (0.3-10 keV)  photon indices (i.e., $\Gamma_{X}^{s}$,
$\Gamma_{X}^{h}$ \& $\Gamma_{X}^{T}$) of NLSy1 and BLSy1 galaxies.  For this, we have
constructed their samples based on the recent large catalog of 11,101
NLSy1s and their redshift matched sample of BLSy1s using their X-rays data
from the {\it ROSAT} and {\it XMM-Newton} (e.g. Sect.~\ref{section
  2.0}). Our sample consist of  221 NLSy1,  154 BLSy1 galaxies in soft (0.1-2.0 keV),  53
NLSy1,  46 BLSy1 galaxies in the hard (2-10 keV), and  139 NLSy1,  97 BLSy1 galaxies in the total (0.3-10 keV) energy bands  (e.g. Sect.~\ref{section 2.0}). A homogeneous analysis is carried
out for the estimations of $\Gamma_{X}^{s}$, $\Gamma_{X}^{h}$, $\Gamma_{X}^{T}$ and other
parameters of nuclear activities such as R$_{\mathrm Edd}$,
L$_{\mathrm bol}$, M$_{\mathrm BH}$, and FWHM(H$\beta$) of the NLSy1
and BLSy1 galaxies. This homogeneous analysis is carried out to
perform a comparative study between these two subclasses of AGN along
with a comparison of them with other classes of luminous AGNs in
soft, hard, and total X-ray energy bands.\par
The advantages of our analysis
are that we have used an enlarged sample of NLSy1s (e.g., see
Table~\ref{sample_summary}).
For comparison, we have used a control sample of BLSy1s, matching (moderately) in the redshift plane (e.g., see
Fig.~\ref{fig:z_lum_match_xmm_rosat} and Sect.~\ref{section 2.0}).  Furthermore, in our analysis,
to compute the soft, hard and  total energy X-ray photon indices, we have used
similar models (mostly) in the soft, hard and total energy X-ray bands.  This extra caution is taken care in our method so as to avoid the variations in the
estimated $\Gamma_{X}^{s}$, $\Gamma_{X}^{h}$ and $\Gamma_{X}^{T}$  due to the use of
different spectral fitting models, as has been the case in many
previous studies as mentioned in Sect.~\ref{section 1.0}.\par
The main results of our systematic homogeneous analysis presented here are
as follows. Firstly, we found a clear significant difference among the
$\Gamma_{X}^{s}$ distribution of NLSy1 and BLSy1 galaxies (being
steeper for the NLSy1 class, e.g., see Fig.~\ref{fig:histo_cpdf_xmm_rosat}) with median values of  2.81 and  2.30 for
the samples of the NLSy1 and BLSy1 galaxies, respectively, having
P$_{null}$ of  4.02$\times10^{-19}$ based on the K-S test.
One reason for this observed difference among the $\Gamma_{X}^{s}$ distribution of NLSy1 and BLSy1 galaxies could be more soft X-ray
excess in NLSy1s.
To lift this degeneracy, we have compared $\Gamma_{X}^{h}$, which are thought to be
free from the soft X-ray excess~\citep[e.g.,
  see][]{Boller1996A&A...305...53B, Brandt1997MNRAS.285L..25B,
  Vaughan1999MNRAS.309..113V,Boller2002MNRAS.329L...1B,
  Czerny2003A&A...412..317C, Vignali2004MNRAS.347..854V} between the
subsamples of  53 NLSy1 and  46 BLSy1 galaxies based on their 2-10 keV {\it XMM-Newton} observations.
In this comparison also we find a  difference in  $\Gamma_{X}^{h}$ with median values of 2.06 and  1.78, having P$_{null}$
of 1.00$\times10^{-3}$ for the subsamples of NLSy1 and BLSy1
galaxies, respectively (e.g., see Fig.~\ref{fig:histo_cpdf_xmm_hard}). 
This confirms that the above result of the
difference in $\Gamma_{X}^{s}$ distribution is unlikely to be solely due to the
soft X-ray excess and rather seems to be intrinsic in their nature.
Furthermore, we noticed that the difference in the median photon indices of hard energy band subsamples of NLSy1 \& BLSy1 galaxies is statistically weaker than the soft energy band samples of NLSy1 and BLSy1 galaxies. This may be due to comparatively about 4 times smaller hard band subsamples of NLSy1 \& BLSy1 galaxies. To lift this degeneracy, we have compared the $\Gamma_{X}^{T}$ between the
subsamples of 139 NLSy1 and 97 BLSy1 galaxies. We again find a significant difference with median values of 2.53 and 2.13, respectively, having $P_{null}$ of 4.50$\times10^{-9}$ which is consistent with the result based on soft X-ray analysis of these NLSy1 and BLSy1 galaxies.\par
Secondly, to ascertain whether there is a bimodality or continuity in
X-ray spectral indices (hereafter $\Gamma_{X}$ will be referring to $\Gamma_{X}^{s}$, $\Gamma_{X}^{h}$ and $\Gamma_{X}^{T}$)
among NLSy1 and BLSy1 galaxies. A detailed correlation analysis of $\Gamma_{X}$
with other physical parameters of AGNs such as R$_{\mathrm Edd}$,
L$_{\mathrm bol}$, M$_{\mathrm BH}$, and FWHM(H$\beta$) is carried
out. This correlation analysis results in the strongest $\Gamma_{X}-\log(R_{\mathrm Edd})$ correlation
for the samples of NLSy1 and BLSy1 galaxies (e.g. Sect.~\ref{section 5.2} and Table~\ref{table_nlsy1_blsy1_corr}),
implying that R$_{\mathrm
  Edd}$ may be the dominant parameter related to $\Gamma_{X}$.
Additionally, the joint analysis of [NLSy1$+$BLSy1] shows that the
variation seems to be continuous rather than a clear significant
bimodality in its distribution (e.g., see last column of
Figs.~\ref{fig:nlsy_blsy_corr_rosat_xmm_soft},~\ref{fig:nlsy_blsy_corr_xmm_hard}~\&~\ref{fig:nlsy_blsy_corr_xmm_total_energy}). This
is also evident in their joint histogram plots of $\Gamma_{X}$ which do not show two
well-separated significant peaks (e.g., see histogram of joint (NLSy1+BLSy1) distribution in
Figs.~\ref{fig:histo_cpdf_xmm_rosat},~\ref{fig:histo_cpdf_xmm_hard}~\&~\ref{fig:histo_cpdf_xmm_total}).
Furthermore, the similarity of the trends and value of Spearman's
correlations for $\Gamma_{X}-\log(R_{\mathrm Edd})$ correlation found
for NLSy1 and BLSy1 galaxies in the soft, hard and total X-ray energy
bands (e.g., see Table~\ref{table_nlsy1_blsy1_corr}) also suggest that
their emission mechanism may be similar.  However, the slopes of the
linear fit of $\Gamma_{X}$ and log(R$_{\mathrm Edd})$ correlations do
differ significantly among the samples of NLSy1 and BLSy1 galaxies
 in soft, hard and total X-ray energy bands (e.g., see
Table~\ref{table_nlsy1_blsy1_corr}) which could probably be due to the
difference in their accretion rates, being higher for the former. \par
We explored this possibility by comparing the distribution of
R$_{\mathrm Edd}$ of NLSy1 and BLSy1 galaxies.
This has resulted in the median values of R$_{\mathrm Edd}$, 0.23 and 0.05 for
221 NLSy1 and 154 BLSy1 galaxies, respectively, in the soft X-ray. 
The R$_{\mathrm Edd}$ distribution differs significantly for the above samples of NLSy1 and
BLSy1 galaxies with K-S test based $P_{null}= 2.66\times10^{-35}$. The above result also holds when we compare
  in 0.3-10 keV energy band. The distributions of R$_{\mathrm Edd}$ of 139 NLSy1 and  97 BLSy1 galaxies, having median values of 0.22 and  0.04 respectively, and found that their K-S test based $P_{null}$ is 3.72$\times10^{-26}$. Similarly, using the analysis of 53 NLSy1 and  46 BLSy1 galaxies in hard energy band, the median values of R$_{\mathrm Edd}$ are found to be 0.25 and 0.03, respectively, resulting in K-S test based $P_{null}$ of 1.72$\times10^{-14}$. In view of the above, negligible dependence of the $P_{null}$ values on the energy bands used in the analysis, allows to concluded that the R$_{\mathrm Edd}$ of NLSy1 and BLSy1 galaxies seem to be  intrinsically significantly different, being it higher for the former.\par
To reconcile this discrepancy one possibility is that the higher R$_{\mathrm Edd}$
in NLSy1s (compared to BLSy1s) can be due to the fact that the inclination
angle of the NLSy1 is lower than that of the BLSy1.  As a result, the
observed FWHM (in km s$^{-1}$) of H$\beta$ line (FWHM$\times sin(\theta$)) of the BLR would have been
underestimated more in the case of NLSy1 (due to smaller inclination)
compared to the BLSy1 ~\citep[e.g., see][]{Baldi2016MNRAS.458L..69B,
  Liu2016IJAA....6..166L, Rakshit2017ApJS..229...39R}. This will
directly impact the under-estimation of their M$_{\mathrm BH}$ (being
proportional to the observed FWHM in the common L-R$_{BLR}$ scaling relationship) and
hence the over-estimation of the R$_{\mathrm Edd}$ value (being inversely
proportional to M$_{\mathrm BH}$). To consider such projection effect of BLR for the NLSy1 and BLSy1
galaxies in our analysis, we corrected the observed FWHM (i.e., projected) values as
$FWHM/sin(\theta)$, by using the median viewing angle ($\theta$)
of 13.6$^{\circ}$ and 27.7$^{\circ}$ as given
by~\citet{Liu2016IJAA....6..166L} for the NLSy1s and BLSy1s,
respectively. This has allowed us to have corrected M$_{\mathrm BH}$ and R$_{\mathrm Edd}$
for our samples of NLSy1 and BLSy1 galaxies. The corrected R$_{\mathrm Edd}$ values still show a difference
(though with less statistical significance) among its
distribution in 221 NLSy1 and 154 BLSy1 galaxies
with K-S test based $P_{null}=3.45\times10^{-2}$. The difference still exists when we even use only the subsamples of 139 NLSy1 and 97 BLSy1 galaxies analyzed in 0.3-10 keV  energy range, giving $P_{null}=2.18\times10^{-2}$.
Here, a possibility also exists that it may also be due to the imperfection of exact
luminosities matching and/or due to our application of average
inclination angle value for the entire sample. Nonetheless, given
that the mismatch in luminosities in our sample is nominal, and the
fact that the R$_{\mathrm Edd}$ difference being very significant, suggest that it
is intrinsically higher in NLSy1 compare to BLSy1.
This could lead to the above measured differences in $\Gamma_{X}^{s}$,
$\Gamma_{X}^{h}$, $\Gamma_{X}^{T}$ distributions and the slopes of the linear fit of the
$\Gamma_{X}$ and $\log(R_{\mathrm Edd})$ correlations.\par
Another possibility for the above difference in the R$_{\mathrm Edd}$ can be
due to under/over estimations of bolometric luminosity, which was
estimated using the scaling relationship of L$_{\mathrm bol}$ and optical
luminosity at 5100 \AA~(e.g. Sect.~\ref{section 4.0}).
To quantify its effect, we have also estimated the R$_{\mathrm Edd}$ values independently by fitting the {\it AGNSED}
model over the 0.3-10 keV band (i.e., R$^{\mathrm SED}_{\mathrm Edd}$)
for the 34 NLSy1 and 30 BLSy1 galaxies subsamples (e.g. Sect.~\ref{section 4.3}). The
resulting distribution of R$^{\mathrm SED}_{\mathrm Edd}$ (e.g., see
Fig.~\ref{fig:histo_cpdf_xmm_AGNSED}) have also
shown a significant difference (with
P$_{null}$= 2.01$\times10^{-7}$) in the subsamples of these 34 NLSy1 and  30 BLSy1 galaxies.
This is consistent with  the conclusion drawn using R$_{\mathrm Edd}$  
distribution, based on L$_{\mathrm bol}$ estimated using the optical spectra (e.g., see Figs.~\ref{fig:histo_cpdf_xmm_rosat},~\ref{fig:histo_cpdf_xmm_hard},~\ref{fig:histo_cpdf_xmm_total}). This is not surprising because there is a significant correlation between R$_{\mathrm Edd}$ (i.e., using optical)
and R$_{\mathrm Edd}^{\mathrm SED}$ (i.e., using X-ray)
for the NLSy1 and BLSy1 galaxies as shown in Fig.~\ref{fig:nlsy_blsy_corr_redd_opt_xray_xmm_hard_new_model_gamaa_hot_agnsed} and tabulated in the last row of Table~\ref{table_nlsy1_blsy1_corr_Gamma_hot_warm}. Furthermore,
the histograms of spectral indices estimated using {\it AGNSED} model
($\Gamma^{\mathrm SED}_{X}$, also presented as $\Gamma^{\mathrm hot}_{X}$ in Fig.~\ref{fig:nlsy_blsy_corr_xmm_hard_new_model_gamaa_hot_agnsed}) also show a significant difference among NLSy1 and BLSy1
galaxies with P$_{null}$ of 8.87$\times10^{-6}$ (e.g., see
Fig.~\ref{fig:histo_cpdf_xmm_AGNSED}). In addition to this, a correlation between
$\Gamma^\mathrm{hot}_{X}$ and $\log(R^{\mathrm SED}_{\mathrm Edd}$) is also found in the above subsamples of NLSy1 and BLSy1 galaxies
(e.g., see Fig.~\ref{fig:nlsy_blsy_corr_xmm_hard_new_model_gamaa_hot_agnsed}) which is similar to the correlations found in 
the soft, hard and total energy bands between $\Gamma_{X}$ and R$_{\mathrm Edd}$.\par
Our above investigations suggest that the R$_{\mathrm Edd}$ of
NLSy1 is unambiguously higher than that of BLSy1 galaxy. This intrinsic difference can explain
our observed significant difference of spectral indices among NLSy1 and BLSy1
galaxies as follow.\par
The higher value of R$_{\mathrm Edd}$ can lead to an increase in the disk
temperature, hence producing more X-ray radiations, and at the same
time, it can also increase the Compton cooling of
corona~\citep{Haardt1991ApJ...380L..51H, Haardt1993ApJ...413..507H,
  Zdziarski2000ApJ...542..703Z, Kawaguchi2001ApJ...546..966K}, which
leads to steepening of the X-ray power-law more in NLSy1 than BLSy1, and hence will
lead to the observed difference we noticed in our spectral indices
in the soft, hard and total 0.3-10 keV energy bands (e.g. Sect.~\ref{section
  5.1}).  Moreover, this also could be the reason for the observed
higher slope of $\Gamma_{X}$ and $\log(R_{\mathrm Edd})$ linear fit for NLSy1
compared to the BLSy1 galaxies (e.g., see
Table~\ref{table_nlsy1_blsy1_corr}), as such stronger dependence in
 case of NLSy1 can be reconciled due to their higher R$_{\mathrm Edd}$ value.\par

As pointed out in Sect.~\ref{section 1.0}, such positive
$\Gamma_{X}^{h} -\log(R_{\mathrm Edd}$) correlation (e.g., see
Table~\ref{table_nlsy1_blsy1_corr}) has also been found for the
luminous AGNs by~\citet{Risaliti2009ApJ...700L...6R}. It may be noted
that they found a $\rho$ of 0.32 based on their sample of 343
AGNs. However, this correlation becomes stronger with the value of
$\rho$ = 0.56 when considering only their subset of 82 objects 
whose black hole masses were estimated using their H$\beta$ lines. This is almost similar
to the Spearman's correlation coefficients found for the
$\Gamma_{X}^{s} - \log(R_{\mathrm Edd}$), $\Gamma_{X}^{h} - \log(R_{\mathrm Edd}$) and
$\Gamma_{X}^{T} - \log(R_{\mathrm Edd}$) correlations in our samples of 
the NLSy1 and BLSy1 galaxies (e.g., see Table~\ref{table_nlsy1_blsy1_corr}). 
On the other hand, the striking contrast to this positive correlation as compared to the corresponding
negative correlation found for the case of LLAGNs
by~\citet{Gu2009MNRAS.399..349G} suggests that emission mechanism in
the NLSy1 and BLSy1 galaxies is different as compared to the LLAGNs but
likely to be similar to the luminous AGNs. \par
Finally, we have also explored the correlation of X-ray spectral slopes (in soft energy band) with that of the
optical plane of Eigenvector 1 (EV1) which is mainly defined by FWHM(H$\beta$)
and the flux ratio of Fe~{\sc ii}  to H$\beta$, R$_{\mathrm FeII}$~\citep {Sulentic2000ApJ...536L...5S}. The FWHM(H$\beta$) is
known to be affected by the inclination angle while the R$_{\mathrm FeII}$ is
driven by the Eddington ratio~\citep{Shen2014Natur.513..210S}. The
R$_{\mathrm FeII}$ values of NLSy1 galaxies are taken from the parent catalog of
NLSy1 galaxies given by~\citet{Rakshit2017ApJS..229...39R} and for the BLSy1
galaxies, it is estimated following the similar procedure as used
in~\citet{Rakshit2017ApJS..229...39R} for the NLSy1 galaxies. In
Fig.~\ref{fig:nlsy_blsy_corr_xmm_rosat_color_codded_FeII}, we have
plotted these quantities color-coded by the $\Gamma_{X}^{s}$. We find
strong Fe~{\sc ii} emitters to have steeper photon indices compared to the
week Fe~{\sc ii} emitters. The Spearman's rank correlation coefficients
between R$_{\mathrm FeII}$ and $\Gamma_{X}^{s}$ are found to be
0.31 and 0.24 for the samples of NLSy1 and BLSy1 galaxies while 0.46 when
both NLSy1 and BLSy1 samples are combined together. This positive
correlation between $\Gamma_{X}^{s}$ and R$_{\mathrm FeII}$ reflects the
strong correlation found between photon indices and Eddington
ratios. Moreover,
Fig.~\ref{fig:nlsy_blsy_corr_xmm_rosat_color_codded_FeII} also corroborate the 
strong anticorrelation between photon indices and FWHM(H$\beta$)
found in the
joint analysis of NLSy1 and BLSy1 samples (e.g. Sect.~\ref{section 5.2} and Figs.~\ref{fig:nlsy_blsy_corr_rosat_xmm_soft},~\ref{fig:nlsy_blsy_corr_xmm_hard},~\ref{fig:nlsy_blsy_corr_xmm_total_energy}).  In view of such a strong
anti-correlation, we may recall
that $\Gamma_{X}$ being related to the X-ray emitting regions
is generally much closer to the central engine of AGN as compared to
BLR clouds whose broadening is measured as the FWHM of H$\beta$ lines. As a result,
such strong anticorrelation between these two seemingly
disconnected regions are worth noting while constructing any models of
the AGNs emission mechanisms along with the observed strong
correlation of the $\Gamma_{X} -\log(R_{\mathrm Edd}$) found both for NLSy1 and
BLSy1 galaxies.\par

\section {SUMMARY}
\label {section 7.0}
In this work, we have quantitatively compared NLSy1 and BLSy1 galaxies
for their X-ray and optical properties along with different
correlations among X-ray and optical parameters, such as
  $\Gamma_{X} - \log(R_{\mathrm Edd}$), $\Gamma_{X} - \log(L_{\mathrm
    bol}$), $\Gamma_{X} - \log(M_{\mathrm BH}$) and $\Gamma_{X}
  -\log(FWHM(H\beta$)) with $\Gamma_{X}$ representing X-ray spectral
  slope either in soft (i.e.  $\Gamma_{X}^{s}$ in 0.1-2.0 keV), in
  hard (i.e.  $\Gamma_{X}^{h}$ in 2-10 keV) and in total (0.3-10 keV) energy bands (i.e. $\Gamma_{X}^{T}$).  For these, we used
  the samples of 221 NLSy1 and  154 BLSy1 galaxies in the soft
    X-ray energy (0.1-2.0 keV) band, while its subsamples of  53 NLSy1, 46 BLSy1 and  139
    NLSy1,  97 BLSy1  galaxies are used also in
    the  hard X-ray and total energy bands, respectively. The summary
  of our main results is as follows.\par

(i) We found the existence of difference in $\Gamma_{X}^{s}$
distribution among the NLSy1 and BLSy1 galaxies, being steeper for the
former in the soft X-ray energy band. Furthermore, the difference is
also found when spectral indices of these two subsets are  compared in
hard and total energy bands (i.e., $\Gamma_{X}^{h}$ and $\Gamma_{X}^{T}$).  In view of the fact that the hard
energy band is generally less prone to the impact of soft X-ray excess, it suggests
that soft X-ray excess is not the main cause of the difference seen here among the
$\Gamma_{X}$ found for the NLSy1 and BLSy1 galaxies.\par

(ii) We found a clear significant difference in R$_{\mathrm Edd}$ among NLSy1 and
BLSy1 galaxies, being larger for the former with P$_{null}$ of
2.66$\times10^{-35}$.  This difference exists even after incorporating
any inclination angle difference among these two subclasses, though with
less significance.  Furthermore, this discrepancy (based on R$_{\mathrm Edd}$ from the optical data) is also reconfirmed
even when we estimated R$_{\mathrm Edd}$ independently based on SED fitting ({\it AGNSED}
model) in the X-ray 0.3-10 keV band (e.g. Sect.~\ref{section 4.3}). This suggests that R$_{\mathrm Edd}$ of the NLSy1 is
intrinsically higher than the BLSy1 galaxy which can be the main reason for
the observed significant difference in $\Gamma_{X}$ among NLSy1 and BLSy1 galaxies
 in soft, hard and  total  energy bands.\par

(iii) Our analysis suggests a significant positive correlation between
 $\Gamma_{X}$ and log(R$_{\mathrm Edd}$) for the samples of NLSy1 and BLSy1
 galaxies in soft, hard and  total energy bands, with stronger
 dependence in the case of NLSy1s. Also, these strong correlations
 between $\Gamma_{X}$ and log(R$_{\mathrm Edd}$) for the NLSy1 and BLSy1
 galaxies show that their X-ray slope can be used as an Eddington
 ratio estimator, which can then also be used to calculate black hole
 mass for a given bolometric luminosity estimate.\par

(iv) Overall correlations such as $\Gamma_{X}-\log(R_{\mathrm Edd}$) have almost similar trends among the NLSy1
 and BLSy1 galaxies.  These correlations are found also consistent
 (qualitatively) with the luminous AGNs (at least in hard X-ray),
 apart from their higher significance compare to NLSy1 and BLSy1
 galaxies. This gives support to the theoretical prediction that, the
 X-ray emissions may also be produced in NLSy1 and BLSy1 galaxies by the
 disc-corona system as proposed for the case of luminous AGNs.
 This model is also consistent with the steeper $\Gamma_{X}$ we have found in our samples of NLSy1 in comparison to the BLSy1 galaxies.\par  
 
  \vspace{0.65cm} We thank the anonymous referee for the constructive
  comments on our manuscript. We gratefully acknowledge Mainpal Ranjan,
   Jeewan C. Pandey and Xinwu Cao for their very useful discussions.

\bibliography{references}{}
\label{lastpage}
\end{document}